\documentclass[format=acmsmall, review=false, screen=true]{acmart}

\acmJournal{JACM}
\acmVolume{00}
\acmNumber{0}
\acmArticle{00}
\acmYear{2019}
\acmMonth{0}
\copyrightyear{2019}
\acmArticleSeq{0}

\setcopyright{acmlicensed}

\received{January 2019}

\pdfoutput=1

\usepackage{subfig}
\usepackage{tabularx}

\usepackage{amsthm}

\usepackage{xcolor}
\usepackage{listings}

\usepackage{draftwatermark}

\definecolor{mGreen}{rgb}{0,0.6,0}
\definecolor{mGray}{rgb}{0.5,0.5,0.5}
\definecolor{mPurple}{rgb}{0.58,0,0.82}
\definecolor{backgroundColour}{rgb}{0.95,0.95,0.95}

\lstdefinestyle{CStyle}{
    commentstyle=\color{mGray},
    numberstyle=\tiny\color{mGray},
    stringstyle=\color{mPurple},
    xleftmargin=0.07\textwidth,
    xrightmargin=0.07\textwidth,
    rulecolor=\color{red},
    frame=single,
    basicstyle=\ttfamily\tiny,
    breakatwhitespace=false,
    breaklines=true,
    captionpos=b,
    keepspaces=true,
    showspaces=false,
    showstringspaces=false,
    showtabs=false,
    tabsize=2,
    language=C
  }

\lstdefinestyle{BashStyle}{
    commentstyle=\color{mGray},
    numberstyle=\tiny\color{mGray},
    stringstyle=\color{mPurple},
    xleftmargin=0.07\textwidth,
    xrightmargin=0.07\textwidth,
    rulecolor=\color{red},
    frame=single,
    basicstyle=\ttfamily\tiny,
    breakatwhitespace=false,
    breaklines=true,
    captionpos=b,
    keepspaces=true,
    showspaces=false,
    showstringspaces=false,
    showtabs=false,
    tabsize=2,
    language=bash,
    deletekeywords={type, local, bind}
}

\usepackage[final]{fixme}

\fxsetup{layout=inline, theme=color, mode=multiuser}
\FXRegisterAuthor{ds}{ads}{\color{red}Dimitri}
\FXRegisterAuthor{sv}{asv}{\color{red}Sander}

\graphicspath{{fig/}}

\begin{document}



\title[Design of the Ouroboros packet network]{Design of the Ouroboros packet network}

\author{Dimitri Staessens}
\affiliation{%
  \institution{Ghent University - imec} %
  \streetaddress{Technologiepark-Zwijnaarde 126} %
  \city{Ghent} %
  \postcode{B-9052} %
  \country{Belgium}} %
\email{dimitri.staessens@ugent.be} %

\author{Sander Vrijders}
\affiliation{%
  \institution{Ghent University - imec} %
  \streetaddress{Technologiepark-Zwijnaarde 126} %
  \city{Ghent} %
  \postcode{B-9052} %
  \country{Belgium}} %
\email{sander.vrijders@ugent.be} %

\begin{abstract}
  The 5-layer TCP and 7-layer OSI models are taught as high-level
  frameworks in which the various protocols that are used in computer
  networks operate. %

  These models provide valid insights in the organization of network
  functionalities and protocols; however, the difficulties to fit some
  crucial technologies within them hints that they don't provide a
  complete model for the organization of -- and relationships between
  -- different mechanisms in a computer network. %

  Recently, a recursive model for computer networks was proposed,
  which organizes networks in layers that conceptually provide the
  same mechanisms through a common interface. %
  Instead of defined by function, these layers are distinguished by
  scope. %

  We report our research on a model for computer networks. %
  Following a rigorous regime alternating design with the evaluation
  of its implications in an implementation, we converged on a
  recursive architecture, named Ouroboros. %
  One of our main main objectives was to disentangle the fundamental
  mechanisms that are found in computer networks as much as
  possible. %
  Its distinguishing feature is the separation of unicast and
  broadcast as different mechanisms, giving rise to two different
  types of layers. %
  These unicast and broadcast layers can easily be spotted in today's
  networks. %

  This article presents the concepts underpinning Ouroboros, details
  its organization and interfaces, and introduces the free software
  prototype. %
  We hope the insights it provides can guide future network design and
  implementation. %
\end{abstract}

\maketitle

\renewcommand{\shortauthors}{D. Staessens and S. Vrijders}

\section{Introduction}
\label{sec:intro}
The goal of a network architecture is to organize network functions
(mechanisms) in such a way that it reduces complexity, simplifies
implementation, operation and management of networks that are designed
according to it. %
This article provides an overview of a new (recursive) network
architecture, motivates the functional organization into different
components, describes the interfaces between these components and
presents a prototype implementation, including some performance
indicators. %

The first distributed computer communications networks were developed
in the 1960s \cite{baran1964distributed, davies1967digcommnet} and,
following the rise of the ARPAnet in the early 1970s, structured
research into internetwork architectures (networks that interconnect
smaller networks) started with the formation of the International
Packet Network Working Group (INWG) \cite{mckenzie2011inwg} - later
designated the International Federation for Information Processing
(IFIP) Working Group 6.1 (WG6.1) \cite{day2016clamor}. %
The final proposal for INWG \cite{cerf1976inwg96} was converging on 3
layers, each with their own transport protocol that included
addressing. %
In the 1980s and into the 1990s, the Open Systems Interconnect (OSI)
7-layer reference model was developed \cite{zimmermann1980osi,
  bachman1982osi, day1983osi, russell2013internetwasnt}. %
The only packet-switched architecture for global networks that is
widely adopted and deployed today is a 5-layer model that underpins
the Internet, functionally defined through the Transmission Control
Protocol / Internet Protocol (TCP/IP) protocol suite
\cite{cerf1983dod, rfc791, rfc793}. %

In contrast to the proposal from INWG, the current Internet follows a
{\em catenet} model using gateways within a single global network
address space \cite{pouzin1974catenet, cerf1978catenet,
  postel1981arpa}. %
Some of the current research questions relating to the TCP/IP Internet
hint at inefficiencies at an architectural level. %
For instance, when TCP segments are sent that are large enough to
require IP fragmentation, the loss of any IP fragment requires the
entire TCP segment to be retransmitted \cite{kent1995harmful}. %
Congestion control is usually implemented in Transport Layer (L4)
protocols \cite{rfc2914}, such as TCP \cite{jacobson1988cac}, Stream
Control Transmission Protocol (SCTP) \cite{rungeler2009ccsctp}, and
Datagram Congestion Control Protocol (DCCP) \cite{rfc4340}. %
This leaves the IP network vulnerable to congestion \cite{nagle1984cc}
caused by (malicious) programs that use L4 protocols that do not
implement congestion control such as the User Datagram Protocol (UDP)
\cite{rfc768}. %
QUIC \cite{langley2017quic, ietf2018quicloss}, which is a reliable
transport protocol that runs on top of UDP, also implements congestion
control. %
Application programming interfaces (APIs) for networking, such as the
{\em socket()} interface \cite{opengroup2017posix}, require quite some
knowledge from the programmer on the protocols used. %
The {\em socket()} interface also requires an update whenever a new
protocol is developed \cite{rfc2553}, and end-user programs that use
the {\em socket()} interface will need to be adapted to be able to use
any new protocol. %
Research into more generic programming interfaces for transport
networks is ongoing: $\emptyset$MQ \cite{hintjens2013zeromq} has been
quite succesful as a message interface, and the NEAT project
\cite{khademi2017neat} and the IETF Transport Services (taps) Working
Group \cite{ietf2018tapsarch} are actively working towards this
goal. %

\begin{figure}[ht]
  \centering
  \Description{Stacking of layers for different network
    architectures}
  \includegraphics[width=0.9\textwidth]{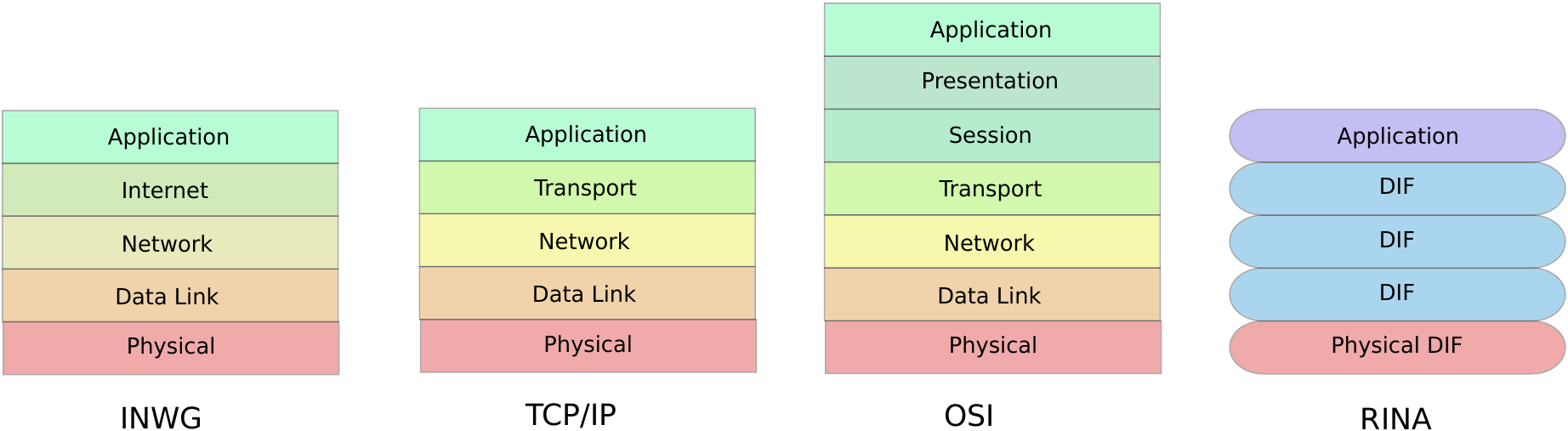}
  \caption{Layers in network architectures}
  \label{fig:layering}
\end{figure}

Recent research into network architectures led to the {\em Recursive
  InterNetwork Architecture} (RINA) \cite{day2008pna, trouva2011rina},
which provides a number of key insights on packet networks. %
RINA reaffirms that {\em communication endpoints are processes}, an
observation that was also made by early designers of packet networks
\cite{carr1970hosthost, cerfkahn1974tcp}. %
It champions a {\em separation of mechanism and policy}
\cite{hansen1970nucleus}, separating {\em what} is done from {\em how}
it is done \cite{dijkstra1969programming}. %
In contrast to defining layers by function \cite{dijkstra1968the,
  day2011lostlayer}, RINA follows an elegant%
\footnote{The lack of an objectively quantifiable measure for elegance
  may be one of the biggest tragedies in science and engineering. %
  This is particularly true for computer science.} %
layering paradigm consisting of {\em network layers of identical
  mechanisms but different policies and/or scope}, called Distributed
Inter-Process Communication (IPC) Facilities or DIFs
\cite{day2008ipc}. %
It defines the concept of a {\em flow} as the loci of resource
allocation necessary to allow the transfer of datagrams from source to
destination, rather than identifying a flow as a sequence of packets
between a source and destination \cite{clark1988darpadesign,
  tanenbaum2010cn}. %
The recursive structure necessitates a {\em common API} to all layers,
which -- since the scope of a layer spans everything from a single
machine to a complete Internet -- blurs the line between IPC and
networking. %
Other architectures which follow a similar high-level structure are
the {\em Recursive Network Architecture} (RNA) \cite{touch2006rna,
  touch2008rnametaprotocol} and the Dynamic Recursive Unified Internet
Design (DRUID) \cite{touch2011druid}. %

A brief comparison of layering in different architectures is shown in
Fig. \ref{fig:layering}. %
Throughout this text, layers that are defined by function (protocol)
are drawn as boxes, while layers defined by scope are drawn as
ovals. %

Ouroboros%
\footnote{The name comes from the recursiveness of the architecture
  and the ubiquitous use of ring buffers in the prototype
  implementation. %
  A ring buffer resembles a snake eating its own tail.} %
is a novel (recursive) network architecture that incorporates the
experience we gained during research collaborations on RINA
\cite{vrijders2014irati}, as well as a thorough revisiting of a lot of
early network research literature that formed the origins for most
computer networking concepts. %
From a high level perspective, Ouroboros looks very similar to RINA,
as evidenced by the Rumba \cite{ vrijders2018rumba} framework, which
can be used to deploy the IRATI \cite {grasa2013irati} and rlite
\cite{maffione2018rlite} RINA implementations as well as our Ouroboros
prototype \cite{staessens2017ouroboros}. %
The system research approach \cite{pike2000sri} to building Ouroboros
adopts the UNIX design philosophy where {\em each program does one
  thing well} \cite{mcilroy1978unix}. %
The end result stems from a number of iterations alternating design
and implementation. %
When compared to RINA, the organization of the mechanisms in the
layers is very different in Ouroboros. %
This warrants considering RINA and Ouroboros as different
architectures. %
We will comment on the most important differences where appropriate and
useful. %

The main objectives for the Ouroboros architecture are as follows. %
First of all, it should have {\em no external assumptions} on any
hardware or operating system environment. %
Second, we aim to {\em reduce protocol headers} to an absolute
minimum%
\footnote{{\em ``In protocol design, perfection has been reached not
    when there is nothing left to add, but when there is nothing left
    to take away''} \cite{rfc1925}.}. %
Finally, we aim for {\em simple and expressive APIs}, hiding
complexity as much as possible. %

We will now outline the organization of this article, highlighting our
contributions. %
Given that this article details a complete system, it necessarily
touches many aspects of computer networks. %
Core concepts that are needed are detailed in Sec. \ref{sec:concepts},
starting off with some basic definitions from graph theory
(Sec. \ref{ssec:graphs}) and computing (Sec. \ref{ssec:progdef}),
which are used extensively in the remainder of the text. %
Next, we touch upon naming and addressing in networks. %
To our taste, an architecture should name every element exactly once,
which has led us to question deeply what exactly names and addresses
are. %
Our solution to this question is provided in Sec. \ref{ssec:naming}. %
In computer science literature, there is a distinction between unicast
and multicast communication based on whether the communication is
one-to-one or many-to-many. %
Additional distinctions are being made for one-to-any (anycast), and
one-to-all, many-to-all and all-to-all (broadcast). %
The Ouroboros architecture avoids the need to make a distinction
between multicast and broadcast, and sees it as a {\em fundamentally}
different mechanism from unicast. %
This is based on the simple observation that both the application and
the network must be aware that multicast is needed. %
Ouroboros is -- to our knowledge -- the first network architecture
that clearly separates how it handles unicast from how it handles
multicast; putting the functionality into distinct types of network
layers. %
This leads to a solution which finds a middle ground between putting
multicast in the network, yet still providing the advantages of using
an overlay \cite{jannotti2000overcast}. %
The key distinction between these types of layers is based on whether
their packet transfer elements implement {\sc forwarding}, which is
defined in Sec.\ref{ssec:routing}, or {\sc broadcast}, which is
defined in Sec. \ref{ssec:multipt}. %
After these more general concepts, we introduce some terminology that
are a bit more specific for recursive networks. %
A lot of these terms were adopted from the RINA architecture to which
-- as we stated before -- Ouroboros is highly indebted. %
The terms {\em bootstrapping} and {\em enrolment} are used for adding
a node to a network layer, and are explained in
Sec. \ref{ssec:bootstr}. %
A concept of paramount importance is a {\em flow}
(Sec. \ref{ssec:flow}), which is an abstract construct for the network
resources that support the unicast transmission of packets between two
processes. %
Flows are provided by (unicast) layers, that consist of a number of
processes working together as a distributed application, whereas
multicast applications are supported by a broadcast layer
(Sec. \ref{ssec:layers}). %
Once a layer is established, it can support flows for higher layers. %
To make an application reachable over a layer, it must be known to a
layer. %
Sec. \ref{ssec:binding} explains how Ouroboros does this by {\em
  binding} a process to a name, which is then {\em registered} in a
layer, queriable by a directory service. %
Processes consist of a collection of logical components (finite state
machines), some of which create {\em connections} to communicate over
a protocol (Sec. \ref{ssec:conn}). %
Exactly how a layer creates a flow to support the connections of
higher level applications, is briefly detailed in Sec. \ref{ssec:fa}
for local communication; and Sec. \ref{ssec:uniipcp} details how
unicast IPCPs create flows over a network layer. %
The broadcast IPCP that supports multicast/broadcast applications is
explained in Sec. \ref{ssec:bcipcp}. %
At this point, all elements are in place to conceptually understand
how Ouroboros creates and manages flows. %
These flows are however not reliable, and are augmented with a
reliable transport protocol, providing fragmentation, retransmission,
flow control and packet ordering (Sec. \ref{ssec:connmgmt}). %
Sec. \ref{sec:concepts} continues with sections briefly dealing with
Quality of Service (Sec. \ref{ssec:qos}) and congestion control
(Sec. \ref{ssec:cgcontrol}). %
An in-depth overview of the packet processing pipeline
(Sec. \ref{ssec:pipeline}) concludes this section., detailing the
functions that a packet is experiencing while being forwarded and
moving from one layer to the next, using two stripped-down protocols:
a data transfer protocol for moving packets between IPCPs, and a
transport protocol for the end-to-end functions. %

To help the reader get a bit familiar with the concepts,
Sec. \ref{sec:examples} runs through two examples. %
The first example (Sec. \ref{ssec:forward}) deals with how scalable
forwarding is achieved within a layer, while the second example deals
with how Ouroboros maintains connectivity for a moving end device
connected to different wireless networks (Sec. \ref{ssec:mobex}). %

We delve into the key interfaces that each of the components
implements in Sec. \ref{sec:interfaces}. %
The most important is the application programming interface
(Sec. \ref{ssec:intf:api}). %
This interface consists of a small number of minimal and intuitive
function calls for establishing and managing flows from within a user
program.  %
While unicast and multicast are definitely distinct mechanisms, and
applications need to be aware of this, the function calls that are
needed are in both cases the same. %
This is illustrated with 2 short code examples, one outlining a simple
unicast application and the other doing the same for a simple
multicast application. %
Currently, programming (IP) multicast applications is far from trivial
\cite{rfc3170}, we hope our simple API can bring benefits for aspiring
multicast application developers. %
Sec. \ref{ssec:intf:fa} details the flow allocation interface that
needs to be implemented by the unicast and broadcast IPCPs, which
consists of 6 function calls, out of which the broadcast IPCP only
needs to implement 3. %
Insights on how Ouroboros interfaces with current networks and
applications are given in Sec. \ref{ssec:intf:legacy}. %

To become more familiar with Ouroboros, the reference implementation
is briefly described in Sec. \ref{sec:impl}. %
Its main components are the library (Sec. \ref{ssec:impl:lib}) that is
used by all Ouroboros applications, the IPC Resource Manager daemon
that is the central active component (Sec. \ref{ssec:impl:irmd}), and
the IPCP daemons (Sec. \ref{ssec:impl:unicast}-\ref{ssec:impl:ipcps})
for creating unicast and broadcast layers and for interfacing with the
physical network. %
To get some insights in how to manage an Ouroboros network, we provide
a small toolkit that implements the management primitives, consisting
of a compact command line interface for creating layers and making
applications available over them, detailed in
Sec. \ref{ssec:impl:irmtools}. %
This section is concluded with the description of the ovpn tool that
allows connecting IP-based applications over the prototype
(Sec. \ref{ssec:impl:ovpn}), a description of other small tools for
network testing and demo-ing (Sec. \ref{ssec:impl:tools}), and a small
example on how to create a 4-node network
(Sec. \ref{ssec:impl:examples}). %

Finally, we summarize our final thoughts and experiences developing
and using Ouroboros in our Conclusions, Sec. \ref{sec:conclusion}. %

\section{Concepts}
\label{sec:concepts}
This section defines the concepts that are implemented in Ouroboros,
which are common to most -- if not all -- packet switched networks. %
This section details a {\em model} and deals with (abstract) programs,
it does not specify an implementation, and it should not be directly
interpreted with respect to any particular implementation of a
computing system or environment%
\footnote{{\em ``\ldots{} but approached with a blank mind,
    consciously refusing to try to link it with what is already
    familiar\ldots''} \cite{dijkstra1988cruelty}.}. %
The definitions in this section may differ (even ever so slightly)
from previously held notions of the term. %
We address how this academic model can be interpreted towards
efficient implementations in Sec. \ref{sec:impl}.

\subsection{Graphs and networks}
\label{ssec:graphs}
A {\em graph} is a pair $G = (V, E)$ consisting of a set of {\em
  vertices} $V$ and a set of {\em edges} $E \subset V^2$ of
two-element subsets of $V$ \cite{jungnickel2007gna}. %
An edge $e = \{v_i,v_j\}$ has (distinct) end vertices $v_i$ and
$v_j$. %
A {\em directed graph} or {\em digraph} is a pair $G = (V, A)$
consisting of a set $V$ of vertices and a set $A$ of ordered pairs
$(v_i, v_j)$ (called {\em arcs}) where $v_i \neq v_j$. %
A {\em network} is a digraph $G = (V,A)$ on which a mapping
$w: A \rightarrow \mathbb{R}$ from the edgeset to the reals is
defined; the number $w(a)$ is called the {\em weight} of the arc $a$%
\footnote{The weight of an edge is often called a {\em metric} in
  computer science literature. %
  To avoid confusion, we will use the term {\em metric} only in its
  mathematical meaning.}. %
If $a = (v_i,v_j) \in A$, then $v_i$ is {\em adjacent} to $v_j$ and
{\em incident} with $a$. %
The set of {\em neighbors} $N$ of a vertex $v$ is the set of
vertices that are adjacent to $v$, $N(v) = \{u \in V: (v, u) \in A\}$. %

A {\em walk} is a sequence of vertices
$\mathcal{W} = (v_0, v_1, \ldots , v_n)$ so that
$a_i = (v_{i-1},v_i) \in A , i = 1, \ldots, n$. %
So each walk also implies a sequence of edges.  We define the weight
of a walk $\mathcal{W}$ as the sum of the weights of its arcs,
$w(\mathcal{W}) = \sum_{i=1}^nw(a_i)$. %
If the {\em source} $v_0$ and {\em destination} $v_n$ are the same
($v_0 = v_n$), the walk is {\em closed}. %
A walk where each of the arcs $a_i$ is distinct is called a {\em
  trail}. %
A {\em path} is a trail for which each of the $v_i$ are distinct. %
A {\em closed trail} for which the $v_i$ are distinct is called a {\em
  cycle}. %
A {\em directed acyclic graph} (DAG) is a digraph that does not
contain any cycles. %
If for every pair of vertices $\{v_s, v_d\} \in V^2$ there is at least
one path $\mathcal{P} = (v_s, \dots, v_d)$, we call the (di)graph {\em
  connected}. %
The distance function defined by the weight of the shortest path
between two vertices in a network (or $\infty$ if no such path exists)
is called the {\em geodesic} distance. %
A {\em metric} is a (distance) function
$d: X^2 \rightarrow [0, {+}\infty[$ fulfilling positivity, symmetry
and triangle inequality:
$\forall x,y,z \in X: d(x, y) \geq 0 \land d(x,y) = d(y,x) \land
d(x,z) \leq d(x,y) + d(y,z)$.
A geodesic distance is in general not a metric since it doesn't always
fulfil the positivity and symmetry requirements in a digraph. %

\subsection{Application, program, process, thread}
\label{ssec:progdef}
A {\em program} is a (finite) sequence of instructions that can be
executed by a computer%
\footnote{An implementation of a Universal Turing Machine.}, %
designed to solve a specific problem. %
A {\em process} is an instantiation of a program
\cite{tanenbaum2014mos}, and is identified by a {\em process id}
(PID). %
Programs may consist of multiple parallel execution {\em threads},
identified by a {\em thread id} (TID). %
An {\em application} is a collection of processes. %
A {\em distributed application} is a collection of processes residing
on different computing systems. %
The term {\em (computing) system} in this text should be interpreted
as an implementation of a Turing Machine \cite{turing1936a} capable of
running the (set of) program(s) that are mentioned with regards to
that system. %
A {\em distributed system} is a collection of computing systems that
is characterized by concurrency of components, lack of a global clock,
and independent failure of components%
\footnote{However, to complicate things further, {\em distributed
    operating systems} present an entire distributed system as a
  single system.} %
\cite{coulouris2011ds}. %
To us, the lack of a global clock is the only distinguishing feature
that really matters. %


\subsection{Naming and addressing}
\label{ssec:naming}
In most literature on networks, there is a distinction between {\em
  names}, {\em addresses} and {\em routes} \cite{rfc791}. %
The 'name' of a resource indicates {\em what} we seek, an 'address'
indicates {\em where} it is, and a 'route' tells us {\em how} to get
there \cite{shoch1978naming, saltzer1978naming}. %

Distinctions have been made between {\em location-independent} names
and {\em location-dependent} addresses, {\em flat} and {\em
  hierarchical} names and -- based on whether the name is semantical,
i.e. conveys information about the object -- {\em pure} and {\em
  impure} names \cite{needham1993names}. %
What distinguishes an address from a name is oftenly expressed in
vague terms, such as {\em ``an intermediate form between a name and a
  route, oriented to machine processing and used to generate the
  route''} \cite{hauzeur1986modelnma}. %
A concise definition for an address, that summarizes the perceived
consensus in current literature, would probably reflect that addresses
are location-dependent, impure synonyms for a resource, usually in the
form of a (hierarchically) structured number. %

To come to a more precise definition of an address, we will now
introduce some concepts using a simple analogy from postal
addresses. %
The authors' offices are located at {\em ``iGent Tower, 126
  Technologiepark-Zwijnaarde, 9052, Ghent, Belgium''}. %
Only readers who already know where this is will be able to derive a
location from this string, all others will need either a {\em map} (an
{\em absolute location}) or a set of directions (a {\em relative
  location} from his or her current position) to find us. %
There are two important things to notice. %

First, the string is a {\em compound name}, composed of names selected
from a (non-strict) {\em partially ordered set of namespaces}, that
are bound to objects from different sets: {\em countries} (Belgium),
{\em cities} (Ghent), {\em postal codes} (9052)%
\footnote{For completenes, we note that Belgian postal codes are
  actually compound names consisting of substrings from a total order
  of three namespaces reflecting a hierarchy in the postal system: the
  number consisting of the first two digits designates the sorting
  center, the third digit designates the postal office, and the final
  digit designates the issuing office. %
  Technically, any {\em number} is also a compound name with a strict
  total order on namespaces that indicate the magnitudes of the digits
  that it consists. %
  The total order relation allows us to see them as a single namespace
  without loss of generality.}%
, {\em roads} (Technologiepark-Zwijnaarde) and {\em housenumbers}
(126). %
{\em ``iGent tower''} is a synonym for the remainder of the string. %
The city name is a (not necessarily unique%
\footnote{For instance, in Belgium there are two towns called {\em
    ``Nieuwerkerken''}.})  synonym for a certain group of unique
(within the country) postal codes. %
This strict partial ordering between the named objects (derived from
the operator {\em ``located within''}) is what gives the string a
semantic of {\em ``location''}: Cities are located within countries,
streets within cities, and so on. %

Second, only one of these substrings is really location-dependent by
itself: the housenumber%
\footnote{This example is true for Ghent. Manhattan is well-known for
  having (most of) its roads named according to a grid plan of
  numbered streets and avenues.} %
``126''. %
That is because there is a metric defined on the housenumber namespace
from which it is selected. 
Once in the street, it is trivial to find the house without a map or
set of directions. %
Even more useful is when the names are taken from a {\em normed vector
  space} or {\em coordinate space}, denoting an absolute location. %

So, in general, we can define an address as a compound name,
consisting of names in a non-strict partially ordered set of
namespaces that are either metrized or have an associated entity that
provides a relative location, allowing to locate an object bound to
this name. %
Elements that have no such metric nor entity associated are synonyms
that can be added for human readability, these are not necessary for
the purpose of locating the object. %
These elements are in spaces that are non-strict in the partial
order. %

In the remainder of this article, when we use the term {\em name}, we
mean a string without any implied internal structure. %
The term {\em compound name} will be used for strings that consist of
names taken from namespaces that have a partial order relationship. %
For convenience and brevity, if the namespace from which a name was
taken has an associated distance function, we will assume it is also
coordinate space and use the term {\em coordinate}. %
The reader should keep in mind that wherever the term {\em coordinate}
is used in the remainder of this article, it can be replaced by {\em
  ``element of a space with associated distance function''}. %

\subsection{Routing and forwarding}
\label{ssec:routing}

Routing is broadly defined as the process of selecting a path for
traffic in a network. %
In computer science literature, there are two main groups of
approaches to routing. %

On the one hand there is the {\em hierarchical routing} solution.
This is the approach taken in IP networks, where a set of subnetworks
is defined using {\em prefixes} or {\em subnet masks}
\cite{rfc4632}. %
A scalability issue with IP stems from not following the partial
ordering implied by the subnetting in the delegation of IP addresses,
causing fragmentation of the IP address space
\cite{sriraman2007ipaddr} and making prefix aggregration in routers
increasingly inefficient%
\footnote{{\em ``Addressing can follow topology or topology can follow
    addressing. Choose one.''} -- Rekhter's Law \cite{rfc4984}.}. %

On the other hand we have {\em geographic} or {\em geometric} routing
\cite{kuhn2003geometric}, where each node is assigned a coordinate so
next hops can be calculated making use of the coordinates of the
direct neighbors%
\footnote{Geographic coordinates are a compound name, consisting of
  latitudes and longitudes, but there is no order implied between
  these two coordinate spaces. %
  Hence the {\em partial} ordering in our definition of an
  address.}. %

The examples above illustrate that the concept of routing encompasses
both the dissemination and gathering of information about the network,
and the algorithms for calculation and selection of the paths. %
We will now define the concepts underpinning ``routing'' more
formally. %
As far as we know the definitions are original, although the reasoning
behind it is similar to the reasoning commonly used in formulating
(integer) linear programming solutions to problems in graph theory. %

Let $G = (V, A)$ be a network. %
{\sc routing} is any algorithm that, given source and destination
vertices $v_s$ and $v_d$, for each vertex $v$ in $V$ returns a subset
$H(v) \subseteq N(v)$ of neighbors with the associated set of arcs
$L(v) = \{(v, u) \in A: u \in H(v))\}$, so that the following 4
conditions are met: (1) the graph
$D = (V', A') = (\cup_{v \in V}(H(v)), \cup_{v \in V}(L(v)))$ is a
directed acyclic subgraph of $G$; (2) $v_s$ is the only vertex in $D$
with only outgoing arcs; (3) $v_d$ is the only vertex in $D$ with only
incoming arcs; and (4) $A' = \emptyset$ if and only if there is no
path $\mathcal{P}$ between $v_s$ and $v_d$ in $G$.

Equivalently%
\footnote{Choose for $d$ the length of the longest path between the
  corresponding vertices in $D$. \qed}, %
{\sc routing} is any algorithm that, given source and destination {\em
  vertices} $v_s$ and $v_d$, for each vertex $v$ in $V$ returns a
subset $H(v) \subseteq N(v)$ of neighbors so that (1)
$\forall{u} \in H(v): d(u, v_d) < d(v, v_d)$ for any chosen distance
function $d$ on $G$; and (2) $\cup_{v \in V}(H(v)) = \emptyset$ if and
only if there is no path $\mathcal{P}$ between $v_s$ and $v_d$ in
$G$. %

In other words, either the distance function bounds the routing
solution, or the routing solution bounds the distance function. %

We define {\sc forwarding} as any algorithm that, for each vertex $v$
in $V$ returns the set of arcs $L(v)$. %
{\sc forwarding} is often implemented as {\sc routing} with an
additional edge selection step. %


The necessary and sufficient condition for {\sc routing} is full
knowledge of the graph $G = (V, A)$ and a valid geodesic distance
$d$. %
For a given vertex $v$, the necessary and sufficient set of
information to obtain $L(v)$ is knowledge of its neighbor set $N(v)$
and the subset of the geodesic distances originating at its neighbors,
$d_{v} \subset d: N({v)} \times V \rightarrow \mathbb{R}$. \qed %

In less formal terms, {\sc routing} and {\sc forwarding} provide a set
of vertices and arcs, respectively, so that there are never loops if
one travels to a next vertex or along an arc in the set. %
If implemented in a centralized way, {\sc routing} and {\sc
  forwarding} roughly need to know the full network. %
When implemented in a distributed way, a node roughly needs to know
its neighbors and the distances to the destination from itself and
from all its neighbors. %

The definitions above show what information needs to be disseminated
in a network to allow {\sc forwarding}. %
Let's assume that vertices know their neighbors or incident outgoing
arcs, then what is needed is a dissemination procedure for the
(geodesic) distance function $d$. %
This is implemented in a class of dissemination protocols, called {\em
  distance vector} protocols, such as Routing Information Protocol
(RIP) \cite{rfc1058}. %
Each vertex $v$ announces the distances it knows to its neighbors
$N(v)$. %
Border Gateway Protocol (BGP) \cite{rfc4271} is a {\em path vector}
protocol as it disseminates paths instead of only a distance. %
This additional information to allow choosing between different
available routes based on intermediate nodes or networks. %

A different approach is for each vertex $v$ to announce its
neighboring links to each of its neighbors $N(v)$, effectively
disseminating the full network graph, which allows for more
sophisticated routing algorithms. %
Protocols that follow this approach, such as Open Shortest Path First
(OSPF) \cite{rfc2328}, are called {\em link state} protocols. %

In Ethernet networks, switches can derive a valid distance function
from {\em MAC learning}. %

The dissemination of information in the network uses network
resources, and the convergence time when the graph changes -- deletion
of a link or vertex due to, for instance, a cable cut or a router
power outage -- can lead to problems, such as the packets getting
stuck in a temporary loop. %
Let $\mathcal{N}$ a set of names with a bijection
$b: \mathcal{N} \leftrightarrow V$ that binds a name $n_i$ to a vertex
$v_i$. %
For some graphs $G = (V, A)$, a distance function (metric) over
namespace $\mathcal{N}$ of vertex names,
$d_\mathcal{N}: \mathcal{N}^2 \rightarrow \mathbb{R}$ , can be used to
fully infer a valid distance function,
$d: V^2 \rightarrow \mathbb{R}$, over the graph, reducing the need for
routing tables and the dissemination of network information \cite{
  papadimitriou2005geometric, kleinberg2007geometric}. %

\subsection{Multipoint communication}
\label{ssec:multipt}

The definition of {\sc routing} above intentionally does not include
multipoint communication. %
Unicast and multicast applications must be aware that they are using
unicast or multicast communications. %
Therefore, Ouroboros sees unicast and broadcast as fundamentally
different mechanisms, each giving rise to a different type of layer. %

{\sc broadcast} is a trivial well-known algorithm that, for an
incoming arc $(i, v) \in A$ of vertex $v$, returns all outgoing arcs
$L'(v) = \{(v, u) \in A: u \neq i)\}$, with an associated set of
vertices $H'(v) \subset N(v) = \{w \in V: (v, w) \in L'(v)\}$. %
$H'(v) = N(v)$ if $v$ is the source.

It is clear that the graph
$G'' = (V'', A'') = (\cup_{v \in V}(H'(v)),\cup_{v \in V}(L'(v))) = (V, A) = G$
does not fulfill the conditions for a solution to routing unless the
input graph already fulfills these conditions. %

{\sc broadcast} is thus mostly used to send information to all network
members on a DAG or tree. %

\subsection{Bootstrapping and enrolment}
\label{ssec:bootstr}

{\em Bootstrapping} is the process of manually configuring a network
element to be the first member of a new network layer so it starts its
internals to start {\sc forwarding} or {\sc broadcast}. %
{\em Enrolment} is the process where an entity contacts (an) existing
member(s) of the network to get the necessary configuration
information to start its internal components so it can start {\sc
  forwarding} or {\sc broadcast}.

Enrolment is not necessarily a clear-cut procedure or protocol, but
may encompass a number of exchanges for getting authentication and
encryption keys and configuration information from different
sources. %

Examples of enrolment in current network are the Wi-Fi association
process in 802.11 \cite{ieee80211}, and the Dynamic Host Configuration
protocol \cite {rfc2131} in IP networks. %
Examples of distributed enrolment are the Spanning Tree Protocol
\cite{perlman1985stp} for Ethernet and the Internet Group Management
Protocol \cite{rfc4604} for joining IP multicast networks. %

We will illustrate the enrolment process for Ouroboros later in a
more practical setting in Sec. \ref{sec:impl} when we discuss our
reference implementation in detail. %

\subsection{Flows between processes}
\label{ssec:flow}

A {\em flow} is the abstraction of a collection of resources within a
network layer that allow bidirectional communications using packets
between two processes that are clients of this layer. %
A flow enables a point-to-point {\em packet delivery service} and can
be viewed as a bidirectional pipe that has a number of observable
quantities associated with it that describe the probability
$p(S,t, \ldots) \in [0, 1]$ of a packet of given size $S$ being
transferred within a certain period of time $[t_x, t_x + t]$. %
The maximum probability for error-free transfer depends on the {\em
  packet-drop-rate} (PDR) and {\em bit-error-rate} (BER) of the
flow. %

The Ouroboros architecture ensures that flows are {\em content
  neutral}, i.e. the probability $p(S, t, \ldots)$ above is independent
of the bits that make up the packets sent along a flow. %

The {\em delay} or {\em latency} describes how long packets take to
traverse the flow, and the variation on the delay is called {\em
  jitter}, or more precisely, {\em packet delay variation}
\cite{rfc3393}. %
The delay for a flow has four main components, {\em propagation} and
{\em queuing} delays, {\em transmission} and {\em processing}
delays. %

There are 2 upper bounds, the {\em Maximum Packet Lifetime} (MPL) is
the maximum amount of time that a packet can take to transfer over the
flow, and the {\em Maximum Packet Size} (MPS) is the maximum length
for a packet to be acceptable for transfer. %
In other words, the probability for a packet to arrive after MPL
expires should be 0%
\footnote{This may be hard to guarantee with 100\% certainty, so MPL
  should be {\em ``large enough''} so that this probability is 0 in
  practice. IP has a bound on the Maximum Datagram Lifetime (MDL) via
  the Time-To-Live or Hop Limit decrement in each router to a maximum
  of 255 seconds \cite{rfc791}, with a recommended value of 64 seconds
  \cite{rfc1700}. %
  In addition, TCP defines the Maximum Segment Lifetime (MSL) as 120s
  \cite{rfc793}.}, %
and the probability for a packet to arrive that exceeds the MPS is
equal to 0. %
Similarly, there can be lower bounds such as Minimum Packet Lifetime
(mPL) and Minimum Packet Size (mPS). %

The resources that make up a layer are finite, limiting the total
number of packets that can traverse the network layer in a given
period of time. %
Flows provide the mechanism to put a network layer fully in control of
its own resources. %
The resources that constitute the flow can either be shared with other
flows or dedicated (reserved) for this particular flow. %

Other externally measureable quantities can be associated with a flow,
such as bandwidth and load for flows with reserved resources. %
The probability function may depend on these quantities (e.g. if the
load reaches a certain threshold, delay could increase). %

A flow endpoint is identified in a system by a {\em flow ID} (FID),
which defines the {\em layer boundary}%
\footnote{In this respect, a flow id is similar to an OSI {\em Service
    Access Point} (SAP) or RINA {\em port id}.}. %
For security reasons, a process has no direct access to the flow, but
rather accesses the flow through a {\em Flow Descriptor (FD)} to read
and write from a flow. %
Flow identifiers are unique within the scope of a system, flow
descriptors are unique within the scope of a process%
\footnote{This is similar in function to a UNIX file descriptor. A
  UNIX kernel space implementation of Ouroboros would probably use
  file descriptors for flow descriptors.}. %

Flows are an important concept for enabling Quality-of-Service (QoS)
in a layer, which we detail later in Sec. \ref{ssec:qos}. %

\begin{figure}[ht]
  \centering
  \includegraphics[width=0.9\textwidth]{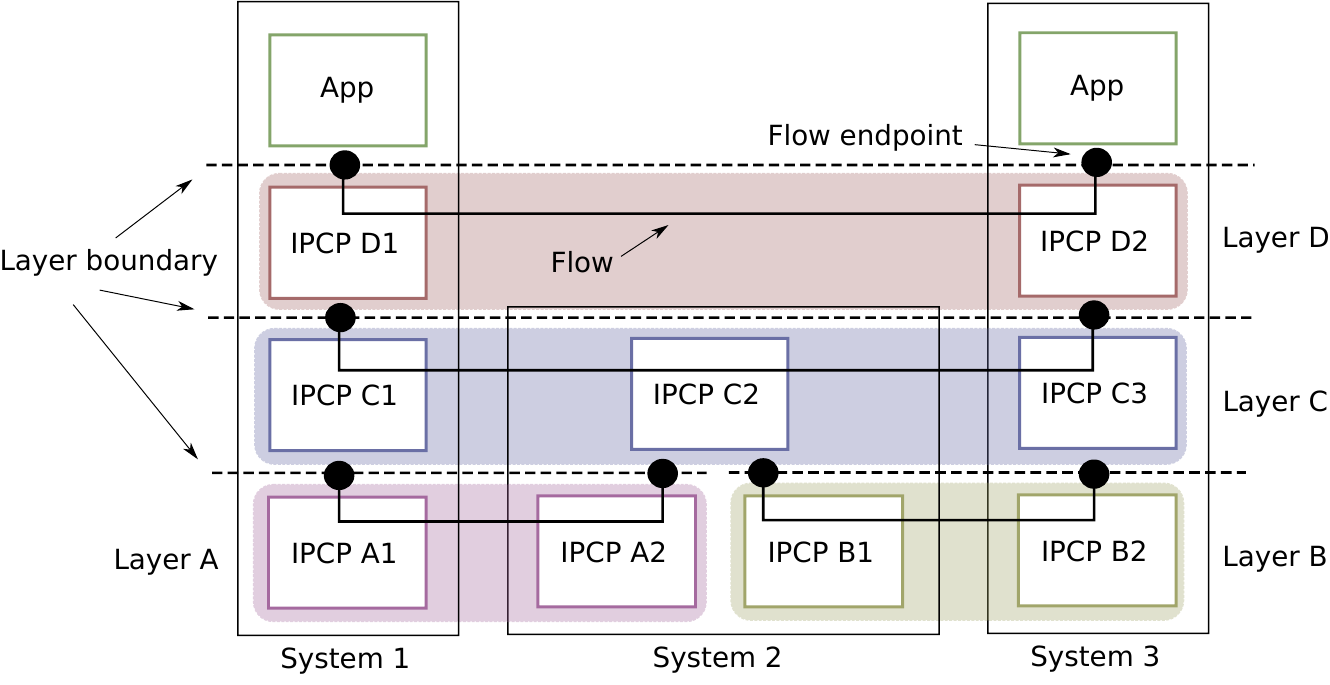}
  \Description{IPC Processes in different layers on different systems}
  \caption{Layers of unicast IPC processes}
  \label{fig:flid}
\end{figure}

\subsection{Layers of processes}
\label{ssec:layers}

The layering in Ouroboros follows RINA in its fundamental observation
that a layer is a {\em collection of processes} that form a
distributed application. %
The processes that make up a {\em network layer} are called {\em IPC
  processes} (IPCPs)%
\footnote{The moniker {\em ``IPC process''} can be traced back to the
  Lawrence Livermore LINCS architecture, which provided an {\em IPC
    service} on top of its transport layer, called LINCS-IPC
  \cite{fletcher1982lincs, watson1981deltat}.}. %
Ouroboros has 2 distinct types of IPCPs to build the recursive
network: unicast IPCPs that implement {\sc forwarding} and provide
flows, and broadcast IPCPs that implement {\sc broadcast}. %
They both provide a {\em packet delivery service} to the processes
above, making use of unicast flows provided by the network layer
below%
\footnote{RINA IPCPs provide an {\em IPC service}. In contrast,
  Ouroboros IPCPs provide a packet delivery service. This nomenclature
  may cause considerable confusion, but we haven't found a better name
  for an IPCP in Ouroboros.}. %
Broadcast layers will usually be the top layer in a certain system. %
This induces (strict) partial order relationships between the IPCPs in
a system and between network layers in a network
(Fig. \ref{fig:flid}), which allows us to represent dependencies
between layers and and dependencies between IPCPs as {\em directed
  acyclic graphs} (DAGs). %

In order to be able to provide flows, a unicast layer has to be
capable of at least two basic functionalities: maintaining a {\em
  directory} that keeps track of the location (an IPCP address) where
a certain destination process is available, and {\em forwarding
  packets} to this location with non-zero probability of arriving
within an appropriate timeframe%
\footnote{Within a datacenter, this can be in the order of
  microseconds; between Earth and a Mars rover, this will be on the
  order of minutes.}. %

\subsection{Binding and registering}
\label{ssec:binding}

The management component of Ouroboros is the {\em IPC Resource
  Manager} (IRM), which provides an interface for administration
and configuration of the IPC resources (IPCPs, flows, \ldots) within a
system. %
The IRM is also in charge for delegating outgoing flow requests to
layers and assigning incoming flow requests to processes. %

Ouroboros {\em binds}%
\footnote{Binding here is used in the sense of an operation, {\em
    bind()}. %
  It should not be confused with ``binding'' in the sense of assigning
  a name to an object \cite{saltzer1978naming}. %
  In that sense, processes are uniquely bound to a PID.} %
processes to a {\em name}, that can be {\em registered} in a layer,
populating its directory. %
These names are given to rather hard-to-define objects -- a set of
flow endpoints that may be created in the future -- therefore we will
consider the {\small \texttt{name}} to be this very abstract object
and use a different font for it. %
{\em Binding} and {\em registering} are operations that are (usually)
performed by an administrator, not the application programmer, which
greatly simplifies writing distributed programs for Ouroboros. %
The IRM keeps a mapping between PIDs and {\small \texttt{names}}, and
the layer keeps a mapping between {\em hashes} of these {\small
  \texttt{names}} and their locations. %
These mappings are many-to-many, so a PID can be bound to multiple
{\small \texttt{names}} and multiple processes can be bound to the
same {\small \texttt{name}}. %
Similarly, multiple {\small \texttt{names}} can be registered at the
same location and a {\small \texttt{name}} can be registered at
multiple locations (anycast). %
These bindings are also dynamic during the process runtime, so it is
easy to create a running server process, test it over a private
network, and then make it available over a public network. %
All without touching the running server. %

The operations do not need to be performed in any particular order,
although binding a process will of course only be possible after the
process is running. %
We also allow binding programs to a {\small \texttt{name}}, that will
cause all future instances of this program to be bound to that {\small
  \texttt{name}}. %

\begin{figure}[ht]
  \centering
  \Description{Two processes containing partially ordered protocol
  machines which use a flow to communicate}
  \includegraphics[width=0.8\textwidth]{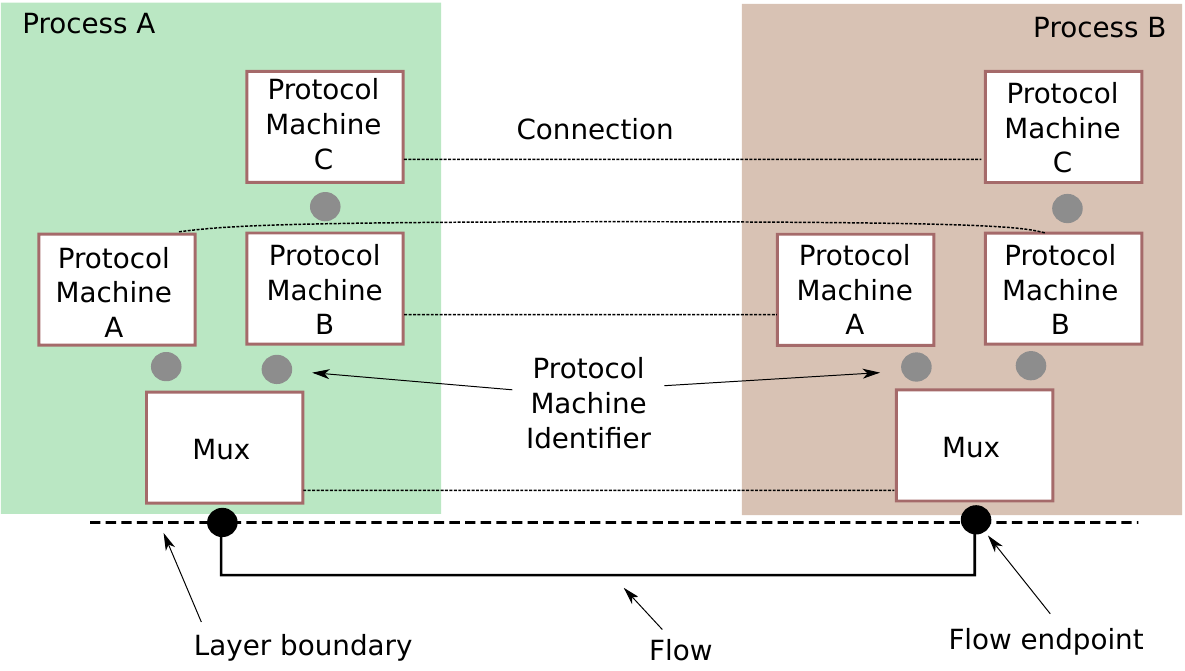}
  \caption{A partially ordered set of protocol machines using a flow
    (simplified)}
  \label{fig:pms}
\end{figure}

\subsection{Connections between protocol machines}
\label{ssec:conn}

Processes can be abstracted mathematically as a set of {\em finite state
  machines} (FSMs). %
A pair of FSMs that communicate using the packet delivery service that
is enabled by a flow are called {\em protocol machines} (PMs). %
The communication between two protocol machines at each end of a flow
is called a {\em connection}. %
A protocol machine is identified by a {\em protocol machine
  identifier} (PMID) (Fig. \ref{fig:pms}). %

In its full generic form, the Ouroboros architecture exactly mirrors
the {\em inter-layer} relationship between processes in a system (a
strict partial order of IPCPs) to the {\em intra-layer} relationship
between protocol machines (a strict partial order of PMs). %
The naming hierarchy of PIDs all the way to FIDs and FDs is copied to
PMIDs all the way to {\em connection identifiers} (CIDs) and {\em
  connection descriptors} (CDs). %
Similar to an IRM that manages flows within a system, the Connection
Manager (CM) manages connections within a process. %
This allows full dynamic creation of protocol machines in processes
without any statically configured state. %

Dynamic allocation of protocol machines in a process is useful only in
the most complex and elaborate distributed applications. %
We will present a simplified version, in line with common practice in
protocol design, assuming a static allocation of these identifiers and
a static order of the protocol headers, and leave the complete
description of this part of the architecture as an exercise to the
reader. %

In this simplified form, protocol machines at the same level in the
partial order are identified using a simple protocol machine called a
{\em multiplexing protocol machine} or {\em multiplexer} using a
static PMID as its invariable field; it is not modified during the
lifetime of the packet. %
A multiplexer consists of two {\em classifier} FSMs that share state%
\footnote{ Multiplexing protocol machines are omnipresent in today's
  network protocols, albeit usually disguised as part of another
  protocol machine instead of being standalone. Version numbers,
  protocol fields, types and ports are all examples of this generic
  PMID hiding in various protocols that are deployed today.}. %

Each protocol machine will process its associated {\em packet header}
and deliver the packet to the next (protocol) state machine
downstream/upstream in the DAG%
\footnote{In architectures like OSI (and RINA), there is a distinction
  between a {\em Protocol Data Unit} (PDU) -- which is a packet
  including the header (or {\em Protocol Control Information}) -- and a
  {\em service data unit} (SDU), which is the opaque data. %
  These names are used to make a distinction {\em between layers}, so
  when an N--PDU is handed to the N--1 layer it becomes the
  (N--1)-SDU, which is encapsulated inside an (N--1)-PDU, and so on. %
  This stems from the notion that each layer has only a single
  protocol that encapsulates higher layer data. %
  The protocol machines in the Ouroboros model {\em may} also perform
  encapsulation {\em within a layer}, so we chose not adopt the
  PDU/SDU nomenclature to avoid confusion.}. %

Some of the protocol machines can be short-lived and bypassed after
certain conditions are met (for instance, protocol machines performing
configuration or authentication). %

It is quite common (and useful) to view a set of protocol machines
that depend on each other in a linear hierarchy as a single entity
during implementation. %
In this article, we will also consider some multiplexers as part of
other protocol machines to avoid overcomplicating the explanation. %

\subsection{Flow allocation}
\label{ssec:fa}

The procedure with which a process requests a unicast flow to another
process is called {\em flow allocation}. %
A flow is (usually) identified by two flow endpoints, created by the
end-hosts' IRMs, and identified by their flow IDs. %
The flow IDs allow access to shared state between the IRM, and the two
processes, and are unique within a system. %
Flow endpoints have some minimal resources associated with them, most
importantly an incoming {\tt rx} FIFO and outgoing {\tt tx} FIFO for
packet processing. %

\begin{figure}[ht]
  \centering
   \Description{The sequence of steps needed for local flow allocation}
   \includegraphics[width=0.6\textwidth]{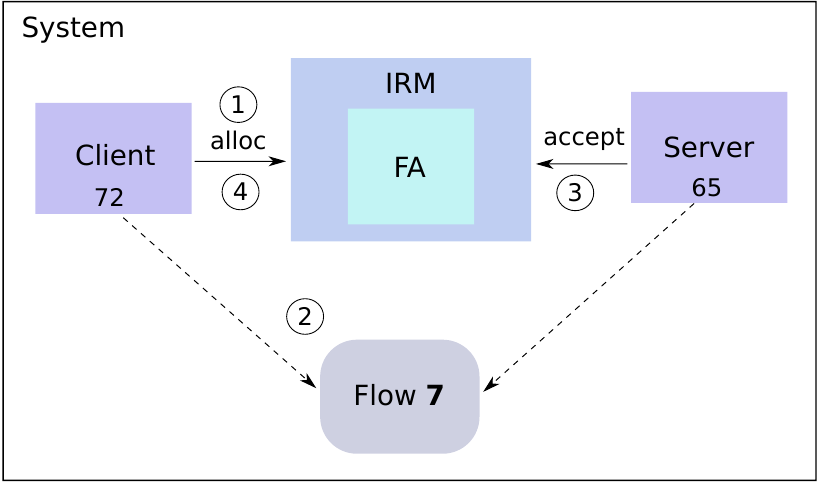}
   \caption{Flow allocation (local)}
   \label{fig:fa-local}
 \end{figure}

Fig. \ref{fig:fa-local} illustrates local flow allocation,
(i.e. between two processes on a single system), which needs only one
flow endpoint, over which the two processes communicate directly. %
Flow allocation over a layer is explained in Sec. \ref{sssec:fa}. %

We assume the server running as PID $2000$, bound to the {\small
  \texttt{name}} {\em ``server''} and is calling {\em accept()}. %
The client first calls {\em alloc()} to request the IRM for a flow to
the {\small \texttt{name}} {\em ``server''} (1). %
The IRM knows a process (PID 2000) that is bound to that {\small
  \texttt{name}}, and creates a flow endpoint for the flow (2). %
The {\em accept()} call and the {\em alloc()} call return at the
server and client respectively (3) and (4). %
The client and server can now communicate over the flow. %

The attentive reader may now ask her- or himself how the client
process communicates with the IRM, because that is also IPC. %
This is where we venture into the realm of operating systems. %
In the ideal implementation, the IRM would be part of the operating
system kernel. %
In that case the calls between processes and the IRM can be either
implemented as a {\em system call} in a monolithic kernel, or as an
IPC mechanism that is bootstrapped between the IPC facility in a
microkernel and each process it spawns. %
Our current (user-space) implementation resorts to UNIX sockets for
IPC with the IRM. %

The use of the Ouroboros paradigm as the single unifying packet
transport technology in (distributed) operating system design is a
particularly interesting area for further study. %
Previously developed distributed operating systems had their own
specific protocol for communication between kernel components, and
supported TCP/IP in userland. %
Examples are the special-purpose transport protocol between kernels in
Sprite \cite{ousterhout1988sprite}, the Fast Local Internet Protocol
(FLIP) \cite{kaashoek1993flip} used in Amoeba, and the Internet Link
(IL) protocol \cite{presotto1995il} used in Plan 9 from Bell Labs. %

\subsection{The basic functions in the unicast IPC process}
\label{ssec:uniipcp}

We will now present the key protocol machines that implement the basic
unicast layer functionality (the directory service, packet forwarding,
enrolment and flow allocation) in an Ouroboros IPCP. %
We further simplified this model in the Ouroboros reference
implementation (Sec. \ref{ssec:impl:unicast}), however it is good to
be aware of this more general structure. %

\begin{figure}[ht]
  \centering
   \Description{The protocol machines in the unicast IPC process}
   \includegraphics[width=0.8\textwidth]{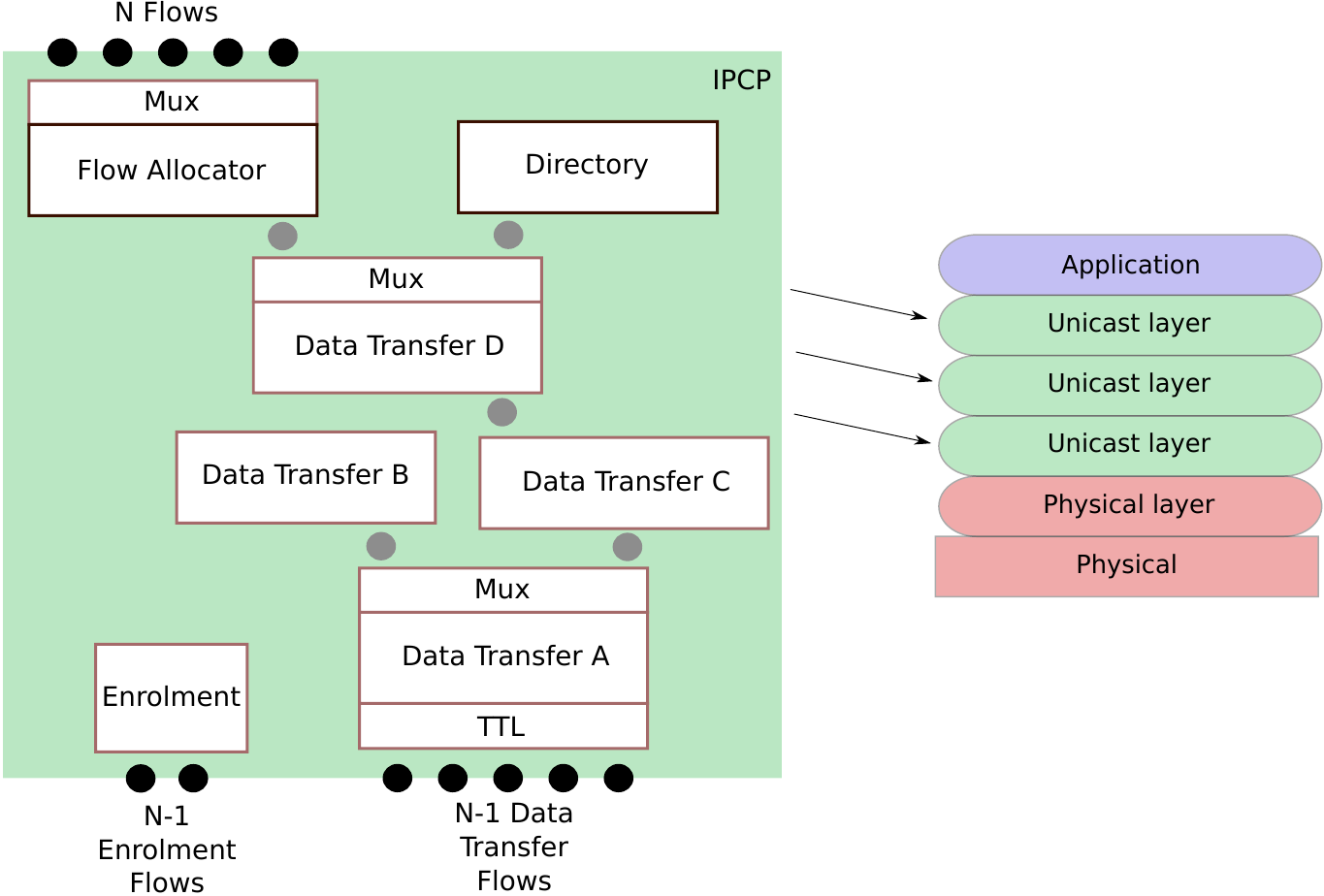}
   \caption{Protocol machines in the unicast IPC process}
   \label{fig:ipcp}
\end{figure}

\subsubsection{The directory}
\label{sssec:dir}

The destination for a flow is a {\small \texttt{name}} that is
registered in the network, it denotes that the IRM of this system is
aware of this {\small \texttt{name}} and that it can delegate flows to
that {\small \texttt{name}} to processes on this system. %
The directory is the entity that maps endpoint {\small \texttt{names}}
to {\em addresses} within a layer. %
A {\small \texttt{name}} can be registered on different addresses, and
different {\small \texttt{names}} can be registered at the same
address. %
The directory resolves a {\small \texttt{name}} to a single address
(unicast or anycast, multicast is detailed in Sec.
\ref{ssec:multipt})%
\footnote{The directory service is similar to the lookup service
  provided by Domain Name System (DNS) \cite{rfc1034, rfc1035} servers
  for IP networks. %
  The directory does not -- nor does it need to -- provide management
  of a global namespace such as provided by ICANN and the registrars
  for IP networks. %
  There is, however, a need for access control to this component in
  public networks, which is a topic for future research.}. %
To improve security, a layer registers hashes instead of plain-text
{\small \texttt{names}} (the hash algorithm used is configurable). %
SHA3-256 \cite{bertoni2013keccak, chang2012sha3} is used as a default
in the implementation, and we will use this algorithm in this article
when we mention hashes. %
Hash values are written as the first 8 characters of their hexadecimal
notation. %
A directory can be implemented as a database (central, replicated,
hierarchical or distributed) or as a (distributed) hash table (Sec.
\ref{ssec:impl:unicast}). %

The directory is also the entity that provides the mechanism to
support {\em anycast}: at the address resolution step, if there are
multiple destination addresses associated with a certain {\small
  \texttt{name}}, it will return one of these {\small
  \texttt{names}}. %

\begin{figure}[ht]
  \centering
  \Description{The sequence of steps in flow allocation on two
    systems}
  \includegraphics[width=0.9\textwidth]{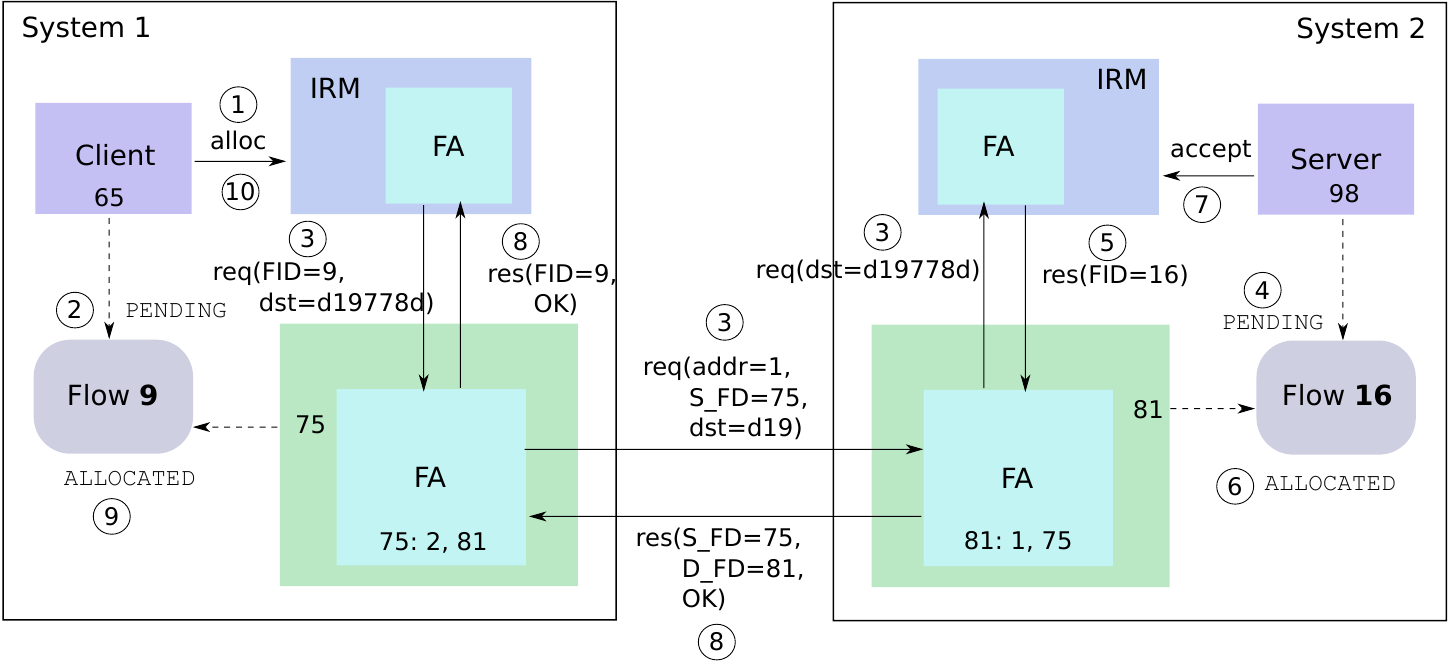}
  \caption{Flow allocation over a layer}
  \label{fig:fa}
\end{figure}

\subsubsection{Flow allocator}
\label{sssec:fa}

We will illustrate flow allocation over a layer using a (simplified)
example where a client on system 1 requests a flow to the server
process on system 2 (Fig \ref{fig:fa}). %
The IPCP on system 1 has address $1$, the IPCP on system 2 has address
$2$, the server is running as PID $2000$ on system 2. %
The server process is registered in the layer under the {\small
  \texttt{name}} {\em ``server''}. %
The directory for the layer maps the hash {\em d19778d}%
\footnote{The full SHA-3 hash of ``server'' is
  d19778d2e34a1e3ddfc04b48c94152cced725d741756b131543616d20f250f31.} %
to the address $2$, the IRM in system 2 will map PID 2000 to the
{\small \texttt{name}} {\em ``server''}. %

The steps for flow allocation are as follows: (1) The client requests
its IRM for a flow to {\em ``server''}. %
(2) The IRM creates a new flow endpoint in {\tt PENDING} state and
assigns it an available {\em flow id} FID 9. %
(3) The IRM sends the request on FID 9 for {\em d19778d} to the local
IPCP, where the FID 9 is mapped to a local FD 75. %
The flow allocator in the IPCP in system 1 will forward the request on
FD 75 to the flow allocator of the IPCP in system 2, which announces
the request to its IRM. %
(4) System 2's IRM receives the request for a flow to {\em d19778d}. %
It checks if a there is a process available (it finds PID 2000) and
creates a flow endpoint with FID 16, in {\tt PENDING} state, on system
2. %
(5) The IPCP is notified of the new flow on FID 16, which is mapped to
internal FD 81. %
The Flow Allocator in the IPCP now contains a mapping that local FD 81
is a flow to system 1, FD 75. %
(6) The IRM sets the flow endpoint in the {\tt ALLOCATED} state, and
the (7) {\em accept} call returns to the server, letting it know it
now has a new flow with FD 98. %
(8) The flow allocator in the IPCP on system 2 concurrently responds
to the flow allocator in the IPCP on system 1 that the flow request
with source endpoint 75 is accepted at endpoint 81 on system 2. %
(9) The IRM puts the flow in {\tt ALLOCATED} state and (10) the
allocation call for the client returns with FD 65. %
At this point the client can write/read to/from FD 65 and the server
can read/write from/to FD 98. %

The procedure illustrates the flow allocation protocol that is used in
Ouroboros. %
The request message contains 4 fields: the source address%
\footnote{Ouroboros doesn't send source addresses in its data transfer
  protocol for security and privacy reasons (See
  Sec. \ref{ssec:pipeline}).}, %
the source endpoint, the QoS specification for the flow (not shown in
the example) and the destination hash. %
The reply message contains 3 fields: the source endpoint, the
destination endpoint and a response code indicating if the flow was
accepted. %
To make it more future-proof, the flow allocation protocol will most
likely benefit from a version field. %
After flow allocation, both flow allocators at the endpoint IPCPs map
a local FD to a remote address, remote FD and QoS. %

\subsubsection{Data transfer protocol machines}
\label{sssec:dt}

The protocol machines that are responsible for forwarding packets are
called {\em data transfer protocol machines} (DTPMs); they implement
{\sc forwarding}. %
The only disseminated state (field in the packet header) associated
with a data transfer protocol machine is a {\em data transfer protocol
  machine name}. %

An example is given in Fig. \ref{fig:ipcp}, for an IPCP with 4 data
transfer protocol machines (with names $A$ to $D$), each chosen from a
namespace, $\mathcal{A}$, $\mathcal{B}$, $\mathcal{C}$ or
$\mathcal{D}$, consisting of uppercase letters, identified by a
multiplexing machine%
\footnote{We could call them all A since they are different
  namespaces, but that would unnecessarily complicate the
  explanation.}. %
This directed acyclic graph structure between DTPMs -- a
generalization of the {\em ``hierarchical address spaces''} commonly
found in literature -- defines the protocol headers shown. %
The operation is simple: if the name in the header matches the name of
the PMID, it delivers the packet to the state machine above, else it
sends the packet on the flow identified by {\sc forwarding}. %
This IPCP functions as a pure {\em router}%
\footnote{This is very important: the only mandatory components that
  an Ouroboros router has to implement are the DTPMs and associated
  dissemination, and enrolment.} %
for the {\em address space} that contains the namespaces
$[A-Z]$.$[A-Z]^\mathcal{B}$ (this system cannot allocate flows over
DTPM B), and as a possible end host over DTPMs for the address space
$[A-Z]$.$[A-Z]^\mathcal{C}$.$[A-Z]$. %
There should not be any packets with address $A$.$B^\mathcal{B}$
(since this IPCP can't provide flows for this address space). %
Packets with address $A$.$C^\mathcal{C}$.$D$ will be delivered to the
directory or flow allocator. %
This sheds some light on the question {\em ``Which network entity
  should be given an address?''}. %
The Ouroboros model defines an address precisely as {\em ``a compound
  name consisting of data transfer protocol machine names, where the
  (non-strict) partial ordering of the transfer protocol machine
  namespaces from which these are taken reflects the strict partial
  ordering of the data transfer protocol machines within an IPC
  Process''}. %

This structure does not imply that only pure hierarchical addresses
are possible. %
For instance, techniques like {\em longest prefix matching} can be
modeled as multiple forwarding FSMs that operate on the same header
space, where some treat a compound namespace as a single flat
namespace. %
In the remainder of this article, when we say that some IPCP has
address X.Y.Z it means that there are data transfer protocol machines
associated with names X, Y and Z inside this IPCP. %

\subsubsection{Enrolment}
\label{sssec:enr}

This component is responsible for enroling into a new layer, and for
accepting enrolment requests from possible new members when already
enroled in a layer. %
Enrolment goes directly over an ephemeral point-to-point N-1 flow
that is usually deallocated after enrolment completes. %
Optimizations may keep this flow for data transfer. %

\subsection{The basic functions in the broadcast IPC process}
\label{ssec:bcipcp}

Multicast applications are supported using a broadcast layer, for
which its IPCPs have a single packet transfer component that
implements {\sc broadcast}, in addition to an enrolment component. %
The scope of the broadcast layer defines the scope for the multicast
application, so usually there will be a broadcast layer for each
multicast ``stream'' (although for instance audio/video streams could
be separated in the application since the scope is the same for
both). %

\begin{figure}[ht]
  \centering
  \Description{The protocol machines in the broadcast IPC Process}
  \includegraphics[width=0.8\textwidth]{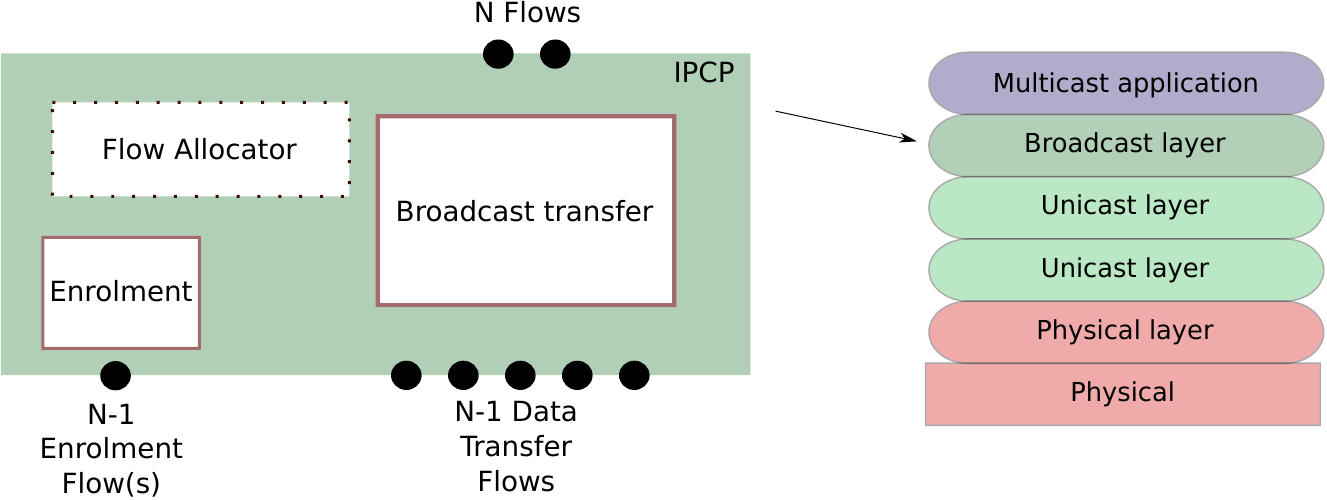}
  \caption{Protocol Machines in the broadcast IPC process}
  \label{fig:bcipcp}
\end{figure}

The flow allocator in the broadcast IPCP is a small proxy that only
implements a local function that responds positively to a flow join
request if the destination name is the layer name. %
Examples of the use of the unicast and broadcast IPCP are shown in
Sec. \ref{ssec:impl:examples}.

There are two main approaches for implementing the broadcast IPCP. %
The first is stateless: the packet transfer component does not require
a unique name within the layer, and the only data that needs to be
exchanged at enrolment is the layer name. %
Stateless broadcast IPCPs do not add any protocol headers; all they do
is read packets from a FD and forward them on all other FDs (this
includes the N-flows). %
Broadcast layers that are stateless must have trees as their
connectivity graph, therefore the enrolment procedure could be
augmented with a policy to ensure that the connectivity graph of the
broadcast layer is a tree; the optimum solution being a Steiner tree
\cite{bezensek2014steiner}. %

The second approach is stateful: the topology can be any graph, and
the packet transfer component implements a policy to prevent packets
to loop in the network. %
A simple such policy is to give the packet transfer component a name,
and add a packet header with a source name and a sequence number. %
Each stateful broadcast IPCP then maintains a table tracking <source,
sequence number> pairs, so that the packets that travel in loops can
be dropped. %
The stateful approach can be more resilient, but typically consumes
more resources in the lower layer. %

Note that differentiating between unicast and broadcast IPCPs in the
model does not preclude implementations to provide an IPCP that
combines the unicast and broadcast mechanisms into one IPCP, but,
since the scope of the layer is defined by its set of IPCPs this may
prove less useful than it sounds. %

\subsection{Reliable connection management}
\label{ssec:connmgmt}

From the definitions above, the packet delivery service offered by a
flow can lead to lost, corrupted, duplicated or missequenced packets;
a flow is -- by nature -- unreliable. %
To provide reliable transfer of information, mechanisms are
implemented on top of the flow to provide ordering, duplicate
detection and retransmission (Automated Repeat-reQuest, ARQ). %

A necessary and sufficient condition to achieve reliable transport is
to bound three timers: the maximum packet lifetime (MPL), the time to
acknowledge a packet ({\em A}) and the time after which there will be
no more retransmissions ({\em R}) for a certain packet
\cite{fletcher1978tbproto, watson1981timerbased}. %
This was implemented in a remarkably simple and elegant protocol,
called Delta-$t$ \cite{watson1981deltat}. %
The OSI TP4 protocol \cite{osi1995x224} and the RINA EFCP protocol
\cite{gurcun2010softstate} are directly based on Delta-$t$. %
It is important to note that $A$ and $R$ are strict bounds; after they
expire, no more acknowledgments (ACKs) or retransmissions may be
sent. %
However, these timers are not deadlines; they do not imply that an ACK
or retransmission must be sent before they expire (the protocol must
be able to deal with lost ACKs). %
In TCP, these three bounds are achieved by the Maximum Segment
Lifetime (MSL, 120s), the Maximum Retransmission Timeout (MAX\_RTO,
240s) and Delayed ACK timeout (500ms) \cite{rfc1122}. %

Ouroboros also implements a protocol based on Delta-$t$, where the MPL
bound is obtained from the MPL characteristic for the flow over which
the reliable connection protocol machine is running. %
The values of $A$ and $R$ are internal to this connection. %

In addition, {\em Fragmentation and reassembly} allows sending pieces
of data that are larger than the MPS of the underlying flow that
supports it. %
Fragmentation and reassembly are enabled by default for flows with
in-order delivery, and can be (independently) enabled/disabled at the
endpoints. %
Ouroboros assigns sequence numbers per-fragment and uses a two-bit
fragmentation scheme, where the {\em first fragment} (FFGM) bit
indicates that a received packet is the first fragment of a fragmented
packet, and a {\em more fragments} (MFGM) bit that is set to $0$ on
the last fragment of the packet ($10$ indicates a non-fragmented
packet, $11$ the first fragment, $01$ an intermediate fragment and
$00$ the last fragment). %
If the flow is unreliable (no retransmission), packets that have lost
fragments are usually discarded. %
If the flow is to support partial delivery of packets (deliver
fragments as soon as they arrive), the MFGM bit is simply ignored at
the destination. %

{\em Flow control} manages the resources at the connection endpoints,
avoiding that the source sends information faster than the destination
can process. %
Ouroboros currently implements a commonly used window-based flow
control scheme by maintaining a right window edge, updated in every
packet, that is controlled by the receiver. %
Other flow control mechanisms exist, such as rate based flow control,
but they are usually tailored to specific situations and are not
widely used (rate based flow control is useful only for applications
for which processing scales linearly w.r.t. input size), but could be
added in the future. %

\begin{figure}[ht]
  \centering
  \Description{The protocol machines involved with connection
    management on the sending and receiving side}
   \includegraphics[width=0.50\textwidth]{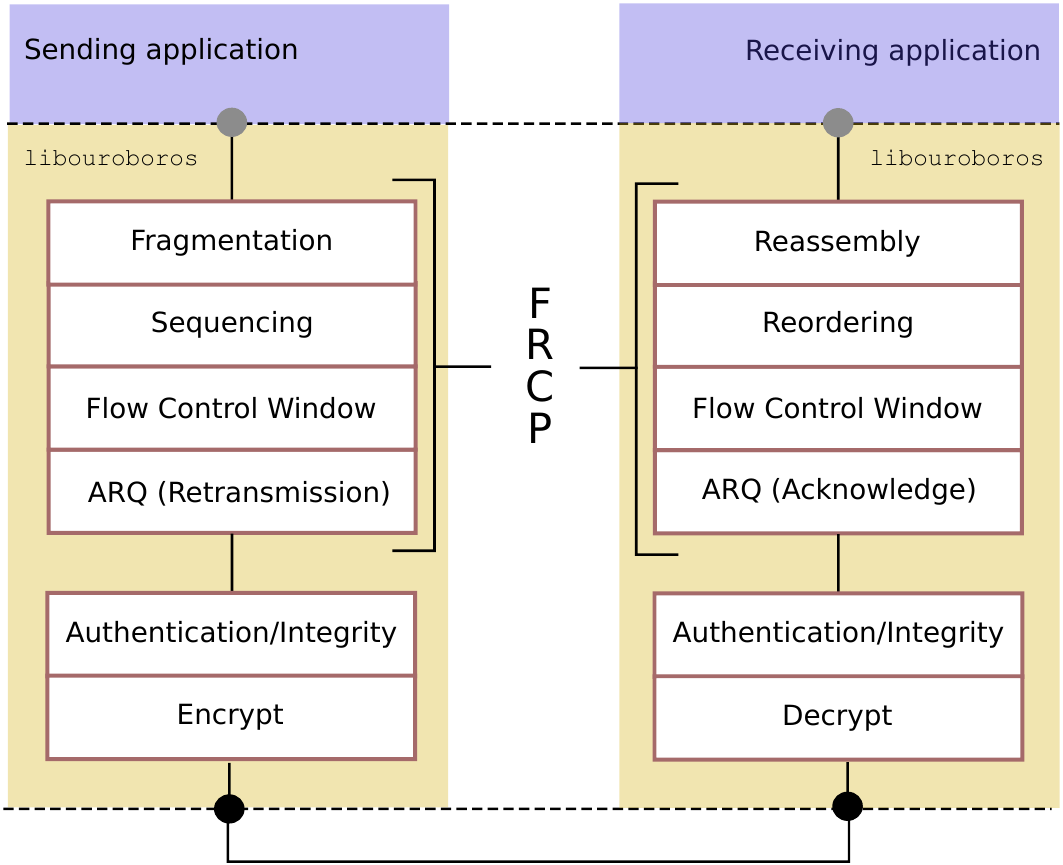}
   \caption{Reliable connection management}
   \label{fig:frct}
\end{figure}

Fragmentation, Automated-Repeat-reQuest (ARQ) and flow control are
combined in a single protocol, the flow and retransmission control
protocol (FRCP) (See Fig. \ref{fig:frct}). %

Ouroboros has the reliable connection state machine as part of every
application; it is part of an application library as opposed to as a
separate (kernel space) program. %
The functions that perform error detection, authentication and/or
encryption are also provided by the library, but separate from FRCP to
allow independent activation and configuration. %
This is clarified in Sec. \ref{ssec:pipeline}. %

\subsection{Quality of service}
\label{ssec:qos}

The design of Ouroboros includes 3 mechanisms that allow maintaining a
certain Quality of Service (QoS) for a flow. %

The first is taking into account QoS requirements in the resource
allocation when a flow is established; for instance, by performing
bandwidth reservation on a per-flow basis during flow allocation%
\footnote{A protocol similar in function to the ReSerVation Protocol
  (RSVP) \cite{zhang1993rsvp} could be introduced.}, %
at step (8) in Fig.\ref{fig:fa}. %
This is a topic for future study. %

The second is differentiated packet handling based on a QoS field in
the packet header. %
Packets with a higher QoS priority can be scheduled accordingly, can
be forwarded along different paths and can be handled differently in
case of congestion. %

The third is the configuration of the FRCP protocol and the
integrity/security mechanism in the library. %
At the application, the specification of the QoS for a flow should be
technology-agnostic \cite{cheong1999qosspec}. %
Ouroboros currently defines its QoS specification in terms of delay
(ms), bandwidth (bits/s), availability (class of 9s), packet loss rate
(packets per billion), bit error rate (errors per billion bits), in
order delivery (boolean) and maximum interruption (ms). %
It should also include authentication and encryption strength; a
technology-agnostic way for specifying the trade-off between
encryption strength and algorithm complexity is needed. %
We discuss the application interface in Sec. \ref{ssec:intf:api} and
the implementation in Sec. \ref{ssec:impl:lib}. %

\subsection{Congestion avoidance}
\label{ssec:cgcontrol}

Flow control prevents the destination process from being flooded by
the sending process, whereas congestion control prevents the network
from being flooded by its users \cite{jain1988congavoid}. %
Usually, the two functions lend themselves to combined implementations
because the underlying mechanism is similar: feedback is used to tune
the rate of a flow \cite{welzl2005cc}. %

Ouroboros separates congestion control from flow control to ensure
that each network layer has full control over its own resources. %
In essence, both flow control and congestion control do work on the
same buffer (the endpoint FIFO to which the application queues
outgoing packets), but with flow control it's the sender application
that will limit sending to the FIFO according to its flow control
window, and with congestion control it's the forwarding IPCP in the
supporting layer that will limit reading from the FIFO according to
its congestion control window. %
A quick glance at Fig. \ref{fig:pipeline} may help the reader; for a
given flow endpoint, the sender {\tt tx} buffer is the IPCP {\tt rx}
buffer and vice versa. %

Congestion avoidance also consists of two different mechanisms:
congestion notification, which is implemented in the DTPMs, and the
congestion management function that is part of the flow allocator. %

In Ouroboros, flow control works on a per-packet basis, while
congestion control works on a per-byte basis. %
The most obvious method for congestion notification is an explicit
congestion notification (ECN) scheme. %
We are currently investigating multi-bit Forward ECN (FECN) approaches
modeled on the operation of Data Center TCP (DCTCP)
\cite{alizadeh2010dctcp}. %

\subsection{Packet pipeline}
\label{ssec:pipeline}

\begin{figure}[t]
  \centering
  \Description{The pipeline for processing packets}
  \includegraphics[width=\textwidth]{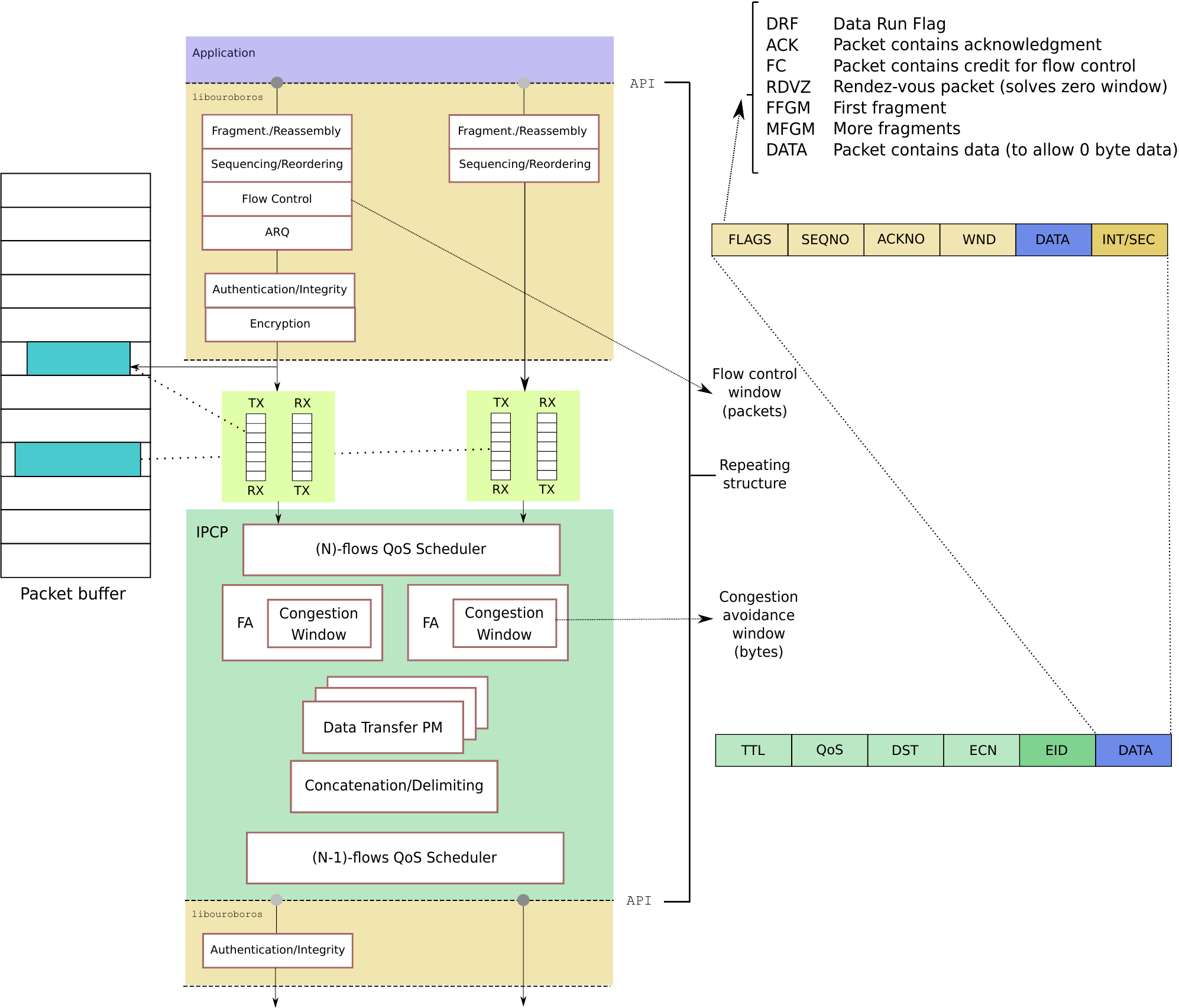}
  \caption{Ouroboros packet processing pipeline}
  \label{fig:pipeline}
\end{figure}

Fig. \ref{fig:pipeline} shows the packet pipeline, representing a
model that can be implemented and optimized in a number of different
ways depending on the requirements and environment: servers, IoT
devices, routers, \ldots will all have very different
implementations. %

We will explain 3 paths through the pipeline: an application writing
to a flow; a packet being forwarded; and an application reading from a
flow. %
The path of the packet is between the endpoints which are a darker
shade of grey. %
We assume the packet passes all steps (most of these steps can be
bypassed if they are not needed to achieve the requested QoS). %
We also assume that there is sufficient credit to send all fragments
and only explain the high-level steps, the details of the Delta-$t$
protocol operation are omitted. %

The system stores packets in a large block of packet buffers to avoid
copies when the packets are processed by the system, which can be
efficiently implemented as a memory pool (also known as a fixed-size
block allocator). %
Each packet buffer allows for head and tail space for adding packet
headers when the packet is encapsulated in a lower layer%
\footnote{Similar in objective as Linux {\em sk\_buff} or BSD {\em
    mbuf}.}. %

When a packet is written to the system using {\em flow\_write()}, it
is first fragmented and the fragments are encapsulated in the FRCP
header and put into the packet buffer. %
The FRCP protocol has a very simple structure, consisting of a
sequence number (SEQNO) field, an acknowledgment number (ACKNO) field,
a field for updating the (flow control) window (WND) and the payload
(DATA), in addition to a FLAGS field containing in total 7 flags, 2
for Delta-$t$: a data run flag (DRF) to indicate there are no
unacknowledged packets, and a rendez-vous (RDVZ) packet that is used
to resolve a zero window at the sender; 2 for fragmentation (the first
fragment FFGM and more fragment (MFGM) bits), and 3 flags that
indicate whether the SEQNO, ACKNO and DATA fields are used. %
If needed, a check or authentication information is appended, let's
assume the packet is encrypted%
\footnote{Fig. \ref{fig:pipeline} shows encryption after
  authentication, so the Message Authentication Code (MAC) is applied
  to the cleartext and then the result is encrypted, which is known as
  MAC-then-encrypt, but this is a just choice we made when drawing the
  illustration. %
  There are other options, such as encrypt-then-MAC and
  encrypt-and-MAC, and each has its strengths and weaknesses
  \cite{bellare2008authenc}. %
}. %

Then, the application releases its ownership of the buffer and passes
a reference to the buffer to the (next) IPCP, via a FIFO queue
(usually implemented as a ring buffer), this is where the layer
boundary is crossed. %

The lower layer IPCP then reads this packet from the FIFO, via a
(N)-flows QoS scheduler that prioritizes the flows%
\footnote{We think this is more efficient than using priority queues
  instead of FIFOs, however, both approaches are valid.}. %
This is also where congestion control is enforced: flows that
experience congestion are throttled via a congestion window. %

The packet is then encapsulated in a minimalistic data transfer
protocol header that consists of only 5 fields%
\footnote{The reader can observe that these fields roughly correspond
  to the {\em mutable} fields in IP, removing the need for security
  mechanisms (akin to IPSec) associated with this protocol.}: %
a Time-To-Live (TTL) that ensures that packets don't live forever in
the network and enforces the Maximum Packet Lifetime of the flows, a
destination address (DST) so it can be forwarded to the destination,
the QoS priority identifier, an ECN field to flag when a flow
experiences congestion, and an endpoint identifier (EID)%
\footnote{Note that according to the model in Sec. \ref{ssec:conn},
  this EID field is not really part of this data transfer protocol,
  but the header of a multiplexer protocol machine. %
  This is important for concatenation and separation (bundling
  multiple packets towards the same destination to avoid the
  processing overhead in intermediate routers), which has to consider
  the EID field part of the payload.}%
\footnote{This EID is already according to our implementation, the
  model would have a PMID field to identify either the FA or
  directory, and the FA would have another mux with PMID for the
  N-flows. %
  The EID field combines these two identifiers into one.} %
to identify the endpoint at the destination (either an internal
component of the IPCP or an N flow, see Sec. \ref{ssec:impl:ipcps} for
more details). %
The fields are initialized to default values by the Flow Allocator
(FA): TTL is set to a certain (configurable) maximum value for the
layer, the default is 60 seconds; DST is filled out with the value
returned from the directory query; the QoS priority is set according
to the QoS specification; and the EID is set according to the value
that was agreed during flow allocation, obtained at step (8) in
Fig. \ref{fig:fa}. %

The packet is then processed according to the set of DTPMs that
corresponds with the QoS priority, possibly concatenated into bigger
units to reduce processing in the intermediate nodes, and forwarded on
the correct outgoing flow%
\footnote{IPCPs use zero-copy versions of the {\em flow\_read()} and
  {\em flow\_write()} operations for efficiency.}. %
If the network is getting congested (the packet experiences
considerable queuing on the outgoing flow), the ECN field is
updated. %
The concatenation step is located just after the forwarding step to
allow concatenation of all packets towards the same destination with
the same QoS, limited by the MPS of the outgoing flow%
\footnote{This is an ideal case which requires quite some state and
  resources (per-destination queues, etc). %
  It is very likely that concatenating packets per-flow may already
  bring sufficient benefit for any implementation, this should be
  evaluated.}. %

Packets coming from a lower layer traverse the pipeline in the
opposite direction. %
They are read from the flow and processed by the (N-1)-flows QoS
scheduler and passed to the set of DTPMs. %
If the packet is for a different destination, it is possibly further
concatenated and forwarded to the destination. %

If the destination address matches the IPCP address, the packet is
delivered to the correct protocol machine according to the EID, in our
case this will identify an N flow. %
If the ECN fields are marked, the FA sends a packet to update the
sender's congestion window. %
The packet is then written to the correct endpoint FIFO, passing the
layer boundary, to be read by the application. %
It passes through decryption and is passed to FRCP, which does
bookkeeping of its reception, puts it in order and possibly
reassembles it into a full packet before finally delivering it to the
application. %

Worth noting is that the protocols in Ouroboros do not require a {\em
  length} field. %
The Heartbleed bug was a well-known example of the potential
consequences that an ill-checked length field can have
\cite{durumeric2014heartbleed}. %

\section{Examples of the model}
\label{sec:examples}

This section presents some examples to aid the reader in understanding
the model presented in Section \ref{sec:concepts}. %
These examples are definitely not the only possible solutions within the
model. %

\subsection{Forwarding within a layer}
\label{ssec:forward}
To illustrate the concepts in the model, we provide a small example of
forwarding in a unicast layer. %

As detailed in Sec. \ref{sssec:dt}, packets are moved through the
network layer via a partially ordered set of DTPMs that implement
{\sc forwarding}. %
In the example, we assume a total order for simplicity, which allows
to use the familiar dotted shorthand notation for the (ternary)
compound name that makes up the node address. %
The node addresses need to be unique within the scope of the layer. %
Upon receipt of a packet (from an internal component or a flow), the
packet will be passed through the DAG of DTPMs. %

\begin{figure}[ht]
  \centering
  \Description{A network of IPCPs and the path taken from one
    IPCP to another as well as passing of the packet between
    protocol machines inside an IPCP}
  \includegraphics[width=0.7\textwidth]{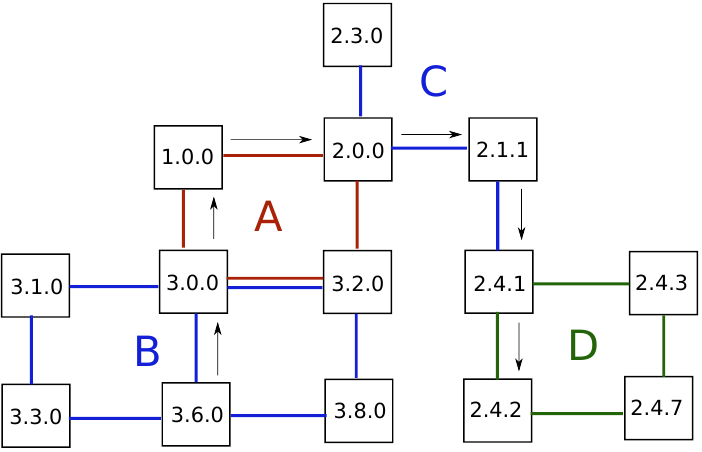}\\
  \vspace{5mm}
  \includegraphics[width=0.6\textwidth]{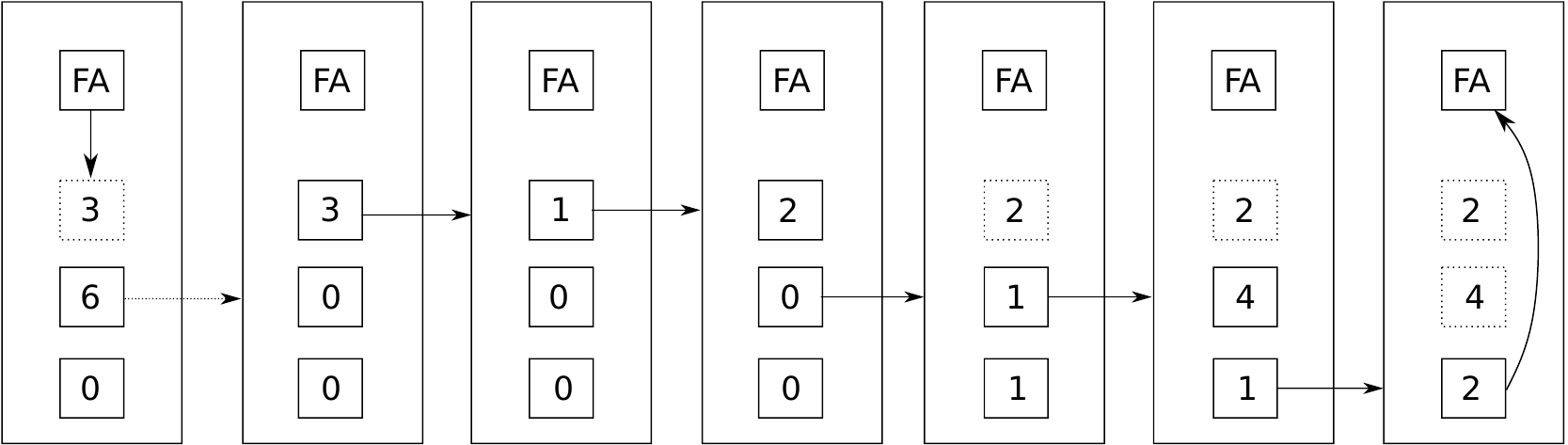}
  \caption{Sending a packet from 3.6.0 to 2.4.2}
  \label{fig:forwarding}
\end{figure}

Fig. \ref{fig:forwarding} shows the example setup. %
The layer has IPCPs that consist of 3 levels of DTPMs. %
Each DTPM has an associated dissemination protocol machine (not shown)
-- for instance, a link state protocol -- at its level. %
For lack of a better term, let's call a group of dissemination
protocol machines that cooperate in a link-state protocol protocol a
{\em routing area}.  %
Routing areas are defined by the scope of underlying broadcast layers,
and the dissemination protocol machines use the broadcast layer for
the area%
\footnote{The reader may be initially puzzled by this statement. %
  Why should an IPCP in a unicast layer need broadcast layers? %
  However, let us gently remind ourselves that for their distribution
  of link-state advertisements, OSPF routers join a multicast group at
  $224.0.0.5$, the scope of which is defined by the IP subnet.}. %
In the case of Fig. \ref{fig:forwarding}, the IPCPs with addresses
$1.0.0$, $2.0.0$, $3.0.0$ and $3.2.0$ participate in a routing area
$A$%
\footnote{The model does not require naming the routing areas,
  although it will often be convenient to do so.} %
at the top level (shown in red). %
At the middle level, there are 2 routing areas: one consisting of the
IPCPs with addresses $3.1.0$, $3.0.0$, $3.2.0$, $3.3.0$, $3.6.0$ and
$3.8.0$ (area $B$), and one containing $2.3.0$, $2.0.0$, $2.1.1$ and
$2.4.1$ (area $C$). %
The nodes with address $3.0.0$ and $3.2.0$ are in both areas $A$ and
$B$, and $2.0.0$ is in both areas $A$ and $C$. %
The dissemination protocol at each DTPM level will announce to the
participating nodes whether a certain node is also participating at
any higher level, which can be done with a single bit. %
Finally, there is a routing area $D$ at level 3, consisting of the
nodes $2.4.1$, $2.4.2$, $2.4.3$ and $2.4.7$. %

We now illustrate sending a packet from $3.6.0$ to $2.4.2$. %
Each DTPM has a complete view of the address, but only makes
forwarding decisions within its own level. %
The IPCP with address $3.6.0$ only participates in the second routing
level, which will cause DTPM $6$ to do a check if the address is local
(matching all higher level names with the IPCP address: dst $2.X.X$ is
not within $3.X.X$). %
The destination is an address outside the area, so DTMP $6$ will
forward to one of the nodes in the area that is marked as
participating in a higher level routing area, for instance node
$3.0.0$. %

Node $3.0.0$ forwards packets at the highest level DTPM, so it sends
it to $1.0.0$, which sends it to $2.0.0$ which has 2 as name for its
highest level DTPM. %
Since the highest level DTPM's name is the same as in the packet, it
is passed to the middle DTPM. %
There the name is different, so it looks in its forwarding table and
the packet is sent to $2.1.1$, which forwards it to $2.4.1$. %
Since the name of the middle DTPM now matches that of the packet,
forwarding continues on the lowest DTPM: $2.4.1$ sends it to
$2.4.2$. %
On the IPCP with address $2.4.2$ all DTPM names are matched, so the
packet is delivered. %



\subsection{End-host mobility}
\label{ssec:mobex}
Maintaining active TCP connections on mobile hosts is quite
challenging%
\footnote{Keep in mind that the connection doesn't break because of
  the interruption, but because switching from one network device to
  the next (or changing from one IP subnet to another when using the
  same network device) causes the IP address to change.}. %
Current solutions -- such as in mobile IP \cite{perkins1997mobile} and
Location-Identifier Separation Protocol (LISP) \cite{meyer2013locator}
-- rely on establishing tunnels, which puts a significant burden on
the tunnel endpoints inside the network, hampering scalability of
these solutions. %

A key advantage of recursive networks is how they can maintain
connectivity for mobile hosts without relying on tunnels, which has
recently been demonstrated for RINA networks
\cite{grasa2018mobility}. %
Our example below will use a slightly different strategy where the
end-host does not participate in a dissemination protocol to avoid
routing updates. %
Both approaches are equally valid and can be used in both RINA and
Ouroboros. %

\begin{figure}[t]
  \centering
  \Description{A figure showing a mobile host moving from WiFi
    to 4G, showing the layers, IPCPs and addresses of the DTPMs
    in the IPCP}
   \includegraphics[width=0.7\textwidth]{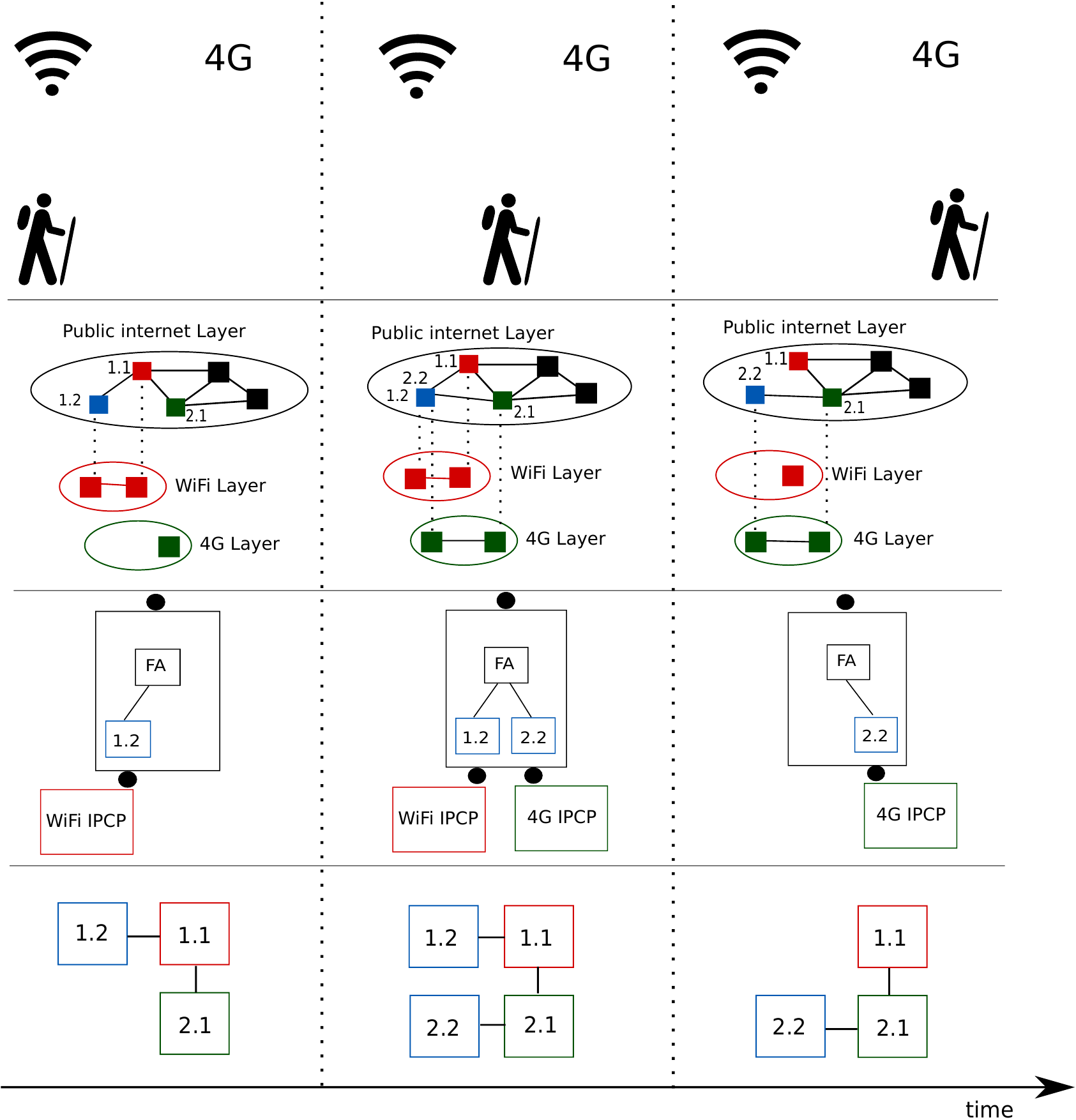}
   \caption{A mobile host switching from WiFi to 4G}
   \label{fig:mobility}
\end{figure}

Fig. \ref{fig:mobility} shows a scenario in which a person, using a
mobile end host that is connected to a wide area internet layer, moves
from an area where access is provided via Wi-Fi technology towards an
area where access is provided through a 4th generation (4G) mobile
network. %

The upper middle section of Fig. \ref{fig:mobility} shows the network
layers that play an important role in mobility. %
From left to right it shows that the mobile device (blue) keeps its
connection in the internet layer by switching from the Wi-Fi layer to
the 4G layer using a make-before-break strategy: during the handover
it is connected to both the Wi-Fi and 4G network%
\footnote{Such a strategy is less feasible moving between different 4G
  networks since the cost of having two antennas may be prohibitively
  high. %
  In this case, a break-before-make strategy would be used that
  minimizes interruption time.}. %
The lower middle section in Fig. \ref{fig:mobility} shows the IPCPs in
the mobile device, with their flow endpoints. %
Initially, the mobile device has two IPCPs, a unicast IPCP with
address $1.2$ that is enrolled in the internet layer, and a
technology-specific WiFi-IPCP%
\footnote{These specific IPCPs are needed to interface with existing
  technologies and at the physical layer. %
  They don't have the usual DTPM but a tailored protocol machine, in
  this case a protocol machines that will transmit packets over
  Ethernet. %
  These are explained further in Sec. \ref{ssec:intf:legacy} and
  Sec. \ref{ssec:impl:ipcps}.} %
that is enrolled in the WiFi layer. %
By now, it should now be clear that address $1.2$ means that there are
actually two DTPMs in a strict order, but we will draw them as a
single DTPM and reference to them as a single DTPM in this example to
simplify the explanation. %

When the host moves, it will detect the 4G network and create another
technology-specific 4G-IPCP that connects to the 4G network. %
The unicast IPCP will be notified of this new 4G IPCP, and allocate a
new flow over the 4G network to the public network layer%
\footnote{A more complete description: the 4G device detects a new 4G
  network in range and informs the IRM in the system. The IRM creates
  the specific 4G IPCP to join that 4G network, and then notifies
  other IPCPs that they may have new options to allocate flows. %
  Subsequent flow allocations are then handled accordingly by the IRM
  to use these new layers.}. %
During the establishment of this new data transfer flow, the mobile
host's public internet IPCP learns that the new gateway has address
$2.1$, so it creates a new DTPM which is assigned a compatible address
$2.2$. %
The flow allocator will notify the end host that it now has 2 DTPMs%
\footnote{The reader may recall from Fig. \ref{fig:fa} that the FA
  keeps a mapping from an N-flow to an address and an N-1 flow
  descriptor, 81: 1, 75 for the IPCP in system 2 in that figure. %
  In the case of this mobility example, the incoming FD (81) will now
  be mapped to 2 <address, FD> pairs.}. %
The flow allocator can now use the 2 DTPMs and may load balance
between them. %
If the Wi-Fi network strength weakens, the unicast IPCP will be
notified, and the flow allocator will take action to stop using the
Wi-Fi flow and use only the DTPMs with address $2.2$ that is using the
4G IPCP. %
When the person moves out of reach of the Wi-Fi network, the 4G
connection takes over all traffic. %
In the right panel, the WiFi signal has dropped completely, all flows
over this Wi-Fi IPCP are deallocated and tthe WiFi IPCP is
destroyed. %

The bottom part of Fig. \ref{fig:mobility} shows a view of the
(combined) DTPMs in the unicast IPCP that is enrolled in the internet
layer. %
The middle section makes clear that the network does not view these
two DTPMs as part of the same IPCP, because the endpoint doesn't take
part in the dissemination protocol to avoid triggering graph adjacency
updates in the network. %

\section{Interfaces}
\label{sec:interfaces}

In this section, we detail the main interfaces in the Ouroboros system
and provide an overview of how Ouroboros can run together with current
network technologies. %

\subsection{Application programming interfaces}
\label{ssec:intf:api}

Ouroboros provides core API interfaces to interact with the system: an
application programming interface (API) for Ouroboros applications
(user programs, and the unicast and broadcast IPCPs) and a management
interface to write tools to manage the networks. %

The application interface for inter-process communication (IPC) is
split into a synchronous API and a scalable event system based on
kqueue \cite{lemon2001kqueue}. %
We provide a brief overview of all the function calls in this API (in
C syntax \cite{kernighan1988cpl}); for a complete listing of all
options, we kindly refer the reader to the man pages that are
installed with the Ouroboros package. %
The API calls for implementing management tools are omitted here
(their details are more technical and less important); instead, we
explain the management tools in Sec. \ref{ssec:impl:irmtools}. %
The benefits of the abstractions provided by Ouroboros are reflected
in the simplicity of these interfaces. %

\begin{center}
  \begin{lstlisting}[caption={Ouroboros application API},
    label={lst:dev-api},
%    float, floatplacement=H,
    captionpos = b, style=CStyle]
/* Returns flow descriptor, qs updates to supplied QoS. */
int     flow_alloc(const char * dst, qosspec_t * qs, const struct timespec * timeo);

/* Returns flow descriptor, qs updates to supplied QoS. */
int     flow_accept(qosspec_t * qs, const struct timespec * timeo);

/* Returns flow descriptor, qs updates to supplied QoS. */
int     flow_join(const char * bc, qosspec_t * qs, const struct timespec * timeo);

int     flow_dealloc(int fd);

ssize_t flow_write(int fd, const void * buf, size_t count);

ssize_t flow_read(int fd, void * buf, size_t count);

int     fccntl(int fd, int cmd, ...);
  \end{lstlisting}
\end{center}

The synchronous API, shown in Lst. \ref{lst:dev-api}, is fully
symmetric. %
It has functions to allocate unicast flows: a client-side {\em
  flow\_alloc()} call and a server-side {\em flow\_accept()} call;
both return a handle to the flow (the {\em flow descriptor}, FD). %
For a broadcast layer, all applications use the {\em flow\_join()}
call, which is similar to {\em flow\_alloc()}. %
The {\em flow\_alloc()} call specifies the QoS for the flow, which is
relayed and returned by the server's {\em flow\_accept()}. %
These calls allow specifying a timeout: a time of 0 seconds results in
asynchronous allocation where the call will return immediately and
allocation will proceed in the background, the result being notified
by the event system; NULL waits indefinitely. %
A {\em flow\_dealloc()} function allows to release the state
associated with a flow. %
Finally there are the self-explanatory {\em flow\_read()} and {\em
  flow\_write()} calls. %
These calls are augmented with a {\em ``flow and connection
  control''}, {\em fccntl()} call that allows configuring some
parameters on the flow, such as specifying whether the read and/or
write calls are blocking, setting timeout delays for the (blocking)
read or write calls, enabling and disabling partial reads, and setting
some configuration options for the transport connection, such as
disabling fragmentation, reassembly or flow control%
\footnote{Similar in use as the POSIX {\em fcntl()}, {\em ioctl()} and
  {\em setsockopt()} calls. %
  The full list of options is given in the man pages.}. %

\begin{center}
  \begin{lstlisting}[caption={Ouroboros ``Hello World'' snippet (unicast)},
    label={lst:hw-uni},
%    float, floatplacement=H,
    captionpos = b, style=CStyle]
static void server_main(void)
{
        int     fd = 0;
        char    buf[BUF_SIZE];
        ssize_t count = 0;

        while (true) {
                fd = flow_accept(NULL, NULL);
                count = flow_read(fd, &buf, BUF_SIZE);
                printf("Message from client is %.*s.\n", (int) count, buf);
                flow_dealloc(fd);
        }
}

static void client_main(void)
{
        int     fd      = 0;
        char *  message = "Hello World!";

        fd = flow_alloc("helloworld-server", NULL, NULL);
        flow_write(fd, message, strlen(message) + 1);
        flow_dealloc(fd);
}
  \end{lstlisting}
\end{center}

An example for a simple {\em ``hello world''} application -- a client
sending the string ``Hello World!'' to a server -- is shown in
Lst. \ref{lst:hw-uni}. %
This application assumes that the server code is registered in some
layer as ``helloworld-server'', which is done outside the application
(we explain these steps in a following paragraph). %
The server (the {\em server\_main()} function) consists of a simple loop
that waits for and accepts a flow, reads a packet from that flow
(which it assumes is a string), prints this sentence, and deallocates
the flow. %
The client (the {\em client\_main()} function) allocates a flow to
``helloworld-server'', writes the message ``Hello World!'' and
deallocates the flow. %

\begin{center}
  \begin{lstlisting}[caption={Ouroboros ``Hello World'' snippet (multicast)},
    label={lst:hw-mc},
%    float, floatplacement=H,
    captionpos = b, style=CStyle]
static void reader_main(void)
{
        int    fd;
        char   buf[BUF_SIZE];

        fd = flow_join("helloworld", NULL, NULL);
        while (true) {
                ssize_t count = flow_read(fd, &buf, BUF_SIZE);
                printf("Message is %.*s.\n", (int) count, buf);
        }
}

static void writer_main(void)
{
        int     fd      = 0;
        char *  message = "Hello World!";

        fd = flow_join("helloworld", NULL, NULL);
        flow_write(fd, message, strlen(message) + 1);/
        flow_dealloc(fd);
}
  \end{lstlisting}
\end{center}

Multicast has a slightly different mode of operation than unicast: the
destination is not a server calling {\em flow\_accept()}, but a {\em
  broadcast layer}, which is reachable using the layer name. %
The broadcast layer provides a {\em many-to-many} packet service,
however client applications can easily be restricted to read-only or
write-only. %
The API uses a {\em flow\_join()} call for broadcast communication. %

A multicast example for a simple {\em ``hello world''} application is
shown in Lst. \ref{lst:hw-mc}. %
The application uses a broadcast layer called ``helloworld'',
accessible via the ``helloworld'' name. %
Writers access the broadcast layer, write the {\em ``Hello World!''}
string and deallocate the resources. %
Readers will read from the layer, displaying the {\em ``Hello
  World!''}  messages from any clients that send it. %

\begin{center}
  \begin{lstlisting}[caption={Ouroboros QoS specification},
    label={lst:qos-api},
%    float, floatplacement=H,
    captionpos = b, style=CStyle]
typedef struct qos_spec {
        uint32_t delay;         /* In ms */
        uint64_t bandwidth;     /* In bits/s */
        uint8_t  availability;  /* Class of 9s */
        uint32_t loss;          /* Packet loss, packets per billion */
        uint32_t ber;           /* Bit error rate, errors per billion bits */
        uint8_t  in_order;      /* In-order delivery, enables FRCT */
        uint32_t max_gap;       /* In ms */
} qosspec_t;

static const qosspec_t qos_raw;
static const qosspec_t qos_raw_no_errors;
static const qosspec_t qos_best_effort;
static const qosspec_t qos_video;
static const qosspec_t qos_voice;
static const qosspec_t qos_data;
  \end{lstlisting}
\end{center}

An important mechanism in the API is the ability to set a Quality of
Service for a flow in a technology-agnostic way, a possible definition
for which is shown in Lst. \ref{lst:qos-api}. %
The struct allows specifying QoS parameters, and a number of these
structs have been predefined for convenience. %
The most important are: two {\em raw} QoS services, one that just
reads/writes the application packets directly to/from the flow and one
that checks integrity (a service similar to UDP); %
a {\em best-effort} QoS that delivers packets in-order; {\em voice}
and {\em video} QoS that add constraints on delay, bandwidth and
interruption time; and a {\em data} QoS that reliably delivers packets
in order. %
When a flow is allocated, the value of the supplied QoS specification
may be updated to the QoS actually supplied to the application. %

\begin{center}
  \begin{lstlisting}[caption={Use of different QoS specifications},
    label={lst:qos-example},
%    float, floatplacement=H,
    captionpos = b, style=BashStyle]
oping --server-name oping --interval 0 --count 10000 --quiet --qos raw

--- oping ping statistics ---
10000 packets transmitted, 10000 received, 2 out-of-order, 0% packet loss, time: 577.336 ms
rtt min/avg/max/mdev = 0.033/0.043/1.213/0.039 ms

oping --server-name oping --interval 0 --count 10000 --quiet --qos voice
Server timed out.

--- oping ping statistics ---
10000 packets transmitted, 9995 received, 0 out-of-order, 1% packet loss, time: 2603.373 ms
rtt min/avg/max/mdev = 0.046/0.071/2.773/0.121 ms

oping --server-name oping --interval 0 --count 10000 --quiet --qos data

--- oping ping statistics ---
10000 packets transmitted, 10000 received, 0 out-of-order, 0% packet loss, time: 609.820 ms
rtt min/avg/max/mdev = 0.025/0.048/0.488/0.028 ms
  \end{lstlisting}
\end{center}

An example is given in Lst. \ref{lst:qos-example}, where a number of
pings (10000) are sent using the Ouroboros ping (oping) application
(see Sec. \ref{ssec:impl:tools}), where the only difference from the
application perspective is that the flow is allocated with different
QoS structs as parameters. %
Because of the multi-threaded nature of the unicast IPCP, packet
reordering is possible when sending very fast pings, but there is no
packet loss. %
With a {\em raw} QoS specification, the flow delivers all packets, but
some arrive out-of-order. %
With a {\em voice} QoS specification, the system will enable in-order
delivery and the out-of-order packets will be dropped. %
A {\em ''Server timed out.''} message is displayed showing that the
oping client gave up waiting (after 2 seconds) for the dropped out-of-order
packets. %
With a {\em data} QoS specification, the system will reorder (and
retransmit) packets, leading to complete in-order delivery of all
pings sent. %

\begin{center}
  \begin{lstlisting}[caption={Ouroboros event system ({\em fset})},
    label={lst:fset-api},
%    float, floatplacement=H,
    captionpos = b, style=CStyle]
struct flow_set;
typedef struct flow_set fset_t;

fset_t *    fset_create(void);

void        fset_destroy(fset_t * set);

void        fset_zero(fset_t * set);

int         fset_add(fset_t * set, int fd);

bool        fset_has(const fset_t * set, int fd);

void        fset_del(fset_t * set, int fd);
  \end{lstlisting}
\end{center}

Scalable applications need to be able to handle a lot of simultaneous
flows, known as the {\em C10k} and {\em C10M} problem for 10 thousand
and 10 million connections to a single server, respectively. %
Ouroboros provides its own event system API, based on kqueue, to
enable writing such applications. %

The first part is for managing flows into groups called flow\_sets or
{\em fset}s, and flows can be added or removed from such a set,
similar to the FD\_SET macros in POSIX, shown in
Lst. \ref{lst:fset-api}. %
There are calls to create and destroy an fset, to add or remove an FD
from a set, an {\em fset\_zero()} call to remove all FDs from the set,
and an {\em fset\_has()} call to check if an FD is part of a set. %
An FD cannot be in multiple fsets at the same time. %

\begin{center}
  \begin{lstlisting}[caption={Ouroboros event system ({\em fqueue})},
    label={lst:fq-api},
%    float, floatplacement=H,
    captionpos = b, style=CStyle]
enum fqtype {
        FLOW_PKT = 0,
        FLOW_DOWN,
        FLOW_UP,
        FLOW_ALLOC,
        FLOW_DEALLOC
};

struct fqueue;
typedef struct fqueue fqueue_t;

fqueue_t *  fqueue_create(void);

void        fqueue_destroy(fqueue_t * fq);

int         fqueue_next(fqueue_t * fq);

enum fqtype fqueue_type(fqueue_t * fq);

int         fevent(fset_t * set, fqueue_t * fq, const struct timespec * timeo);
  \end{lstlisting}
\end{center}

The event system itself operates on these sets. %
The current events that are supported are: the arrival of a packet on
a flow, flow up or down (a failure happened) events, the flow is
allocated (for asynchronous flow allocations) or deallocated. %
The {\em fevent()} call waits for events on any flow in a certain
fset. %
It requires an opaque struct, called an {\em fqueue}, that upon return
of {\em fevent()} will hold all FDs for that have some activity on
them%
\footnote{fqueue (and kqueue on which it is based) is more scalable
  than the well-known {\em select()} call, which requires looping over
  all file descriptors to check for activity.}. %
The {\em fqueue\_next()} call traverses these active flows, and the
{\em fqueue\_type()} call can be used to check the type of the current
event on the current active FD. %



\subsection{Flow Allocation Interface}
\label{ssec:intf:fa}

Every IPCP has to implement a set of primitives that constitute the
flow allocation API and the management operations. %
We restrict ourselves to the flow allocation API, shown in
Lst. \ref{lst:fa-api}, as it provides some insights into what an IPCP
has to implement. %

\begin{center}
  \begin{lstlisting}[caption={Flow Allocation Interface},
    label={lst:fa-api},
%    float, floatplacement=H,
    captionpos = b, style=CStyle]
int   (* ipcp_reg)(const uint8_t * hash);

int   (* ipcp_unreg)(const uint8_t * hash);

int   (* ipcp_query)(const uint8_t * hash);

int   (* ipcp_flow_alloc)(int fd, const uint8_t * dst, qosspec_t qs);

int   (* ipcp_flow_join)(int fd, const uint8_t * dst, qosspec_t qs);

int   (* ipcp_flow_alloc_resp)(int fd, int response);

int   (* ipcp_flow_dealloc)(int fd);
  \end{lstlisting}
\end{center}

The first two calls are to register and unregister a {\small
  \texttt{name}} at the IPCP, adding or removing a directory entry of
<hash, IPCP address>. %
Ouroboros sends the {\small \texttt{name}} information over the API as
a hash for security purposes. %
The IRM of the system learns the hash algorithm for each layer on
bootstrap or enrolment. %
Only unicast IPCPs have to implement the first two functions, in the
broadcast IPCP these functions are NULL. %
The third call queries if a certain {\small \texttt{name}} is
reachable over a layer, it provides an interface into the directory
system, that returns. %
In a broadcast IPCP, this just checks if the hash matches the
broadcast destination. %

Then there are three calls that are used to create the actual flow,
corresponding to the application API calls, of which the first, {\em
  ipcp\_flow\_alloc()}, is the most important. %
Here, the request for a flow is performed, corresponding to step (3)
in Fig. \ref{fig:fa}. %
The {\em fd} parameter is the pending flow that was created by the
IRM, the {\em dst} parameter is the {\small \texttt{name}} of the
destination application, (which is resolved to an address from a
directory query). %
The third parameter, the QoS specification {\em qs} is the most
interesting, as it provides the network with the information that
allows it to reserve some network resources for this particular
flow. %
It will also assign a QoS priority that is used in the forwarding
(per-priority tables) and scheduling (e.g. to prioritize video over
data). %
The {\em ipcp\_flow\_alloc\_resp()} call is invoked when the IRM
responds to an incoming requests request, corresponding to step (5) in
Fig.  \ref{fig:fa}. %
This call is not implemented by broadcast IPCPs. %
Finally, {\em ipcp\_flow\_dealloc()} call releases the local
resources, and any network resources associated with the flow. %
It is an asymmetric function, in that both endpoints of the flow call
it, but only the endpoint that allocated the flow will release network
resources%
\footnote{This could be reversed so that the endpoint that accepted
  the flow releases the network resources. However, this function has
  to be asymmetric to avoid race conditions between both
  endpoints.}. %

\subsection{Interfacing with current network technologies}
\label{ssec:intf:legacy}

We will now detail how Ouroboros plugs into existing technologies,
which follows similar adoption strategies as outlined for RINA
\cite{vrijders2013migr}. %

There are (at least) 4 pieces of technology that allow Ouroboros to
interface with current network infrastructure and applications. %

\begin{figure}[t]
  \centering
  \Description{Stacking of Ouroboros layers over OSI layers shown
    for L1, L2 and L4}
  \includegraphics[width=.8\textwidth]{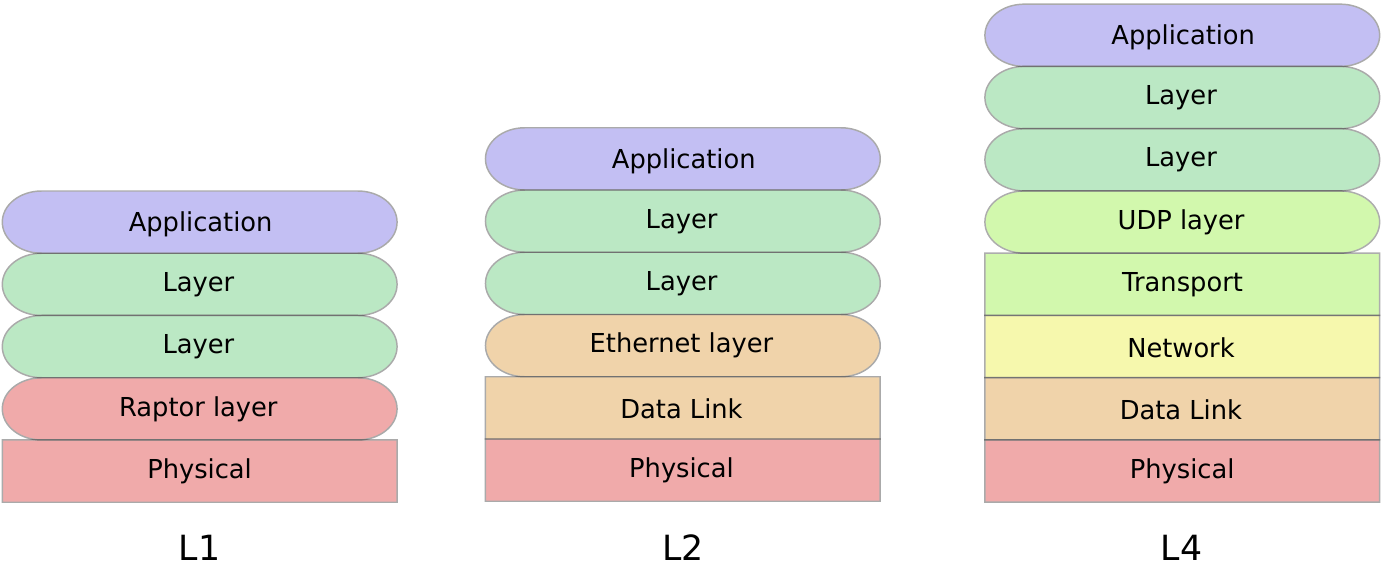}
  \caption{Ouroboros over OSI L1, L2 and L4}
  \label{fig:our-over-x}
\end{figure}

First, we can augment current network protocols with a flow allocator
to implement the Flow Allocation interface in specific IPCPs which
effectively create the lowest layer for the recursive network%
\footnote{Similar IPCPs exist for RINA; however, they do not implement
  a complete flow allocator, but only provide a thin software
  interface that translates calls, dubbing them ``shim'' IPCPs
  \cite{vrijders2013migr}.}. %
This allows to potentially plug Ouroboros into any layer of the OSI
model (Fig. \ref{fig:our-over-x}). %
We currently have implemented a number of IPCPs to run Ouroboros on
top of Layer 1 (a NetFPGA-10G \cite{lockwood2007netfpga}
proof-of-concept called {\em raptor}), Layer 2 (Ethernet), and Layer 4
(UDP). %
These Ouroboros IPCPs are further detailed in Sec. \ref{ssec:impl:eth}
-- \ref{ssec:impl:ipcps}. %

Second, gateways can be implemented between Ouroboros and TCP/IP
networks. %
Such a gateway then listens to the TCP/IP network interface and
translates the destination IP address to a {\small \texttt{name}} that
is known in some layer. %
Such a gateway application has already been demonstrated for RINA
networks \cite{maffione2018rlite, grasa2013irati}. %


A third piece of technology is to provide an application that exposes
a virtual interface and tunnels all traffic on that interface into an
Ouroboros network, creating an Ouroboros VPN, similar to the {\em
  iporinad} tool from the rlite project \cite{maffione2018rlite}. %
This is implemented as the ovpn tool, see Sec.
\ref{ssec:impl:ovpn}. %

Finally, a software adaptation layer is needed to run existing
software over Ouroboros without the need for recompilation. %
A lot of programs are written for the POSIX sockets library, and
higher level input/output (I/O) libraries often make use of it, so
this is a prime target.  A proof-of-concept adaptation layer for POSIX
sockets has already been demonstrated for RINA
\cite{williams2017interposer}. %

\section{Prototype implementation}
\label{sec:impl}


Much of the architectural work was performed in tandem with prototype
development, to incrementally validate our assumptions and assess the
impact of architectural decisions on what we perceived as leading to
unnecessary implementation complexity.  %

The Ouroboros prototype is a user-space implementation (targeted at
Portable Operating System Interface (POSIX) compliant operating
systems) that aims to be accessible and easy-to-use. %
It is mostly targeted towards researchers that want to get acquainted
with the architecture, as well as free software hobbyists and
enthusiasts that are willing to explore new technologies. %
This choice is further motivated by the ongoing IRATI and rlite
efforts to implement RINA%
\footnote{We didn't expect that we would diverge so far from the RINA
  architecture when we started developing the prototype.}, %
which were already targeting kernel-space implementations for Linux. %

\begin{figure}[ht]
  \centering
  \Description{The RPC between and linkage of the different
     software packages}
   \includegraphics[width=0.9\textwidth]{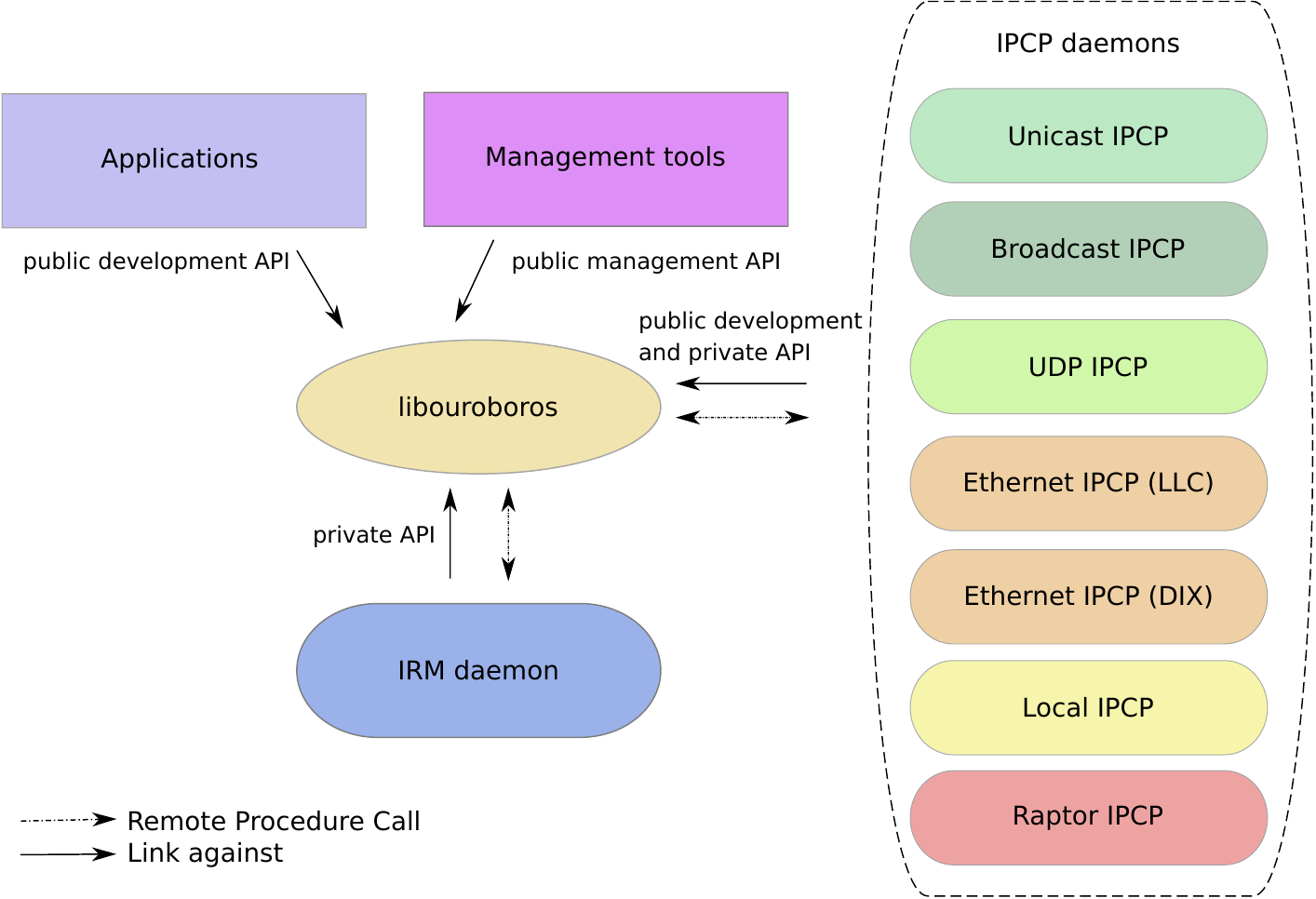}
   \caption{High level overview of the implementation}
   \label{fig:hla}
\end{figure}

The implementation consists of a library, the IPC Resource Manager
daemon (IRMd), a set of IPCP daemons and a set of tools, as shown in
Fig. \ref{fig:hla}. %

The library allows (distributed) applications to communicate over
Ouroboros, allows instructing the IRMd with management tools, and
implements various utility functions as well as the Remote Procedure
Calls (RPC) between the different processes and IRMd in the system%
\footnote{The current prototype uses UNIX sockets for RPC between the
  different components. %
  In a future version this will be removed by bootstrapping all
  communication over the IRMd.}. %

Then there is the central IPC Resource Manager daemon, which should be
considered a part of the operating system. %
It is the back-end for the management toolset, and also responsible
for managing and monitoring applications and IPCPs within the
system. %

7 IPCPs are currently provided, which are also implemented as
daemons. %
First there are the unicast and broadcast IPCPs that constitute the
unicast and broadcast layers, respectively. %
Then there are 3 IPCPs that allow running Ouroboros over the most used
current technologies. %
One for IP networks that encapsulates in UDP datagrams, and two for
Ethernet networks: one using Ethernet II framing and another using
Ethernet with Logical Link Control Type 1. %
A {\em local} IPCP is also provided that is similar to a loopback
interface and is useful for testing complete networks on a single
system. %
Finally, there is the proof-of-concept Layer 1 IPCP, which we named
{\em raptor}, that interfaces with a NetFPGA implementation via a
(Linux-only) kernel driver. %

We also provide a set of tools to manage Ouroboros, that allow users
(typically system administrators) to create/destroy IPCPs and
create/join network layers, and make applications available over
certain layers in the network. %
Some demonstration applications and general tools are also available
as part of the prototype. %


\subsection{Library}
\label{ssec:impl:lib}

Internally, the library is composed of three parts. %
First, it contains the common code shared by the prototype. %
The most important things are the definitions for the RPC messages
between the processes and the UNIX sockets interface currently used
for relaying these messages. %
Apart from that, it holds the data structures that are used elsewhere
in the implementation, such as a hashtable, a bitmap, various hashing
functions and interfaces to external libraries. %

Second, it implements the interface for the management tools to send
commands towards the IRMd over the UNIX sockets. %
As before, we don't go into the details of the management interface%
\footnote{The curious reader can find it in the {\tiny
    \texttt{include/ouroboros/irm.h}} header in the repository.}, %
but limit ourselves to describing the tools that are implemented using
them (see Sec. \ref{ssec:impl:irmtools}).

Third, it provides interfaces for end-user applications, and IPCPs. %
The interfaces described in Sec. \ref{ssec:intf:api} are part of this
library. %
During the execution of the {\em flow\_alloc()} and {\em
  flow\_accept()} calls, the QoS specification is translated into a
configuration for FRCP and a decision is made whether or not to add an
integrity check. %
The configuration for FRCP works as follows. %
If the packets can be delivered out of order, FRCP is completely
disabled, which is useful for some simple applications that do not
need to run any end-to-end protocol over their links, and also for
routers that are designed to handle traffic at line rate. %
When FRCP is enabled, packets will always be delivered in order%
\footnote{This is a feature of the current implementation, not a
  restriction of the architecture. %
  The reason is mainly that implementing out-of-order delivery without
  loss requires a bit of additional bookkeeping.}. %
If there can be no packet loss, ARQ is enabled, if there can be no bit
errors in a packet, an integrity check is added. %
These parameters are all signaled end-to-end during flow
allocation. %
If FRCP is enabled, fragmentation, reassembly and flow control are
enabled by default, and can be disabled by the end application when
needed. %
The library also takes care of the translation between FID and FD,
both for the N flows and the N-1 flows.

One of the key implementation challenges was to implement the
functions that are part of the packet processing pipeline, such as the
flow and retransmission protocol, so that they are passive functions
that are executed in tandem with the I/O or event calls, as opposed to
implementing them as an always-active component of the operating
system. %

To hide most of the implementation details from the API, the RPC
interface and the access to data structures in shared memory (the main
packet buffer and the \texttt{tx}/\texttt{rx} FIFOs in the flow
endpoints) is hidden using {\em init()} and {\em fini()} functions
that are executed automatically when the program initializes and
exits, respectively. %
Since the IRMd creates these data structures, it needs to be started
before any applications that use Ouroboros. %

\subsection{IPC Resource Manager daemon}
\label{ssec:impl:irmd}

The IPC Resource Manager daemon (IRMd) is the central management
component of Ouroboros. %
It can be instructed via the management interface for users (typically
these would be network administrators). %
It performs a number of key management and monitoring tasks, such as
memory management, application and IPCP management, and flow
management. %

\begin{figure}[ht]
  \centering
  \includegraphics[width=0.95\textwidth]{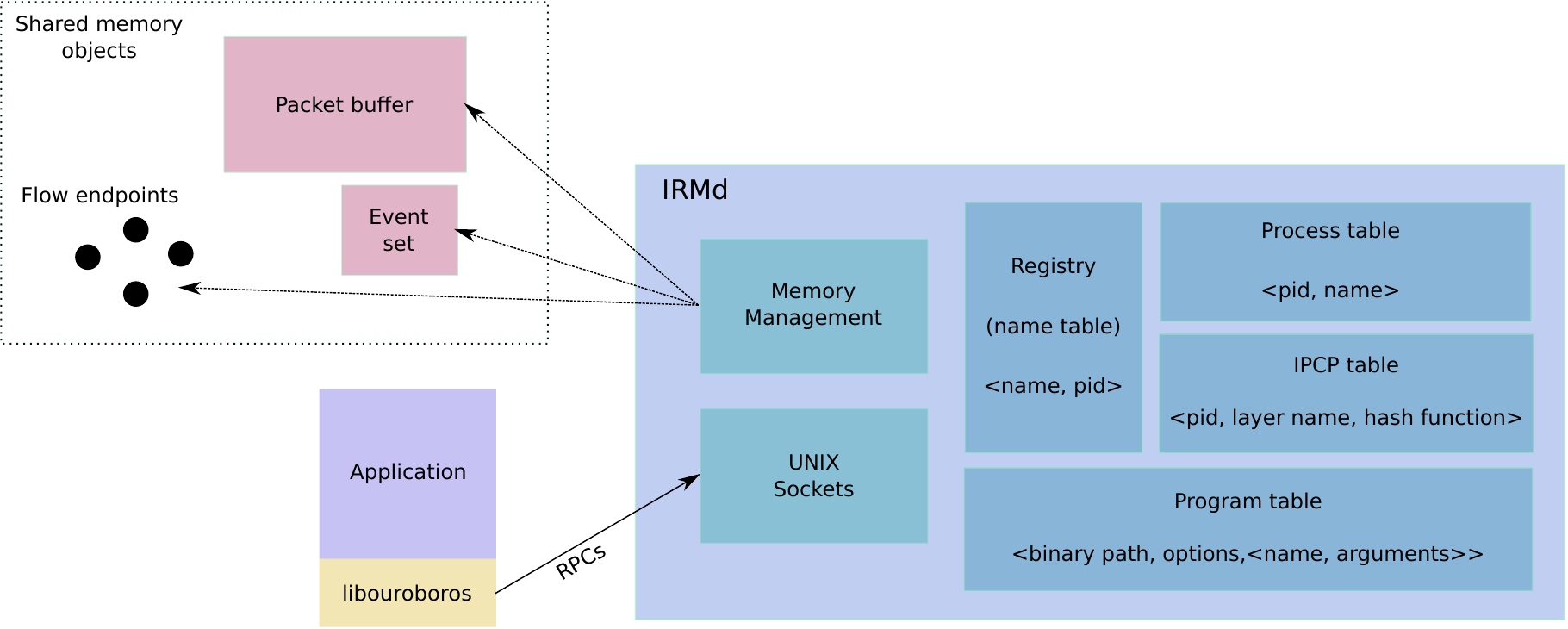}
  \Description{IRMd}
  \caption{The IPC Resource Management daemon}
  \label{fig:irmd}
\end{figure}

Its memory management task mainly comprises creating the shared data
structures that are used for the packet pipeline within the system. %
When the IRMd boots, it creates a (configurable) fixed-block packet
buffer where all packet data is written to. %
It creates shared datastructures for the event system and flow
endpoint FIFOs when needed. %
Currently%
\footnote{At the time of writing, Ouroboros version 0.15.0.}, %
it also creates a UNIX socket over which it listens for new
connections for RPC from programs that use Ouroboros. %
This UNIX socket will be removed in future versions when we bootstrap
the RPC over the Ouroboros pipeline. %

The IRMd also tracks running processes that use Ouroboros. %
When these applications start, they send some basic information to the
IRMd, such as their PID and the name of the binary (during the {\em
  init()} function in the library). %
The IRMd needs this information for {\em binding} (server) programs to
{\small \texttt{names}}, so they can accept flows for these {\small
  \texttt{names}}. %
When a process (identified by its pid%
\footnote{We use the lowercase {\em pid} since this is an actual POSIX
  process ID ({\em pid\_t}) in the implementation.}) %
is {\em bound} to a {\small \texttt{name}}, the IRMd will keep this
mapping in its {\em process table}. %
As a shortcut, it also keeps a {\em program table} which maps programs
to a certain {\small \texttt{name}}, so that all future instantiations
of that program will be bound for that {\small \texttt{name}}. %
The IRMd can also act as a so-called {\em super-server} and start
programs for certain registered {\small \texttt{names}} if the program
that has to accept the flow is known, but not yet running. %
For this, the program table tracks whether it is allowed to act as a
super-server for a particular program and stores the arguments to use
when it instantiates a program for a certain {\small \texttt{name}}. %
The IRMd periodically checks if known processes are still alive, so it
can clean up stale resources (shared data structures for flows and the
event system) when they exit. %
IPCPs are also processes, so they are also present in the process
table, and the IRMd keeps another table for IPCPs that stores some
information such as the names of the layers that they belong to and
the hash function that each of the IPCPs uses. %

For efficiency, it keeps a {\em registry} which keeps track of all
{\small \texttt{names}} known to the system for quick lookups,
avoiding the need to search through the process and program tables. %
Names are added to the registry whenever a new {\small \texttt{name}}
is registered in a layer or bound to a process or program. %
This minimizes latency between flow allocations. %

The IRMd acts as a broker for outbound and incoming flow allocation
requests. %
For outgoing requests (step (1) in Fig. \ref{fig:fa}), the IRMd will
query the IPCPs in the system for the destination {\small
  \texttt{name}} (using the appropriate hashes) and, if that {\small
  \texttt{name}} is known in the layer, request the IPCP to allocate a
flow to that {\small \texttt{name}}. %
For incoming requests (step (3) in system 2 in Fig. \ref{fig:fa}), the
IRMd searches the registry for the {\small \texttt{name}}, and if an
active process is present in the process table, it will wake up this
process (the {\em flow\_accept()} called by the process will return). %
If a program for that {\small \texttt{name}} is known, the IRMd will
first create a process. %

Last but not least, the IRMd also acts as a back-end that executes the
RPC calls from the management tools that the system administrator uses
to create and destroy IPCPs, enrol IPCPs in an existing layer or
bootstrap them to create a new layer, list all IPCPs in a system,
register and unregister {\small \texttt{names}} in a layer, and bind
and unbind processes to a {\small \texttt{name}}. %
We detail these operations further in the tools and examples
sections. %

\subsection{Unicast IPC Process daemon}
\label{ssec:impl:unicast}

\begin{figure}[ht]
  \centering
  \Description{The protocol machines in the implementation of
    the unicast IPC process}
   \includegraphics[width=0.6\textwidth]{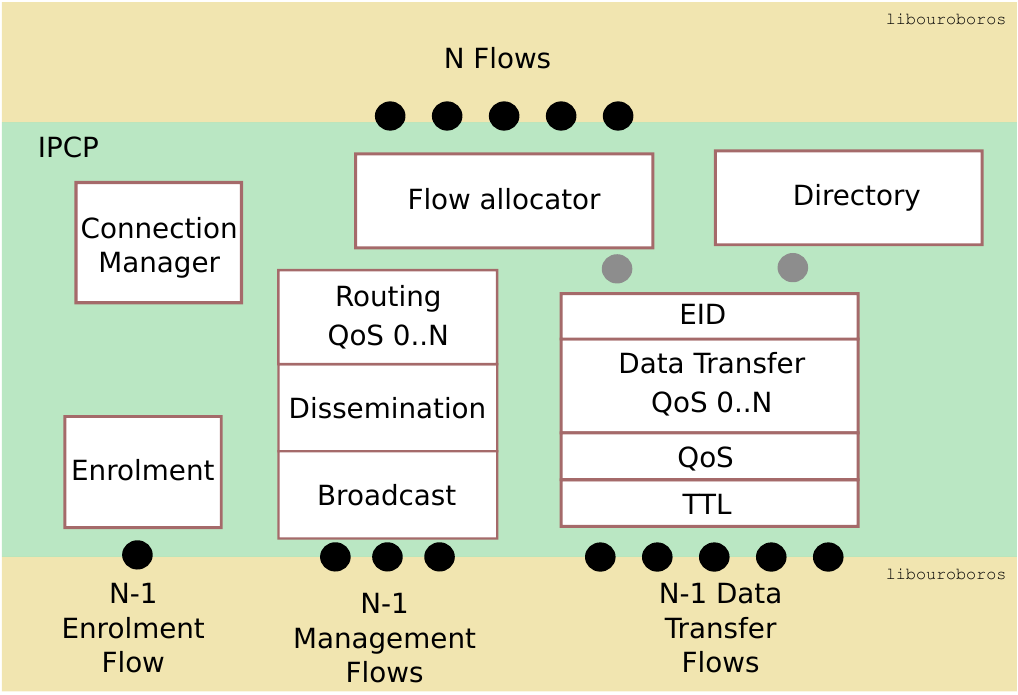}
   \caption{Protocol Machines in the IPC process (implementation)}
   \label{fig:ipcp-impl}
\end{figure}

The unicast IPCP daemon performs the tasks needed in order to allocate
flows to a certain {\small \texttt{name}} making use of existing
layers. %
It provides the flow allocation API towards higher ranked processes,
and makes use of the flow allocation API from lower ranked unicast
IPCPs. %
We took two notable shortcuts in the implementation shown in Fig.
\ref{fig:ipcp-impl} versus the model in Fig. \ref{fig:ipcp}. %
First, we combined the multiplexing protocol machine on top of the
flow allocator (multiplexing to the N-flows) with the multiplexer of
the DTPM (multiplexing to the flow allocator and directory). %
FDs 0-63%
\footnote{To avoid unnecessary arithmetic, the system assigns FDs
  starting from a configurable value to accomodate this internal use
  -- basically using FDs as PMIDs -- the default for this configurable
  value is 64.}%
are reserved to be used for these internal components, larger values
indicate N-flows. %
Second, the TTL protocol machine, QoS multiplexing machine and EID
multiplexer are all implemented as a single Data Transfer component. %
We will now go into a bit more detail of the design decisions for each
of the protocol machines in the unicast IPCP. %
For some of the components (mechanisms), different implementations
(policies) are available, to allow tailoring layers to their operating
environment. %

The connection manager is the component that is in charge of
establishing flows for the IPCP (using the flow allocation API). %
There are three types of flows: enrolment flows, which are used by the
enrolment protocol machine during the enrolment procedure, and torn
down when enrolment is finished; management flows, which are used by
the dissemination protocol machine; and data transfer flows, which are
used by the DTPM. %
After a flow is allocated, the connection managers at the two
endpoints will perform a message exchange%
\footnote{This procedure can be likened in function to the OSI {\em
    Association Control Service Element} (ACSE) protocol
  \cite{iso8650}. %
  The Common Application Connection Establishment Phase in the RINA
  architecture is directly based on ACSE.} %
to determine which component the flow is intended for (flow allocator,
data transfer or enrolment), and what common protocol these components
will use. %
Currently, the information that is exchanged between the connection
managers contains: the name of the protocol machine, a name for the
specific instance (e.g. the DTPM name), and the protocol(s) the
instance supports (the protocol machine passes this information as a
blob to the connection manager when it starts). %
Authentication at the process level should take place at this
exchange. %
If the exchange is successful, the connection manager passes the FD
for the flow to the correct protocol machine, else the flow is
deallocated. %
The component may authenticate again at the component level. %
Note that the flow allocator and directory do not have their own
flows, but communicate over the Data Transfer component. %

The unicast IPCP defaults to a link-state dissemination policy to
spread adjacency information in the layer, so that each IPCP can build
the graph that it needs for its implementation of {\sc routing}%
\footnote{The dissemination PM sits on top of an implementation of a
  {\sc broadcast} transfer PM for distributing the Link State Messages
  (LSMs) to the other members of the layer. %
  The reader will notice that this is a duplication of the stateful
  broadcast layer function, so we could use a broadcast layer to
  support the dissemination of the LSMs. %
  The main reason for not having done this, is that the unicast IPCP
  was implemented before we split the unicast and broadcast mechanisms
  and noticed they basically are different layers. %
  In the current implementation, all unicast IPCPs participate in the
  dissemination protocol, and there is only one level of DTPMs. %
  This would mean we always have to create a broadcast layer having
  exactly the same scope as the unicast layer to support the
  dissemination PM. %
  This is -- when observed from a practical implementation perspective
  -- not all that efficient. %
  When we have different levels of dissemination, like in the example
  in Sec. \ref{ssec:forward}, we will follow the model in the
  implementation.}. %
These messages contain the link endpoint addresses, the QoS
specification of the link, and a sequence number to avoid infinite
forwarding and ordering consistency. %
The collection of LSMs is called the Link State DataBase (LSDB). %

The routing instances use the LSDB to calculate next hop neighbors
(via a shortest path tree) for each supported QoS priority, which are
then used to determine the outgoing FD for each destination address as
entries in the forwarding tables used by the DTPM%
\footnote{Just like modern IP routers make a distinction between the
  Routing Information Base (RIB) and Forwarding Information Base
  (FIB).}. %
The routing instances use link weights according to the QoS
specifications for the links. %
There is an option also to calculate Loop-Free Alternates (LFAs)
\cite{atlas2008basic} for the different next hops. %
The forwarding table in the DTPM has two implementations, one that can
store a single next hop, and another one that can store alternate next
hops for resiliency (such as when using the LFA routing policy). %

The DTPM includes the QoS selection PM (to differentiate between
different cubes) and on top of a TTL (Time to Live) PM, to discard
packets whose TTL is zero. %
Upon receipt of a packet that is addressed to that DTPM, it will check
the endpoint identifier (EID) to see what internal component it is
intended for (flow allocator or directory), or write the packet to an
N-flow. %
In the current implementation, DTPMs are assigned a random 32-bit
address, as the address space is quite large and the chance at
collisions is quite low. %
Efficient distributed address assignment policies are an interesting
topic for future work. %

The directory protocol machine implements the resolver for {\small
  \texttt{names}} to IPCP addresses. %
The default directory implementation is a Distributed Hash Table (DHT)
based on the Bittorrent DHT Protocol specification
\cite{loewenstern2008bep5}, which in turn is based on Kademlia
\cite{maymounkov2002kademlia}. %
The main modifications are support for hash algorithm configuration,
which is exchanged during enrolment of the directory (the Kademlia
JOIN operation), and that the DHT runs on top of the DTPMs (Kademlia
is written specifically for IP networks). %
The DHT is bootstrapped together with the first IPCP, and new IPCPs
try to enrol in the DHT when new Data Transfer flows are created. %

In order to monitor the operation of the unicast IPCPs, they can
export their full LSDB and a number of statistics for each flow. %
With statistics output enabled, the unicast IPCP creates a directory
in {\Small \texttt{/tmp/ouroboros/}} with its name using Filesystems in
Userspace (FUSE). %
For every endpoint (either a flow or internal protocol machine,
represented by its FD), it will display some general information, such
as the remote endpoint address (in case of a data transfer flow), and
a number of statistics such as the number of queued packets and the
number of sent and received packets per QoS priority (See Lst.
\ref{lst:ipcp-fuse}). %

\begin{center}
  \begin{lstlisting}[caption={Flow statistics (partial output)},
    label={lst:ipcp-fuse},
    captionpos = b, style=CStyle]
    Flow established at:       2018-12-30 15:08:17
    Endpoint address:                    949965593
    Queued packets (rx):                         0
    Queued packets (tx):                         0

    Qos cube   0:
     sent (packets):                           213
     sent (bytes):                           72643
     rcvd (packets):                           213
     rcvd (bytes):                           28144
     ...
  \end{lstlisting}
\end{center}

The enrolment protocol machine enrols new members in a layer. %
In case of an enrolment into an existing layer, the enrolment
component of the new member will ask its connection manager for a
connection to the enrolment component a member of the layer. %
Once the connection is established, the existing member will send the
high-level configuration parameters of the layer. %
These include the selected policies for the protocol machines, the
hashing algorithm of the directory, and the layer name. %

\subsection{Ethernet IPC Process daemon}
\label{ssec:impl:eth}

Ouroboros provides a number of options to interface with a physical
(network) device at the lowest level of the recursive network. %

\begin{figure}[ht]
  \centering
  \Description{The internal structure of the Ethernet IPCP
    and its headers}
   \includegraphics[width=0.9\textwidth]{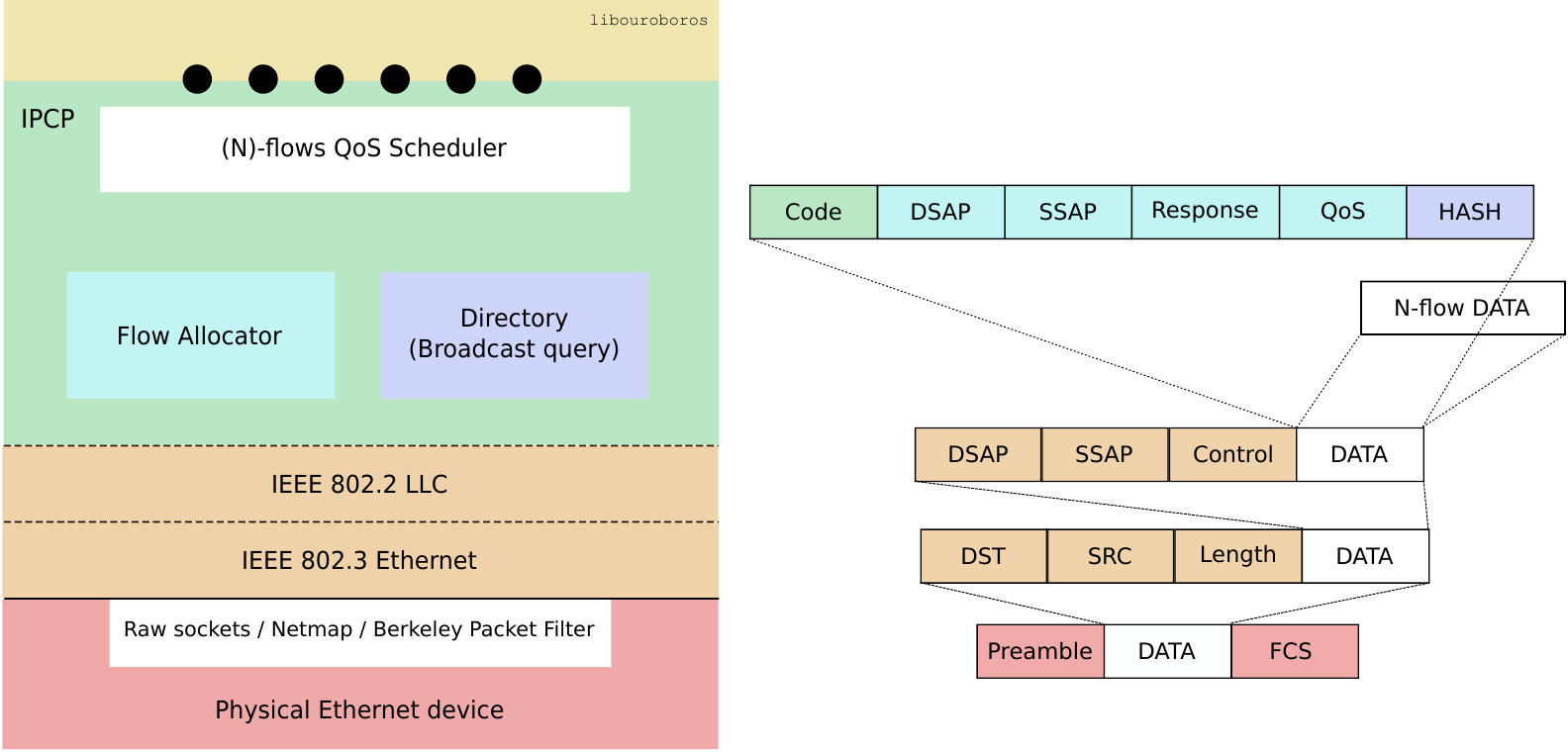}
   \caption{The IPCP over LLC over Ethernet}
   \label{fig:eth}
\end{figure}

The IPCPs that allow running Ouroboros over Ethernet networks are
currently the most used in our lab setup. %
The first Ethernet IPCP (Fig. \ref{fig:eth}), referred to as the
eth-llc, supports IEEE 802.2 Logical Link Control (LLC) type I over
the IEEE 802.3 Ethernet standard. %
The frame format of LLC is close to the Ouroboros (and RINA) model
but we ran into so many issues with LLC support in available equipment
-- WiFi routers in particular -- that we also developed an IPCP over
Ethernet DIX, referred to as the eth-dix, which is the preferred
choice for easy deployment. %

Both these types of Ethernet IPCPs implement a flow allocator and a
directory, which are using a combined (custom) management protocol. %

The directory maps hashes to Ethernet MAC addresses, and each IPCP
only stores its locally registered hashes. %
For resolving a hash, the directory function broadcasts a {\small
  \texttt{query request}} message (MAC address {\footnotesize
  \texttt{FF:FF:FF:FF:FF:FF}}) containing the requested destination
hash. %
Every Ethernet IPCP (of the same type) that has this hash registered
will send a unicast {\small \texttt{query reply}} message back to the
sender with a positive response; in case of multiple responses, the
first one is chosen (anycast). %
This is similar in function to the Address Resolution Protocol (ARP)
\cite{rfc826}, which resolves a generic network protocol address to an
Ethernet hardware address. %
The reason for using a custom protocol instead of ARP is that a lot of
Ethernet equipment seems to drop ARP frames if the protocol type is
not IPv4 (i.e. the PTYPE field is not {\Small \texttt{0x0800}})%
\footnote{The implementation in the Linux kernel also assumes IPv4 as
  the protocol address type. %
  We wrote a kernel module that was compliant with RFC 826 when we
  developed a similar IPCP for RINA \cite{vrijders2013migr}.}. %

The flow allocators are specific to the flavor of Ethernet. %
The eth-llc reuses the Service Access Points (SAPs) from IEEE 802.2 as
the endpoint ID. %
The flow allocation requests are sent to a specific Destination SAP
(DSAP 0), and flows are then established between two SAPs chosen at
the endpoints. %
The eth-dix IPCP uses an EID, for which it adds an inner header%
\footnote{The inner header for the eth-dix also adds a {\em length}
  field to deal with Ethernet padding in runt frames.}. %
A configurable non-standardized Ethertype (default 0xA000) is used to
identify that the traffic is an Ouroboros Ethernet DIX layer, and that
signals that there is an extra inner header present before the actual
payload. %
This Ethertype also serves to run different Ethernet layers in
parallel on the same (Virtual) Local Area Network, which is not
possible with eth-llc. %

To interface with the OS kernels, the Ethernet IPCPs support the raw
sockets API of GNU/Linux, and the Berkeley Packet Filter
\cite{mccanne1993bsd} for BSD and macOS X. %
They also support netmap \cite{rizzo2012netmap}. %

\subsection{UDP IPC Process daemon}
\label{ssec:impl:udp}

While the Ethernet IPCPs are quite convenient for deploying the
prototype in a lab environment, interconnections over larger areas
require interworking with IP networks. %

Similar to the Ethernet IPCPs, we provide a UDP IPCP, that can
transport Ouroboros packets over IP. %
Technically, we could run Ouroboros-over-IP, but the reason why we use
UDP is similar to the reason for creating the eth-llc. %
Ouroboros-over-IP would need the use of a non-standardized IP
protocol, and a lot of available equipment would probably drop such
traffic. %

The directory function in the UDP IPCPs uses the POSIX interface to
resolve hostnames to IP addresses, so it can reuse the existing Domain
Name System. %
Due to restrictions of the lengths of (sub)strings in the DNS
standard, the hash function for the directory in the current UDP IPCP
is limited to an MD5 hash. %
Querying for a name is performing a lookup using the hosts file or a
(configurable) DNS server. %
If Dynamic DNS support for the UDP IPCP is enabled, registering or
unregistering a {\small \texttt{name}} from the UDP IPCP will send a
Dynamic DNS (DDNS) request to a DDNS server that adds or removes an
entry for the associated MD5 hash. %

The flow allocator for the UDP IPCP is similar in function as the flow
allocator in the Ethernet IPCPs. %
Our original implementation of the UDP IPCP reused UDP ports as
endpoint identifiers (just like the eth-llc reused SAPs), with a
dedicated UDP port to initiate the flow allocation. %
This, however, led to various deployment issues due to the ubiquitous
presence of firewalls and Network Address Translation, which block
almost all traffic except for HTTP(S) traffic. %
For easier NAT and firewall traversal, the UDP IPCP was modified to
send all traffic on a (configurable) UDP port, adding an extra inner
header containing endpoint identifiers and a length field like the
eth-dix. %
UDP port 443, which is standardized for the QUIC protocol (which is
encrypted), is a good candidate to use for tunneling Ouroboros, and
therefore our default choice%
\footnote{Tests show 93\% of the time QUIC is successfully used for
  communication by Chrome \cite{langley2017quic}.}. %


\subsection{Other IPC Processes}
\label{ssec:impl:ipcps}

\subsubsection{Broadcast IPCP}
\label{ssec:impl:broadcast}

The broadcast IPCP provided in the implementation is a straightforward
implementation of a stateless broadcast layer. %
A flow join to the name of the broadcast layer will return a flow
endpoint to the requesting application from which it can read and to
which it can write. %
The IPCP itself will forward any received packet on all other flows
(this includes N-flows for broadcasting to other processes in the same
system). %

\subsubsection{Local IPCP}
\label{sssec:impl:local}

The local IPCP provides a local loopback function that allows two
applications on a single system to communicate over a dedicated
link. %
The local IPCP is very useful for testing, allowing the creation of
different networks on a single system. %

\subsubsection{Raptor}
\label{ssec:impl:raptor}

Raptor is a proof-of-concept FPGA demonstrator for running Ouroboros
directly over Ethernet PHY (OSI L1). %
For this, it uses the NetFPGA 10G platform, which has 4 Broadcom
AEL2005 PHY 10G Ethernet devices. %
Raptor consists of three parts: the FPGA design, a Linux kernel module
implementing the device driver and a raptor IPCP in userspace. %
This means that currently raptor only works on GNU/Linux. %
The raptor IPCP is a very minimal IPCP that is similar to the eth-llc
IPCP in its design. %
However, since Raptor runs point-to-point over Ethernet PHY, and we
don't pursue compatibility with Ethernet networks, it does not need
addresses, so raptor's header consists of a single field: the
destination endpoint identifier. %

\subsection{IPC Resource Management toolkit}
\label{ssec:impl:irmtools}

The management of recursive networks revolves around the ability to
create IPCPs and configuring them to enroll into a certain distributed
application -- a network layer -- so that it can start forwarding or
broadcasting packets for client applications. %

\begin{figure}[ht]
  \centering
  \Description{The hierarchy of IRM tool commands}
   \includegraphics[width=0.9\textwidth]{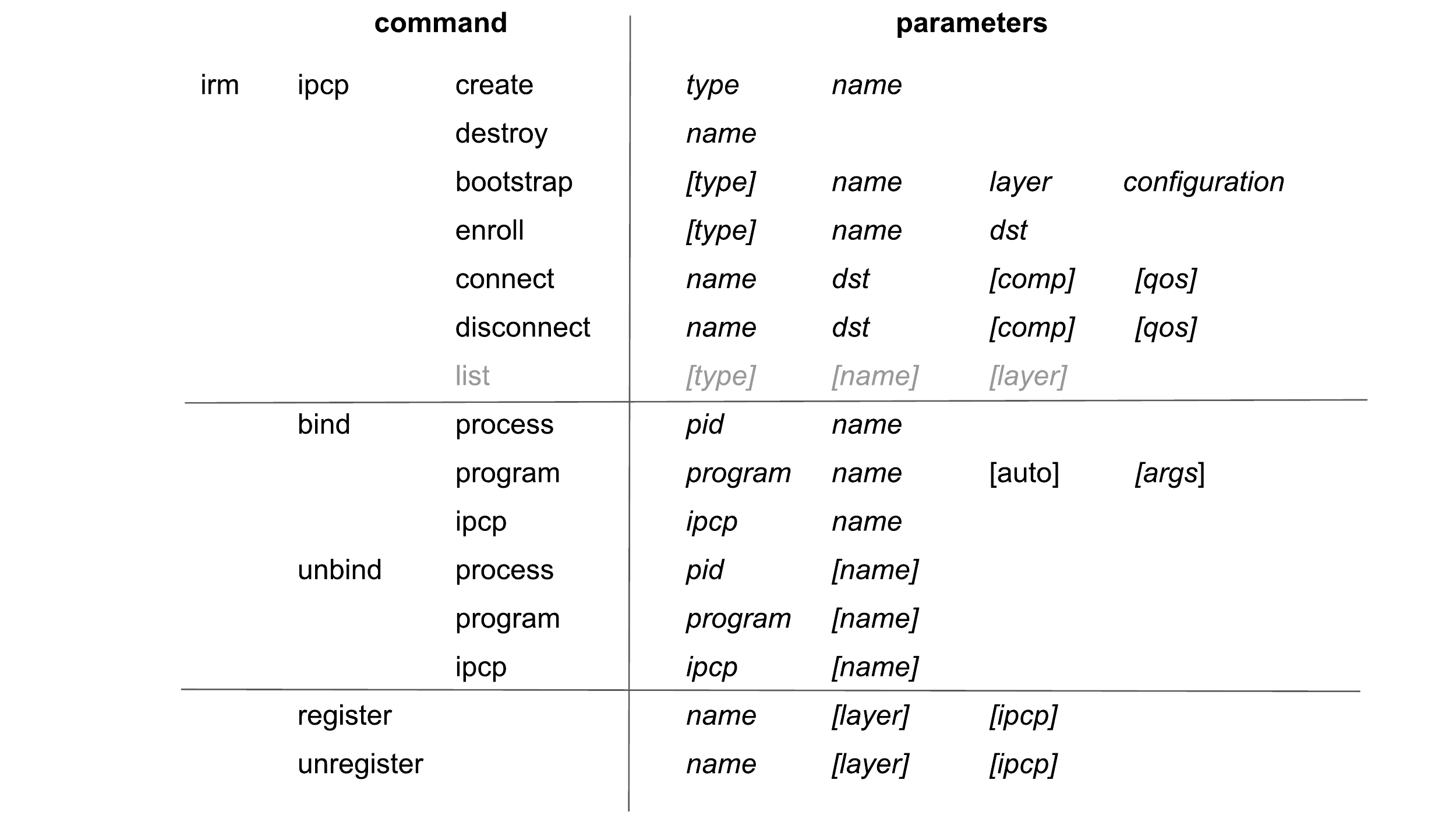}
   \caption{Overview of the command line interface for management}
   \label{fig:irmtool}
\end{figure}

The Ouroboros prototype provides a toolkit that implements the current
management interface to create and manage network layers. %
It provides a command-line interface%
\footnote{Complete details are available in the manual pages. %
  We will provide some step-by-step examples in
  Sec. \ref{ssec:impl:examples}.} %
(CLI) that allows a user (network administrator) to create and destroy
IPCPs, list the IPCPs in a system, enroll in network layers, bootstrap
new layers, and manage adjacencies in unicast and broadcast layers
(Fig. \ref{fig:irmtool}). %
Parameters that are shown between brackets are optional. %

IPCPs of a certain type can be created or destroyed with the %
{\Small \texttt{irm ipcp create name <name> type <type>}} %
command. %
The name of the IPCP should be unique within the layer, and can be
used within the system to reference this IPCP. %
The %
{\Small \texttt{irm ipcp destroy name <name>}} %
command will destroy the IPCP with that name.%

To create a new network layer, an IPCP should be bootstrapped,
specifying the correct configuration for the layer. %
Default options are provided to ease configuration. %
Providing an IPCP {\em type} to the bootstrap command is a shorthand
to create the IPCP if an IPCP with that name is not yet present in the
system. %
As an example, to create a new layer connected to Wi-Fi on device
{\Small \texttt{wlp2s0}}, the command is %
{\Small \texttt{irm ipcp bootstrap type eth-dix layer wifi name wifi.0
    dev wlp2s0}}. %
This will create an Ethernet II IPCP that will capture all traffic on
that interface with (default) Ethertype {\Small \texttt{0xA000}} as
Ouroboros traffic. %
To create a new unicast layer named myvpn with default options%
\footnote{Currently this is: 32-bit address space, random address
  assignment, a starting TTL of 60, simple link-state routing, and a
  SHA3-256 hash function for the DHT directory implementation.}, %
the command is %
{\Small \texttt{irm ipcp bootstrap type unicast layer myvpn name
    myvpn.0}}. %
A listing of these IPCPs on a system is shown in
Lst. \ref{lst:list-ipcps}. %

\begin{center}
  \begin{lstlisting}[caption={IPCPs in a system},
    label={lst:list-ipcps},
    captionpos = b, style=CStyle]
    #irm ipcp list
    +---------+----------------------+------------+----------------------+
    |     pid |                 name |       type |                layer |
    +---------+----------------------+------------+----------------------+
    |   31185 |              myvpn.0 |    unicast |                myvpn |
    |   31158 |               wifi.0 |    eth-dix |                 wifi |
    +---------+----------------------+------------+----------------------+
  \end{lstlisting}
\end{center}

To expand the scope of a layer, new IPCPs can be added to it. %
For the raptor, eth-dix, eth-llc and udp layers, all IPCPs need to be
bootstrapped as the underlying technologies don't have an enrolment
procedure for end hosts. %
Correctly configuring these IPCPs at bootstrap will add them to the
new layer. %
To expand the scope of a unicast or broadcast layer, new IPCPs need to
be enrolled with an existing member. %
The %
{\Small \texttt{irm ipcp enrol type unicast name myvpn.1 dst
    myvpn.0}} %
command will try to enrol the IPCP with name myvpn.1 with an existing
member accepting flows for the {\small \texttt{name}} myvpn.0. %
To simplify enrolment, we usually register ipcps with two names: a
unique name (e.g. myvpn.0) and the layer name (myvpn), so we can use
the layer name (myvpn) as the destination for enrolment
(i.e. anycast). %
As with bootstrapping, providing a type for the enrolment command is a
shorthand to create the IPCP if an IPCP with that name does not yet
exist in the system. %
The bootstrap and enrol commands both have an {\em autobind} option
that will bind the IPCP to two {\small \texttt{names}}: the layer name
and its IPCP name. %

After an IPCP is enrolled, it needs flows to some existing members
(adjacencies) over which it can transfer packets. %
The %
{\Small \texttt{irm ipcp connect name myvpn.1 dst myvpn.0}} %
command would create two flows between myvpn.1 and myvpn.0, one for
data transfer (the adjacency), and one for management%
\footnote{The unicast IPCP currently integrates its own broadcast
  layer (Sec. \ref{ssec:impl:unicast}). %
  The management flow is for that ``internal'' broadcast layer.} (for
sending link-state messages). %
The %
{\Small \texttt{irm ipcp disconnect name myvpn.1 dst myvpn.0}} %
command can be used to remove these flows. %

As a handy tool, the %
{\Small \texttt{irm ipcp list}} %
command can be used to list IPCPs on the system. %
The optional parameters are to filter the list if many IPCPs are on
the same system. %

The above commands allow the management of network layers. %
To make a process reachable over a layer, two steps must be taken:
binding that process to a {\small \texttt{name}} and registering the
{\small \texttt{name}} in a layer. %
The %
{\Small \texttt{irm bind process 2765 name server}} %
command would bind a process with process ID 2765 to the {\small
  \texttt{name}} ``server''. %
We provide a shorthand to bind a certain program to a name: %
{\Small \texttt{irm bind program /usr/bin/oping name oping.server}} %
would bind all future instances of the oping executable to the {\small
  \texttt{name}} oping.server. %
A similar shorthand is provided for IPCPs so we can use the name that
we provided when the IPCP was created. %
This is to avoid the need to look for the {\em pid} of every process
that is started. %
The %
{\Small \texttt{irm unbind}} %
command works the same way. %
The name is optional: all names for a certain process or binary are
unbound if no name is specified. %

Finally, {\small \texttt{names}} can be registered and unregistered in
a layer using the %
{\Small \texttt{irm register}} %
and {\Small \texttt{irm unregister}} %
command. %
These commands usually take a layer name, but if multiple IPCPs of the
same layer are present on a single system (for scalability or simply
for testing), the {\small \texttt{name}} can be registered or
unregistered with a specific IPCP. %

Note that the order of the commands is not fixed. %
There is no need to first register and then bind, or vice versa. %
This is needed to provide as much flexibility as possible%
\footnote{As an advanced exercise, we encourage the reader to try
  this: First start a server process, then create the network layers,
  then register a name in that network, then bind the server to that
  name and then connect a client to that server.}. %

\subsection{Ouroboros Virtual Private Network tool}
\label{ssec:impl:ovpn}

An important tool is the Ouroboros VPN (ovpn) tool, which allows
tunneling IP traffic over an Ouroboros network, similar in operation
to the {\em iporinad} daemon for RINA networks. %

\begin{figure}[ht]
  \centering
  \Description{Three systems with their layering and the ovpn
    application on the two end systems. How it works on a single
    system is also shown, from a software design point of view.}
   \includegraphics[width=\textwidth]{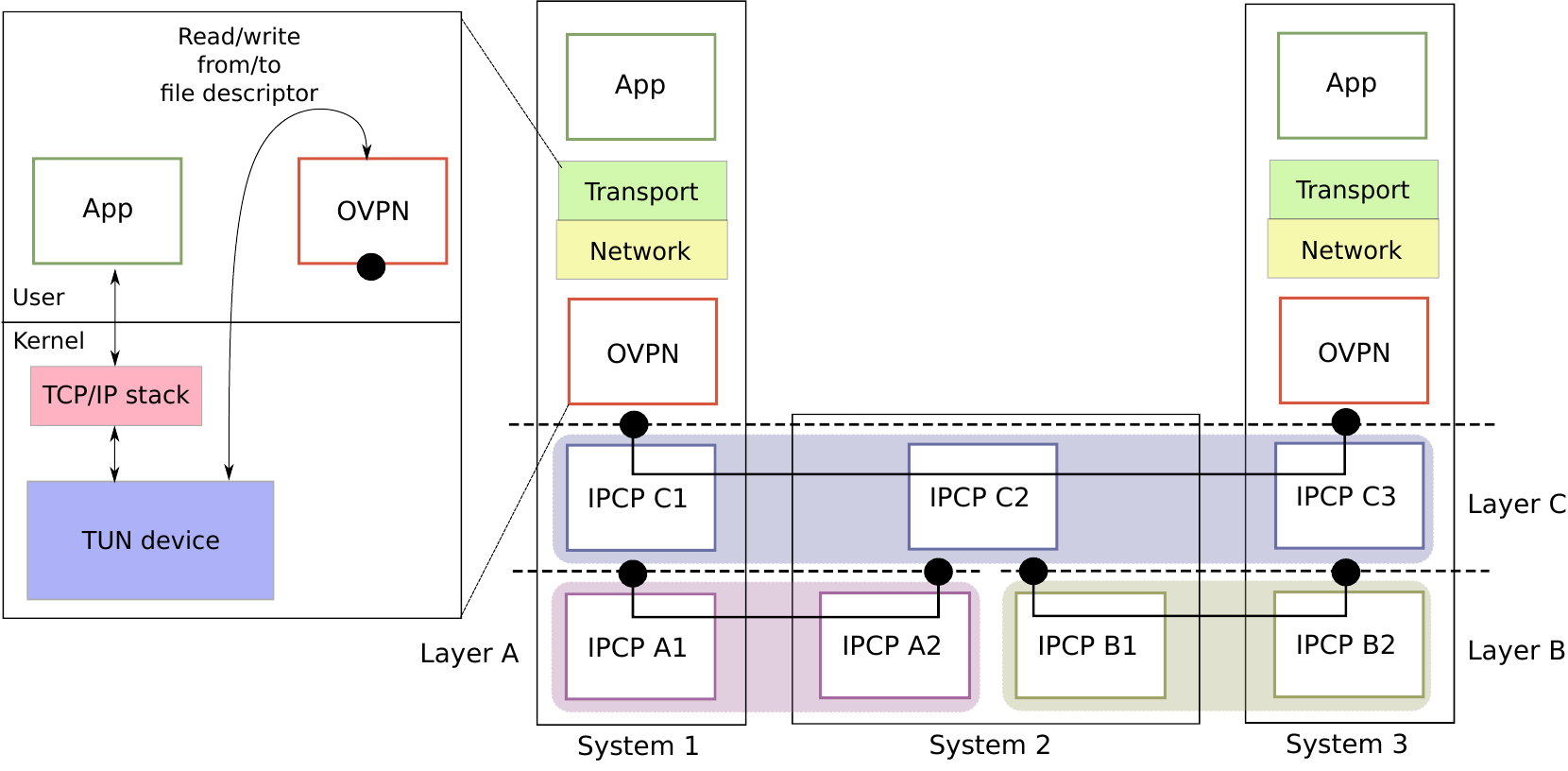}
   \caption{Running IP over Ouroboros with ovpn}
   \label{fig:ovpn}
\end{figure}

The operation of OVPN is shown in Fig. \ref{fig:ovpn}. %
It creates a virtual interface in the host operating system, from
which it reads the packets using the POSIX sockets API and writes them
to a layer using the flow allocation API provided by Ouorboros. %
Currently, the OVPN tool creates a so-called {\em tun} device%
\footnote{A {\em tap} device could easily be added should a need for
  it arise, allowing native Ethernet tools to run over Ouroboros.},
which allows TCP/IP applications to simply use this device like any
other IP interface. %
Administrators can either add a specific routing entry to tunnel the
traffic over Ouroboros, or end users can configure their applications
to use the tunnel interface's IP address. %
The tool allows specifying an IP address and subnet mask to automate
the tunnel configuration process. %

\subsection{Tools}
\label{ssec:impl:tools}

Several small tools are provided together with the Ouroboros
implementation. %

The oecho tool is a simple application that sends a message to a
server which echoes it back. %
The ``hello world'' example in Lst. \ref{lst:hw-uni} is based on this
tool. %
The ocbr and operf tools allow to perform some unidirectional and
bidirectional throughput tests of the prototype. %
The oping tool allows measuring the round trip time between a client
and a server, similar to the well-known ping tool. %
The obc tool is a simple example of a multicast application running
over Ouroboros and is either started as a reader where it receives
messages sent to the multicast flow, or as a writer, where it sends
messages on the multicast flow. %
We also provide a patch for the ioq3 engine \cite{id2012q3a,
  ioquake2012ioq3} to run games that use this engine over Ouroboros
for demo purposes. %

\subsection{Examples}
\label{ssec:impl:examples}

To conclude the implementation section, we'll walk through two
examples for creating 4-node network layers, one example demonstrating
a unicast layer and one example demonstrating a broadcast layer. %
We will use the oping application to demonstrate unicast, and the obc
application to demonstrate broadcast.%

\begin{figure}[ht]
  \centering
  \Description{The IPCPs on the different systems and which
    layers they are using}
  \includegraphics[width=\textwidth]{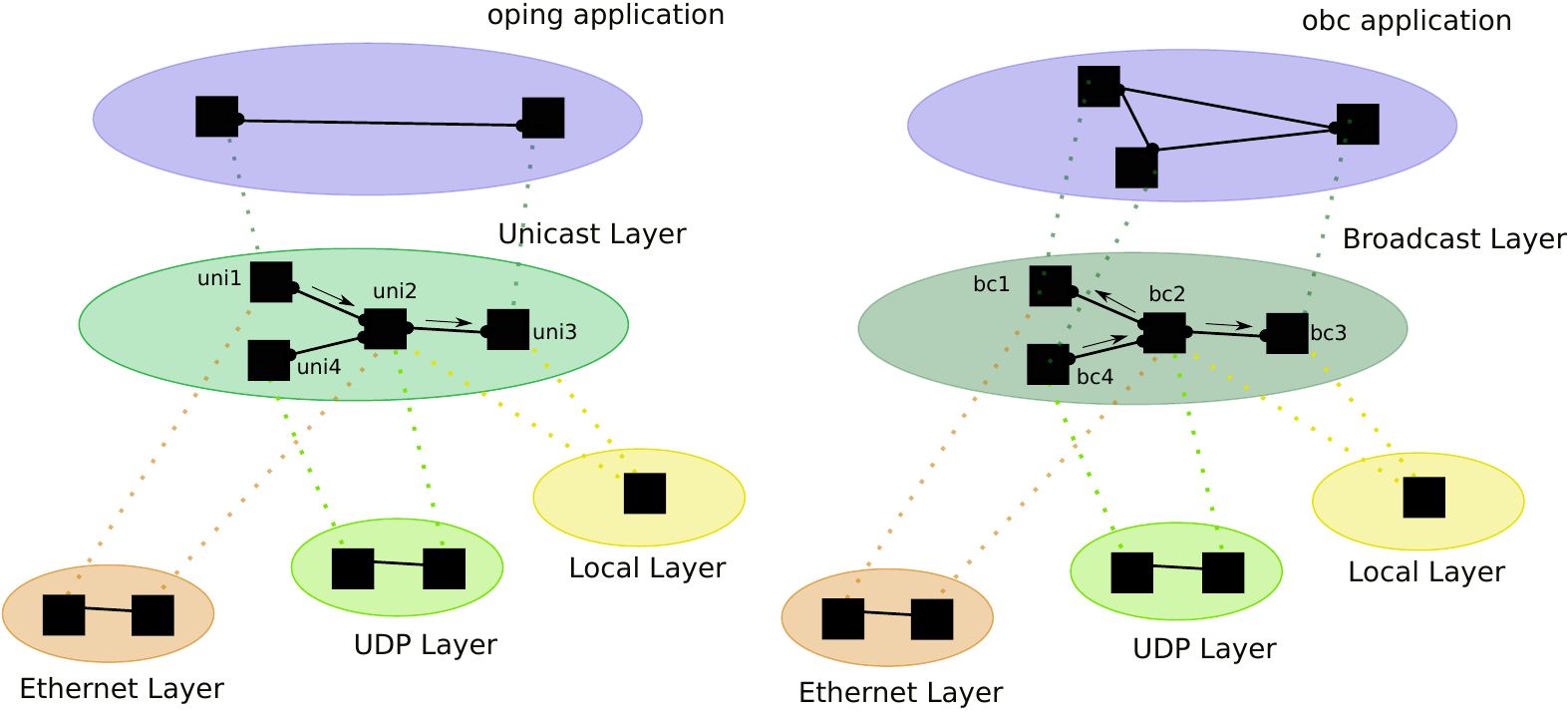}
  \caption{Example scenario for unicast and broadcast}
  \label{fig:simpexs}
\end{figure}

These examples are illustrated in Fig. \ref{fig:simpexs}, unicast on
the left and broadcast on the right. %
To demonstrate the use of different layer types for point-to-point
connectivity, we will use an Ethernet layer, a UDP layer and a local
layer. %
Using a local layer, we need three systems for this example, labelled
$a$, $b$ and $c$. %
System $a$ is connected to system $b$ over the pure Ethernet layer,
system $c$ is connected to system $b$ over a UDP layer on IP subnet
10.1.1.0/24. %
We will number the IPCPs in the top layer $1$ to $4$, with $1$
residing in system $a$ and $4$ residing in system $c$. %
The oping server application will be registered at unicast IPCP $3$. %

\begin{figure}[ht]
  \centering
  \Description{The IPCPs on the different systems and which
    layers they are using}
  \includegraphics[width=0.9\textwidth]{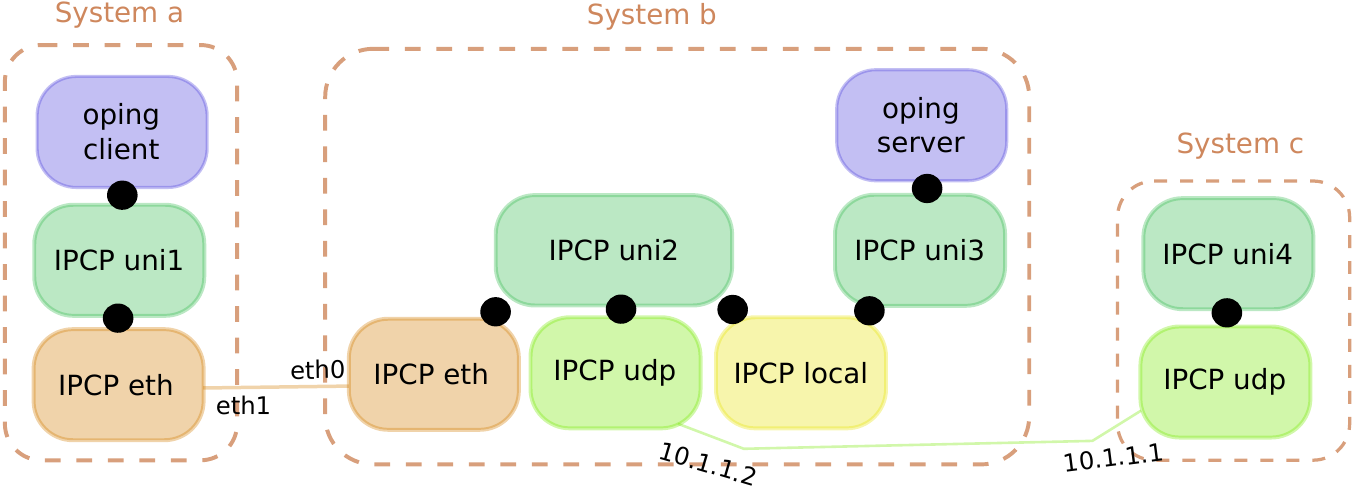}
  \caption{Example scenario (unicast)}
  \label{fig:simpex-uni}
\end{figure}

This is shown in a bit more detail for the unicast example in
Fig. \ref{fig:simpex-uni}. %
In the following, we assume each system has a running IRMd instance. %
We will bootstrap the network from system $a$, which connects to
system $b$ via a pure Ethernet link via its eth1 Ethernet device. %
The commands to create the IPCPs in system $a$ are shown in
Lst. \ref{lst:cmdsa}. %

\begin{center}
  \begin{lstlisting}[caption={Bootstrapping IPCPs on system $a$},
    label={lst:cmdsa},
    captionpos = b, style=BashStyle]
    irm ipcp bootstrap name eth layer ethernet type eth-dix dev eth1
    irm ipcp bootstrap name uni1 layer unicast type unicast autobind
    irm register name uni1 layer ethernet
    irm register name unicast layer ethernet
  \end{lstlisting}
\end{center}

The first command creates and bootstraps an eth-dix IPCP called
``eth'' in an Ethernet layer on the eth1 interface. %
This IPCP will provide the data transfer and management flows for the
unicast IPCP. %
The second command creates this unicast IPCP, named ``uni1'', and
bootstraps it in a layer called ``unicast''. %
The autobind option will bind this IPCP to the {\small \texttt{names}}
uni1 and unicast (so it is equivalent to adding two {\Small
  \texttt{irm bind}} commands).  %
Then both these {\small \texttt{names}} are registered in the eth
IPCP. %
After executing these commands, we have two 1-node layers, and uni1
can accept flows to both ``unicast'' and ``uni1'' over the Ethernet
layer. %

\begin{center}
  \begin{lstlisting}[caption={Bootstrapping IPCPs on system $b$},
    label={lst:cmdsb},
    captionpos = b, style=BashStyle]
    irm ipcp bootstrap name eth layer ethernet type eth-dix dev eth0
    irm ipcp bootstrap name udp layer ipnet type udp ip 10.1.1.2
    irm ipcp bootstrap name local type local layer local
    irm ipcp enrol name uni2 dst unicast autobind
    irm ipcp connect name uni2 dst uni1
    irm register name uni2 layer ethernet layer ipnet layer local
    irm register name unicast layer ethernet layer ipnet layer local
    irm ipcp enrol name uni3 dst unicast
    irm ipcp connect name uni3 dst uni2
    irm bind program oping name oping auto -- -l
    irm register name oping ipcp uni3
  \end{lstlisting}
\end{center}

We will now add the second system, $b$, to the network, and begin with
creating the three lower layers. %
System $b$ connects to the same Ethernet as system $a$ via its eth0
device. %
The first command in Lst. \ref{lst:cmdsb} bootstraps an eth-dix IPCP
on the eth0 interface, allowing applications in system $b$ to
communicate over Ethernet to system $a$. %
System $b$ is connected to system $c$ over an IP network, and has IP
address $10.1.1.1$ assigned to the connected interface. %
The second command bootstraps an IPCP over UDP/IP into a layer called
ipnet on this interface. %
Third, it bootstraps a local IPCP for loopback that emulates an
internal link. %
The local layer consist of only this one IPCP. %

Now that the lower layers are created in IPCP $b$, the unicast layer
can be extended with a new node. %
To do this, we create a unicast IPCP and enrol it with the IPCP in
system $a$. %
Again, this can be done with a single command, shown on line 4, which
creates an IPCP named uni2 and not only enrols the IPCP in the unicast
layer, but also binds it to the {\small \texttt{names}} ``uni2'' and
``unicast''. %
After the uni2 IPCP is enroled, we create an adjacency with uni1 so
it can forward packets to it. %
To be able to further extend the unicast layer from system $b$, we
register the {\small \texttt{names}} ``uni1'' and ``unicast'' in all
three lower layers. %
Now that these {\small \texttt{names}} are known in the local layer,
it is possible to bootstrap the second unicast IPCP in system $b$ and
add it to the unicast layer. %
Note that the enrolment command doesn't specify which layer it needs
to use, the IRMd will select this, and it will first try any IPCPs of
type ``local'', ensuring the situation in Fig. \ref{fig:simpex-uni}. %

We will make the oping application available via the uni3 IPCP. %
To do this, we bind the program to the {\small \texttt{name}} oping
and register that in the uni3 IPCP. %
The bind command uses the auto option, which will make the IRMd in
system $b$ act as a super-server for the oping application: if we
don't start an oping server ourselves, the IRMd will start one for us
-- adding the {\Small \texttt{-l} argument} -- whenever it receives a
flow allocation request for the {\small \texttt{name}} oping. %
The unicast layer now has 3 active members: uni1 on system $a$,
connected to uni2 on system $b$, in turn connected to uni3 in the same
system. %

\begin{center}
  \begin{lstlisting}[caption={Bootstrapping IPCPs on system $c$},
    label={lst:cmdsc},
    captionpos = b, style=BashStyle]
    irm ipcp bootstrap name udp layer ipnet type udp ip 10.1.1.1
    echo '10.1.1.2     869bab4dfd0c0fbedecd374c7089c7a0' | sudo tee --append /etc/hosts
    echo '10.1.1.2     eebe0851cda7f85aeef4229567441a2d' | sudo tee --append /etc/hosts
    irm ipcp enrol name uni4 dst unicast
    irm ipcp connect name uni4 dst uni2
  \end{lstlisting}
\end{center}

The last step needed in creating our 4-node unicast layer is adding
system $c$ to the network (Lst. \ref{lst:cmdsc}). %
System $c$ connects to system $b$ over an IP network via the interface
with IP address $10.1.1.2$, over which we bootstrap the UDP/IP layer,
similar to what we did on system $b$. %
The unicast IPCP on system $c$ will need to enrol with uni2 on system
$b$, which means that the UDP IPCP needs to resolve an address for
``uni2'' and ``unicast''. %
To keep this example simple to execute we use the hosts file instead
of a DDNS server, adding the destination $10.1.1.1$ for the MD5 hashes
of the {\small \texttt{names}} ``uni2'' and ``unicast''. %
After adding these two entries, we enrol the new IPCP uni4 and create
an adjacency with uni2. %
The 4-node unicast network layer we set out to create is now fully
operational. %

\begin{center}
  \begin{lstlisting}[caption={Ping from system $a$ to the oping server},
    label={lst:pinga},
    captionpos = b, style=BashStyle]
    oping -n oping -c 4

    Pinging oping with 64 bytes of data (4 packets):

    64 bytes from oping: seq=0 time=1.166 ms
    64 bytes from oping: seq=1 time=1.083 ms
    64 bytes from oping: seq=2 time=1.158 ms
    64 bytes from oping: seq=3 time=1.136 ms

    --- oping ping statistics ---
    4 packets transmitted, 4 received, 0 out-of-order, 0% packet loss, time: 4002.018 ms
    rtt min/avg/max/mdev = 1.083/1.136/1.166/0.037 ms
  \end{lstlisting}
\end{center}

All that remains is to test the network, to do this, we send 4 pings
using the oping tool to the oping server (which will be automatically
executed on system $b$ once the IRMd on system $b$ receives an
allocation request from uni3. %
The output of executing the oping client is shown in
Lst. \ref{lst:pinga}. %

\begin{figure}[ht]
  \centering
  \Description{The IPCPs on the different systems and which
    layers they are using}
  \includegraphics[width=0.9\textwidth]{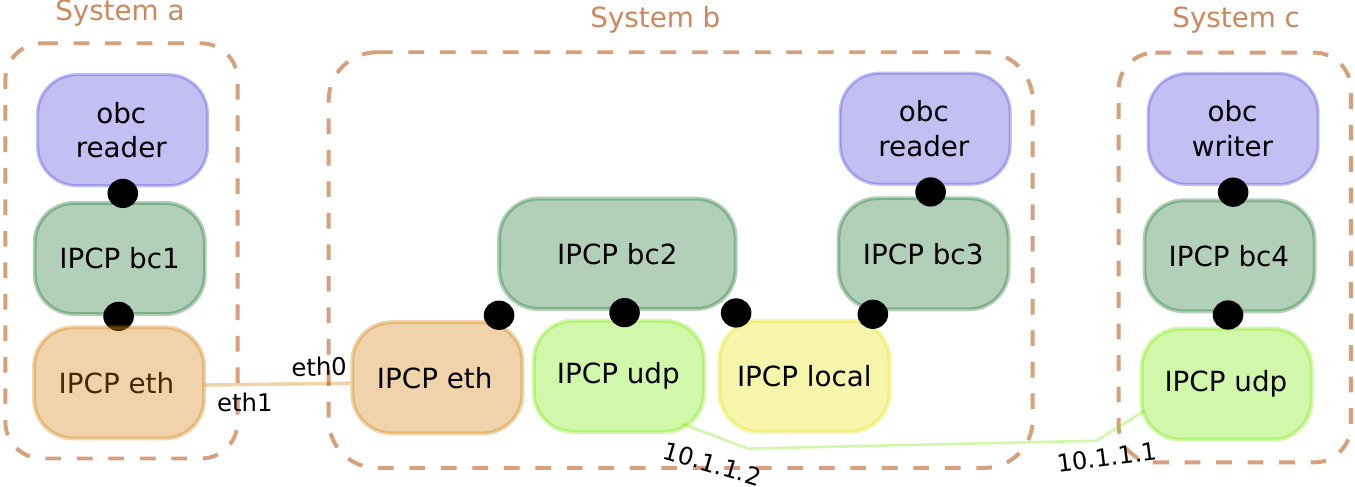}
  \caption{Example scenario (broadcast)}
  \label{fig:simpex-bc}
\end{figure}

We will now create a similar scenario, but with a 4-node broadcast
layer supporting the obc (Ouroboros broadcast) application. %
There is a reader for the broadcast on system $a$ at IPCP bc1 and on
system $b$ at bc3; and a source for the broadcast at IPCP bc4 in system $c$.
The scenario is shown in Fig. \ref{fig:simpex-bc}. %

\begin{center}
  \begin{lstlisting}[caption={Commands on system $a$},
    label={lst:cmdsabc},
    captionpos = b, style=BashStyle]
    irm ipcp bootstrap name eth layer ethernet type eth-dix dev eth1
    irm ipcp bootstrap name bc1 layer broadcast type broadcast autobind
    irm register name bc1 layer ethernet
    irm register name broadcast layer ethernet
    obc -l -n broadcast

    Starting a reader.
    New flow.
    Message is Hello multiple worlds.
  \end{lstlisting}
\end{center}

The commands are very similar as for the unicast example, so we will
just comment on the differences. %
On node $a$, we create a broadcast IPCP with name ``bc1'' and
bootstrap it into a layer named ``broadcast''. %
The {\small \texttt{names}} bc1 and broadcast are then registered in
the Ethernet layer. %
We can already start the receiver at system $a$ in anticipation of the
broadcaster. %

\begin{center}
  \begin{lstlisting}[caption={Commands on system $b$},
    label={lst:cmdsbbc},
    captionpos = b, style=BashStyle]
    irm ipcp bootstrap name eth layer ethernet type eth-dix dev eth3
    irm ipcp bootstrap name udp layer ipnet type udp ip 10.1.1.2
    irm ipcp bootstrap name local type local layer local
    irm ipcp enrol name bc2 layer broadcast type broadcast autobind
    irm ipcp connect name bc2 dst bc1 component dt
    irm register name bc2 layer ethernet layer ipnet layer local
    irm register name broadcast layer ethernet layer ipnet layer local
    irm ipcp enrol name bc3 layer broadcast type broadcast
    irm ipcp connect name bc3 dst bc2 component dt
    obc -l -n broadcast

    Starting a reader.
    New flow.
    Message is Hello multiple worlds.
  \end{lstlisting}
\end{center}

On system $b$, we bootstrap the lower layers like in the unicast
example, enrol and connect a broadcast IPCP bc2 with the bc1 IPCP on
system $a$, register the bc2 IPCP in the local layer, and enrol the
third bc3 IPCP with bc2. %
After creating an adjacency between bc3 and bc2, we start the
Ouroboros broadcast application to listen to the anticipated
broadcast. %

Finally, on system $c$, we bootstrap the IPCP over UDP, enrol the
final broadcast IPCP with IPCP bc2 on system $b$ and create an
adjacency. %
When we send a message ``Hello multiple worlds'' using the obc tool,
it will be received by the obc clients on systems $a$ and $b$.

\begin{center}
  \begin{lstlisting}[caption={Commands on system $c$},
    label={lst:cmdscbc},
    captionpos = b, style=BashStyle]
    irm ipcp bootstrap name udp layer ipnet type udp ip 10.1.1.1
    echo '10.1.1.2     b605ea609b9dfcf295cfd5a7803dcd37' | sudo tee --append /etc/hosts
    echo '10.1.1.2     3fecb40eaa447da8f09e6951c7b483f9' | sudo tee --append /etc/hosts
    irm ipcp enrol name bc4 layer broadcast type broadcast
    irm ipcp connect name bc4 dst bc2
    obc -n broadcast -m "Hello multiple worlds"
  \end{lstlisting}
\end{center}


\section{Conclusions}
\label{sec:conclusion}

Today's computer network technology has sprouted from years of
experience in developing and deploying different solutions and making
them interoperable through standardization efforts. %
While the individual mechanisms (congestion control, routing,
addressing, ...) that are used to build networks are fairly well
understood, putting these mechanisms together in a single consistent
architecture remains a challenge. %

One of the objectives we set when taking on this challenge was to try
to contain complexity; minimize as best as we can the impact that some
change in one element of the architecture can have on other elements
of the architecture; to disentangle as much as possible the elements
that make up a computer network. %
This meant carefully evaluating the implications that making small
changes in the architecture have on the implementation and -- most
importantly -- the interfaces between the elements. %
This approach, combined with the application of a UNIX-like philosophy
led to the Ouroboros architecture we detailed here. %

The Ouroboros architecture brought us a novel perspective on computer
networks; seeing them as flexible ecosystems consisting of 2 distinct
types of distributed programs: unicast layers and broadcast layers. %
These 2 types of layers can easily be observed in todays networks,
albeit not isolation: they are often found within a single layer of
the Internet model. %
On the one hand, they can be found in symbiosis: in addition to
unicast, Ethernet and IP also support broadcast addresses. %
On the other hand, they can also be found some intricate entanglement,
especially for the purposes of restricting scope: IP multicast groups
form a broadcast layer within a unicast layer. %
It's interesting to ponder the questions: how much of the complexity
is inherent in the layers and how much complexity is added (or
subtracted) by entangling them? %

The reference implementation has been primarily developed to validate
and fine-tune the architecture. %
However, it is also engineered to present the research community with
a set of useful tools to continue exploring various aspects of
computer networks. %

The architecture and prototype presented in this article are far from
complete, they are merely first steps in a direction that -- to us --
seemed interesting and worthwhile exploring. %
While we tried to paint as complete a picture as we can from our
current vantage point, a lot of it is still distant on the horizon. %

\section*{Acknowledgments}
This work received some partial funding from the Flemish Government
through grant G045315N. %
Substantial effort was put in unfunded during the authors' spare
time. %

The Rumba framework was used extensively during the development and
testing of Ouroboros. %
It was funded by the European Commission through grant agreement
687871 (ARCFIRE), part of the Future Internet Research and
Experimentation (FIRE) objective of the Eighth Framework Programme
(Horizon 2020). %

We would like to extend our utmost gratitude to John Day, whose
seminal work on RINA has set us off on this path of discovery into the
nature of computer networks and inter-process communication. %

We would like to thank John Day, Dr. Eduard Grasa, Dr. Sven van der
Meer, Miguel Ponce de Leon, Leonardo Bergesio and Miquel Tarzan for
discussions and insights into RINA. %

We would like to thank Francesco Salvestrini and Vincenzo Maffione for
discussions on RINA and their invaluable insights into implementation
and performance issues for network stacks. %

We thank prof Dr. Jeroen Schillewaert for taking the time to
double-check our definitions. %

The raptor IPCP was designed and implemented by Alexander D'hoore as
part of his master thesis. %

We would like to thank our thesis students: Douwe De Bock, Mathieu
Devos, Addy Bombeke, Frederik Vanderstraeten for their thesis work on
recursive networks; and Lo\"ic Vervaeke, Alexander D'hoore, Nick Aerts
and Friedl Rubrecht for their contributions and their invaluable
feedback on using and extending the Ouroboros software prototype. %

We would like to thank prof. Dr. Mario Pickavet and prof. Dr. Didier
Colle for their valuable suggestions for improving this manuscript. %


\bibliographystyle{ACM-Reference-Format}
\bibliography{biblio}


\begin{thebibliography}{116}


\ifx \showCODEN    \undefined \def \showCODEN     #1{\unskip}     \fi
\ifx \showDOI      \undefined \def \showDOI       #1{#1}\fi
\ifx \showISBNx    \undefined \def \showISBNx     #1{\unskip}     \fi
\ifx \showISBNxiii \undefined \def \showISBNxiii  #1{\unskip}     \fi
\ifx \showISSN     \undefined \def \showISSN      #1{\unskip}     \fi
\ifx \showLCCN     \undefined \def \showLCCN      #1{\unskip}     \fi
\ifx \shownote     \undefined \def \shownote      #1{#1}          \fi
\ifx \showarticletitle \undefined \def \showarticletitle #1{#1}   \fi
\ifx \showURL      \undefined \def \showURL       {\relax}        \fi
\providecommand\bibfield[2]{#2}
\providecommand\bibinfo[2]{#2}
\providecommand\natexlab[1]{#1}
\providecommand\showeprint[2][]{arXiv:#2}

\bibitem[\protect\citeauthoryear{Alizadeh, Greenberg, Maltz, Padhye, Patel,
  Prabhakar, Sengupta, and Sridharan}{Alizadeh et~al\mbox{.}}{2010}]%
        {alizadeh2010dctcp}
\bibfield{author}{\bibinfo{person}{Mohammad Alizadeh}, \bibinfo{person}{Albert
  Greenberg}, \bibinfo{person}{David~A. Maltz}, \bibinfo{person}{Jitendra
  Padhye}, \bibinfo{person}{Parveen Patel}, \bibinfo{person}{Balaji Prabhakar},
  \bibinfo{person}{Sudipta Sengupta}, {and} \bibinfo{person}{Murari
  Sridharan}.} \bibinfo{year}{2010}\natexlab{}.
\newblock \showarticletitle{Data Center TCP (DCTCP)}.
\newblock \bibinfo{journal}{\emph{SIGCOMM Comput. Commun. Rev.}}
  \bibinfo{volume}{40}, \bibinfo{number}{4} (\bibinfo{date}{Aug.}
  \bibinfo{year}{2010}), \bibinfo{pages}{63--74}.
\newblock
\showISSN{0146-4833}
\urldef\tempurl%
\url{https://doi.org/10.1145/1851275.1851192}
\showDOI{\tempurl}


\bibitem[\protect\citeauthoryear{Atlas and Zinin}{Atlas and Zinin}{2008}]%
        {atlas2008basic}
\bibfield{author}{\bibinfo{person}{Alia Atlas} {and} \bibinfo{person}{Alex
  Zinin}.} \bibinfo{year}{2008}\natexlab{}.
\newblock \bibinfo{booktitle}{\emph{{Basic specification for IP fast reroute:
  Loop-Free Alternates}}}.
\newblock \bibinfo{type}{{RFC}} 5286. \bibinfo{institution}{IETF}.
\newblock


\bibitem[\protect\citeauthoryear{Bachman and Ross}{Bachman and Ross}{1982}]%
        {bachman1982osi}
\bibfield{author}{\bibinfo{person}{Charles~W. Bachman} {and}
  \bibinfo{person}{Ronald~G. Ross}.} \bibinfo{year}{1982}\natexlab{}.
\newblock \showarticletitle{Toward a more complete reference model of
  computer-based information systems}.
\newblock \bibinfo{journal}{\emph{Computer Networks (1976)}}
  \bibinfo{volume}{6}, \bibinfo{number}{5} (\bibinfo{year}{1982}),
  \bibinfo{pages}{331--343}.
\newblock
\showISSN{0376-5075}
\urldef\tempurl%
\url{https://doi.org/10.1016/0376-5075(82)90103-9}
\showDOI{\tempurl}


\bibitem[\protect\citeauthoryear{Baran}{Baran}{1964}]%
        {baran1964distributed}
\bibfield{author}{\bibinfo{person}{Paul Baran}.}
  \bibinfo{year}{1964}\natexlab{}.
\newblock \showarticletitle{On distributed communications networks}.
\newblock \bibinfo{journal}{\emph{IEEE transactions on Communications Systems}}
  \bibinfo{volume}{12}, \bibinfo{number}{1} (\bibinfo{year}{1964}),
  \bibinfo{pages}{1--9}.
\newblock
\urldef\tempurl%
\url{https://doi.org/10.1109/TCOM.1964.1088883}
\showDOI{\tempurl}


\bibitem[\protect\citeauthoryear{Bellare and Namprempre}{Bellare and
  Namprempre}{2008}]%
        {bellare2008authenc}
\bibfield{author}{\bibinfo{person}{Mihir Bellare} {and}
  \bibinfo{person}{Chanathip Namprempre}.} \bibinfo{year}{2008}\natexlab{}.
\newblock \showarticletitle{Authenticated Encryption: Relations among Notions
  and Analysis of the Generic Composition Paradigm}.
\newblock \bibinfo{journal}{\emph{Journal of Cryptology}} \bibinfo{volume}{21},
  \bibinfo{number}{4} (\bibinfo{date}{01 Oct} \bibinfo{year}{2008}),
  \bibinfo{pages}{469--491}.
\newblock
\showISSN{1432-1378}
\urldef\tempurl%
\url{https://doi.org/10.1007/s00145-008-9026-x}
\showDOI{\tempurl}


\bibitem[\protect\citeauthoryear{Bertoni, Daemen, Peeters, and
  Van~Assche}{Bertoni et~al\mbox{.}}{2013}]%
        {bertoni2013keccak}
\bibfield{author}{\bibinfo{person}{Guido Bertoni}, \bibinfo{person}{Joan
  Daemen}, \bibinfo{person}{Micha{\"e}l Peeters}, {and} \bibinfo{person}{Gilles
  Van~Assche}.} \bibinfo{year}{2013}\natexlab{}.
\newblock \showarticletitle{Keccak}. In \bibinfo{booktitle}{\emph{Advances in
  Cryptology -- EUROCRYPT 2013}}, \bibfield{editor}{\bibinfo{person}{Thomas
  Johansson} {and} \bibinfo{person}{Phong~Q. Nguyen}} (Eds.).
  \bibinfo{publisher}{Springer Berlin Heidelberg}, \bibinfo{address}{Berlin,
  Heidelberg}, \bibinfo{pages}{313--314}.
\newblock
\showISBNx{978-3-642-38348-9}
\urldef\tempurl%
\url{https://doi.org/10.1007/978-3-642-38348-9_19}
\showDOI{\tempurl}


\bibitem[\protect\citeauthoryear{Bezen{\v{s}}ek and Robi{\v{c}}}{Bezen{\v{s}}ek
  and Robi{\v{c}}}{2014}]%
        {bezensek2014steiner}
\bibfield{author}{\bibinfo{person}{Mitja Bezen{\v{s}}ek} {and}
  \bibinfo{person}{Borut Robi{\v{c}}}.} \bibinfo{year}{2014}\natexlab{}.
\newblock \showarticletitle{A Survey of Parallel and Distributed Algorithms for
  the Steiner Tree Problem}.
\newblock \bibinfo{journal}{\emph{International Journal of Parallel
  Programming}} \bibinfo{volume}{42}, \bibinfo{number}{2} (\bibinfo{date}{01
  Apr} \bibinfo{year}{2014}), \bibinfo{pages}{287--319}.
\newblock
\showISSN{1573-7640}
\urldef\tempurl%
\url{https://doi.org/10.1007/s10766-013-0243-z}
\showDOI{\tempurl}


\bibitem[\protect\citeauthoryear{Braden}{Braden}{1989}]%
        {rfc1122}
\bibfield{author}{\bibinfo{person}{Robert~T. Braden}.}
  \bibinfo{year}{1989}\natexlab{}.
\newblock \bibinfo{title}{{Requirements for Internet Hosts - Communication
  Layers}}.
\newblock \bibinfo{howpublished}{RFC 1122}.
\newblock
\urldef\tempurl%
\url{https://doi.org/10.17487/RFC1122}
\showDOI{\tempurl}


\bibitem[\protect\citeauthoryear{Callon}{Callon}{1996}]%
        {rfc1925}
\bibfield{author}{\bibinfo{person}{Ross Callon}.}
  \bibinfo{year}{1996}\natexlab{}.
\newblock \bibinfo{title}{{The Twelve Networking Truths}}.
\newblock \bibinfo{howpublished}{RFC 1925}.
\newblock
\urldef\tempurl%
\url{https://doi.org/10.17487/RFC1925}
\showDOI{\tempurl}


\bibitem[\protect\citeauthoryear{Carr, Crocker, and Cerf}{Carr
  et~al\mbox{.}}{1970}]%
        {carr1970hosthost}
\bibfield{author}{\bibinfo{person}{C.~Stephen Carr},
  \bibinfo{person}{Stephen~D. Crocker}, {and} \bibinfo{person}{Vinton~G.
  Cerf}.} \bibinfo{year}{1970}\natexlab{}.
\newblock \showarticletitle{HOST-HOST Communication Protocol in the ARPA
  Network}. In \bibinfo{booktitle}{\emph{Proceedings of the May 5-7, 1970,
  Spring Joint Computer Conference}} \emph{(\bibinfo{series}{AFIPS '70
  (Spring)})}. \bibinfo{publisher}{ACM}, \bibinfo{address}{New York, NY, USA},
  \bibinfo{pages}{589--597}.
\newblock
\urldef\tempurl%
\url{https://doi.org/10.1145/1476936.1477024}
\showDOI{\tempurl}


\bibitem[\protect\citeauthoryear{Cerf}{Cerf}{1978}]%
        {cerf1978catenet}
\bibfield{author}{\bibinfo{person}{Vinton~G. Cerf}.}
  \bibinfo{year}{1978}\natexlab{}.
\newblock \bibinfo{booktitle}{\emph{The catenet model for internetworking}}.
\newblock \bibinfo{type}{{T}echnical {R}eport}.
  \bibinfo{institution}{DARPA/IPTO}.
\newblock


\bibitem[\protect\citeauthoryear{Cerf and Cain}{Cerf and Cain}{1983}]%
        {cerf1983dod}
\bibfield{author}{\bibinfo{person}{Vinton~G. Cerf} {and}
  \bibinfo{person}{Edward Cain}.} \bibinfo{year}{1983}\natexlab{}.
\newblock \showarticletitle{{The DoD internet architecture model}}.
\newblock \bibinfo{journal}{\emph{Computer Networks (1976)}}
  \bibinfo{volume}{7}, \bibinfo{number}{5} (\bibinfo{year}{1983}),
  \bibinfo{pages}{307--318}.
\newblock
\urldef\tempurl%
\url{https://doi.org/10.1016/0376-5075(83)90042-9}
\showDOI{\tempurl}


\bibitem[\protect\citeauthoryear{Cerf and Kahn}{Cerf and Kahn}{1974}]%
        {cerfkahn1974tcp}
\bibfield{author}{\bibinfo{person}{Vinton~G. Cerf} {and}
  \bibinfo{person}{Robert~E. Kahn}.} \bibinfo{year}{1974}\natexlab{}.
\newblock \showarticletitle{A Protocol for Packet Network Intercommunication}.
\newblock \bibinfo{journal}{\emph{IEEE Transactions on Communications}}
  \bibinfo{volume}{22}, \bibinfo{number}{5} (\bibinfo{date}{May}
  \bibinfo{year}{1974}), \bibinfo{pages}{637--648}.
\newblock
\showISSN{0090-6778}
\urldef\tempurl%
\url{https://doi.org/10.1109/TCOM.1974.1092259}
\showDOI{\tempurl}


\bibitem[\protect\citeauthoryear{Cerf, McKenzie, Scantlebury, and
  Zimmermann}{Cerf et~al\mbox{.}}{1976}]%
        {cerf1976inwg96}
\bibfield{author}{\bibinfo{person}{Vinton~G. Cerf}, \bibinfo{person}{Alexander
  McKenzie}, \bibinfo{person}{Roger~A. Scantlebury}, {and}
  \bibinfo{person}{Hubert Zimmermann}.} \bibinfo{year}{1976}\natexlab{}.
\newblock \showarticletitle{Proposal for an International End to End Protocol}.
\newblock \bibinfo{journal}{\emph{SIGCOMM Comput. Commun. Rev.}}
  \bibinfo{volume}{6}, \bibinfo{number}{1} (\bibinfo{date}{Jan.}
  \bibinfo{year}{1976}), \bibinfo{pages}{63--89}.
\newblock
\showISSN{0146-4833}
\urldef\tempurl%
\url{https://doi.org/10.1145/1015828.1015832}
\showDOI{\tempurl}


\bibitem[\protect\citeauthoryear{Chang, Perlner, Burr, Turan, Kelsey,
  Souradyuti, Bassham, and Blank}{Chang et~al\mbox{.}}{2012}]%
        {chang2012sha3}
\bibfield{author}{\bibinfo{person}{Shu-Jen Chang}, \bibinfo{person}{Ray
  Perlner}, \bibinfo{person}{William~E. Burr}, \bibinfo{person}{Meltem~Sönmez
  Turan}, \bibinfo{person}{John~M. Kelsey}, \bibinfo{person}{Paul. Souradyuti},
  \bibinfo{person}{Lawrence~E. Bassham}, {and} \bibinfo{person}{Rebecca~M.
  Blank}.} \bibinfo{year}{2012}\natexlab{}.
\newblock \bibinfo{title}{Third-Round Report of the SHA-3 Cryptographic Hash
  Algorithm Competition}.
\newblock
\newblock


\bibitem[\protect\citeauthoryear{Cheong and Lai}{Cheong and Lai}{1999}]%
        {cheong1999qosspec}
\bibfield{author}{\bibinfo{person}{F. Cheong} {and} \bibinfo{person}{R. Lai}.}
  \bibinfo{year}{1999}\natexlab{}.
\newblock \showarticletitle{QoS specification and mapping for distributed
  multimedia systems: A survey of issues}.
\newblock \bibinfo{journal}{\emph{Journal of Systems and Software}}
  \bibinfo{volume}{45}, \bibinfo{number}{2} (\bibinfo{year}{1999}),
  \bibinfo{pages}{127 -- 139}.
\newblock
\showISSN{0164-1212}
\urldef\tempurl%
\url{https://doi.org/10.1016/S0164-1212(98)10073-0}
\showDOI{\tempurl}


\bibitem[\protect\citeauthoryear{Clark}{Clark}{1988}]%
        {clark1988darpadesign}
\bibfield{author}{\bibinfo{person}{David~D. Clark}.}
  \bibinfo{year}{1988}\natexlab{}.
\newblock \showarticletitle{The Design Philosophy of the DARPA Internet
  Protocols}.
\newblock \bibinfo{journal}{\emph{SIGCOMM Comput. Commun. Rev.}}
  \bibinfo{volume}{18}, \bibinfo{number}{4} (\bibinfo{date}{Aug.}
  \bibinfo{year}{1988}), \bibinfo{pages}{106--114}.
\newblock
\showISSN{0146-4833}
\urldef\tempurl%
\url{https://doi.org/10.1145/52325.52336}
\showDOI{\tempurl}


\bibitem[\protect\citeauthoryear{Coulouris, Dollimore, Kindberg, and
  Blair}{Coulouris et~al\mbox{.}}{2011}]%
        {coulouris2011ds}
\bibfield{author}{\bibinfo{person}{George Coulouris}, \bibinfo{person}{Jean
  Dollimore}, \bibinfo{person}{Tim Kindberg}, {and} \bibinfo{person}{Gordon
  Blair}.} \bibinfo{year}{2011}\natexlab{}.
\newblock \bibinfo{booktitle}{\emph{Distributed Systems: Concepts and Design}
  (\bibinfo{edition}{5th} ed.)}.
\newblock \bibinfo{publisher}{Addison-Wesley Publishing Company},
  \bibinfo{address}{USA}.
\newblock
\showISBNx{0132143011, 9780132143011}


\bibitem[\protect\citeauthoryear{Davies, Bartlett, Scantlebury, and
  Wilkinson}{Davies et~al\mbox{.}}{1967}]%
        {davies1967digcommnet}
\bibfield{author}{\bibinfo{person}{D.~W. Davies}, \bibinfo{person}{K.~A.
  Bartlett}, \bibinfo{person}{R.~A. Scantlebury}, {and} \bibinfo{person}{P.~T.
  Wilkinson}.} \bibinfo{year}{1967}\natexlab{}.
\newblock \showarticletitle{A Digital Communication Network for Computers
  Giving Rapid Response at Remote Terminals}. In
  \bibinfo{booktitle}{\emph{Proceedings of the First ACM Symposium on Operating
  System Principles}} \emph{(\bibinfo{series}{SOSP '67})}.
  \bibinfo{publisher}{ACM}, \bibinfo{address}{New York, NY, USA},
  \bibinfo{pages}{2.1--2.17}.
\newblock
\urldef\tempurl%
\url{https://doi.org/10.1145/800001.811669}
\showDOI{\tempurl}


\bibitem[\protect\citeauthoryear{Day}{Day}{2008}]%
        {day2008pna}
\bibfield{author}{\bibinfo{person}{John Day}.} \bibinfo{year}{2008}\natexlab{}.
\newblock \bibinfo{booktitle}{\emph{Patterns in Network Architecture: A Return
  to Fundamentals}}.
\newblock \bibinfo{publisher}{Pearson Education}, \bibinfo{address}{New York
  City, New York}.
\newblock


\bibitem[\protect\citeauthoryear{Day}{Day}{2011}]%
        {day2011lostlayer}
\bibfield{author}{\bibinfo{person}{John Day}.} \bibinfo{year}{2011}\natexlab{}.
\newblock \showarticletitle{How in the Heck do you lose a layer!?}. In
  \bibinfo{booktitle}{\emph{2011 International Conference on the Network of the
  Future}}. \bibinfo{pages}{135--143}.
\newblock
\urldef\tempurl%
\url{https://doi.org/10.1109/NOF.2011.6126673}
\showDOI{\tempurl}


\bibitem[\protect\citeauthoryear{Day}{Day}{2016}]%
        {day2016clamor}
\bibfield{author}{\bibinfo{person}{John Day}.} \bibinfo{year}{2016}\natexlab{}.
\newblock \showarticletitle{The Clamor Outside as INWG Debated: Economic War
  Comes to Networking}.
\newblock \bibinfo{journal}{\emph{IEEE Annals of the History of Computing}}
  \bibinfo{volume}{38}, \bibinfo{number}{3} (\bibinfo{date}{July}
  \bibinfo{year}{2016}), \bibinfo{pages}{58--77}.
\newblock
\showISSN{1058-6180}
\urldef\tempurl%
\url{https://doi.org/10.1109/MAHC.2015.70}
\showDOI{\tempurl}


\bibitem[\protect\citeauthoryear{Day, Matta, and Mattar}{Day
  et~al\mbox{.}}{2008}]%
        {day2008ipc}
\bibfield{author}{\bibinfo{person}{John Day}, \bibinfo{person}{Ibrahim Matta},
  {and} \bibinfo{person}{Karim Mattar}.} \bibinfo{year}{2008}\natexlab{}.
\newblock \showarticletitle{{Networking is IPC: a guiding principle to a better
  internet}}. In \bibinfo{booktitle}{\emph{Proceedings of the 2008 ACM CoNEXT
  Conference}}. ACM, \bibinfo{pages}{67}.
\newblock


\bibitem[\protect\citeauthoryear{Day and Zimmermann}{Day and
  Zimmermann}{1995}]%
        {day1983osi}
\bibfield{author}{\bibinfo{person}{John~D. Day} {and} \bibinfo{person}{Hubert
  Zimmermann}.} \bibinfo{year}{1995}\natexlab{}.
\newblock \showarticletitle{Conformance Testing Methodologies and Architectures
  for OSI Protocols}.
\newblock \bibinfo{publisher}{IEEE Computer Society Press},
  \bibinfo{address}{Los Alamitos, CA, USA}, Chapter The OSI Reference Model,
  \bibinfo{pages}{38--44}.
\newblock
\showISBNx{0-8186-5352-3}
\urldef\tempurl%
\url{http://dl.acm.org/citation.cfm?id=202035.202039}
\showURL{%
\tempurl}


\bibitem[\protect\citeauthoryear{Demichelis and Chimento}{Demichelis and
  Chimento}{2002}]%
        {rfc3393}
\bibfield{author}{\bibinfo{person}{Carlo~M. Demichelis} {and}
  \bibinfo{person}{Philip Chimento}.} \bibinfo{year}{2002}\natexlab{}.
\newblock \bibinfo{title}{{IP Packet Delay Variation Metric for IP Performance
  Metrics (IPPM)}}.
\newblock \bibinfo{howpublished}{RFC 3393}.
\newblock
\urldef\tempurl%
\url{https://doi.org/10.17487/RFC3393}
\showDOI{\tempurl}


\bibitem[\protect\citeauthoryear{Dijkstra}{Dijkstra}{1968}]%
        {dijkstra1968the}
\bibfield{author}{\bibinfo{person}{Edsger~W. Dijkstra}.}
  \bibinfo{year}{1968}\natexlab{}.
\newblock \showarticletitle{The Structure of the "THE" multiprogramming
  System}.
\newblock \bibinfo{journal}{\emph{Commun. ACM}} \bibinfo{volume}{11},
  \bibinfo{number}{5} (\bibinfo{date}{May} \bibinfo{year}{1968}),
  \bibinfo{pages}{341--346}.
\newblock
\showISSN{0001-0782}
\urldef\tempurl%
\url{https://doi.org/10.1145/363095.363143}
\showDOI{\tempurl}


\bibitem[\protect\citeauthoryear{Dijkstra}{Dijkstra}{1969}]%
        {dijkstra1969programming}
\bibfield{author}{\bibinfo{person}{Edsger~W. Dijkstra}.}
  \bibinfo{year}{1969}\natexlab{}.
\newblock \bibinfo{title}{The Programming Task Considered as an Intellectual
  Challenge}.
\newblock
\newblock
\urldef\tempurl%
\url{https://www.cs.utexas.edu/users/EWD/transcriptions/EWD02xx/EWD273.html}
\showURL{%
\tempurl}


\bibitem[\protect\citeauthoryear{Dijkstra}{Dijkstra}{1988}]%
        {dijkstra1988cruelty}
\bibfield{author}{\bibinfo{person}{Edsger~W. Dijkstra}.}
  \bibinfo{year}{1988}\natexlab{}.
\newblock \bibinfo{title}{On the cruelty of really teaching computing science}.
\newblock
\newblock
\urldef\tempurl%
\url{http://www.cs.utexas.edu/users/EWD/ewd10xx/EWD1036.PDF}
\showURL{%
\tempurl}


\bibitem[\protect\citeauthoryear{Droms}{Droms}{1997}]%
        {rfc2131}
\bibfield{author}{\bibinfo{person}{Ralph Droms}.}
  \bibinfo{year}{1997}\natexlab{}.
\newblock \bibinfo{title}{{Dynamic Host Configuration Protocol}}.
\newblock \bibinfo{howpublished}{RFC 2131}.
\newblock
\urldef\tempurl%
\url{https://doi.org/10.17487/RFC2131}
\showDOI{\tempurl}


\bibitem[\protect\citeauthoryear{Durumeric, Li, Kasten, Amann, Beekman, Payer,
  Weaver, Adrian, Paxson, Bailey, and Halderman}{Durumeric
  et~al\mbox{.}}{2014}]%
        {durumeric2014heartbleed}
\bibfield{author}{\bibinfo{person}{Zakir Durumeric}, \bibinfo{person}{Frank
  Li}, \bibinfo{person}{James Kasten}, \bibinfo{person}{Johanna Amann},
  \bibinfo{person}{Jethro Beekman}, \bibinfo{person}{Mathias Payer},
  \bibinfo{person}{Nicolas Weaver}, \bibinfo{person}{David Adrian},
  \bibinfo{person}{Vern Paxson}, \bibinfo{person}{Michael Bailey}, {and}
  \bibinfo{person}{J.~Alex Halderman}.} \bibinfo{year}{2014}\natexlab{}.
\newblock \showarticletitle{The Matter of Heartbleed}. In
  \bibinfo{booktitle}{\emph{Proceedings of the 2014 Conference on Internet
  Measurement Conference}} \emph{(\bibinfo{series}{IMC '14})}.
  \bibinfo{publisher}{ACM}, \bibinfo{address}{New York, NY, USA},
  \bibinfo{pages}{475--488}.
\newblock
\showISBNx{978-1-4503-3213-2}
\urldef\tempurl%
\url{https://doi.org/10.1145/2663716.2663755}
\showDOI{\tempurl}


\bibitem[\protect\citeauthoryear{Fletcher}{Fletcher}{1982}]%
        {fletcher1982lincs}
\bibfield{author}{\bibinfo{person}{John~G. Fletcher}.}
  \bibinfo{year}{1982}\natexlab{}.
\newblock \bibinfo{booktitle}{\emph{LINCS: Livermore's network architecture}}.
\newblock \bibinfo{type}{{T}echnical {R}eport}. \bibinfo{institution}{Lawrence
  Livermore Laboratories}.
\newblock


\bibitem[\protect\citeauthoryear{Fletcher and Watson}{Fletcher and
  Watson}{1978}]%
        {fletcher1978tbproto}
\bibfield{author}{\bibinfo{person}{John~G. Fletcher} {and}
  \bibinfo{person}{Richard~W. Watson}.} \bibinfo{year}{1978}\natexlab{}.
\newblock \showarticletitle{Mechanisms for a reliable timer-based protocol}.
\newblock \bibinfo{journal}{\emph{Computer Networks (1976)}}
  \bibinfo{volume}{2}, \bibinfo{number}{4} (\bibinfo{year}{1978}),
  \bibinfo{pages}{271 -- 290}.
\newblock
\showISSN{0376-5075}
\urldef\tempurl%
\url{https://doi.org/10.1016/0376-5075(78)90006-5}
\showDOI{\tempurl}


\bibitem[\protect\citeauthoryear{Floyd}{Floyd}{2000}]%
        {rfc2914}
\bibfield{author}{\bibinfo{person}{Sally Floyd}.}
  \bibinfo{year}{2000}\natexlab{}.
\newblock \bibinfo{title}{{Congestion Control Principles}}.
\newblock \bibinfo{howpublished}{RFC 2914}.
\newblock
\urldef\tempurl%
\url{https://doi.org/10.17487/RFC2914}
\showDOI{\tempurl}


\bibitem[\protect\citeauthoryear{Floyd, Handley, and Kohler}{Floyd
  et~al\mbox{.}}{2006}]%
        {rfc4340}
\bibfield{author}{\bibinfo{person}{Sally Floyd}, \bibinfo{person}{Mark~J.
  Handley}, {and} \bibinfo{person}{Eddie Kohler}.}
  \bibinfo{year}{2006}\natexlab{}.
\newblock \bibinfo{title}{{Datagram Congestion Control Protocol (DCCP)}}.
\newblock \bibinfo{howpublished}{RFC 4340}.
\newblock
\urldef\tempurl%
\url{https://doi.org/10.17487/RFC4340}
\showDOI{\tempurl}


\bibitem[\protect\citeauthoryear{Fuller and Li}{Fuller and Li}{2006}]%
        {rfc4632}
\bibfield{author}{\bibinfo{person}{Vince Fuller} {and} \bibinfo{person}{Tony
  Li}.} \bibinfo{year}{2006}\natexlab{}.
\newblock \bibinfo{title}{{Classless Inter-domain Routing (CIDR): The Internet
  Address Assignment and Aggregation Plan}}.
\newblock \bibinfo{howpublished}{RFC 4632}.
\newblock
\urldef\tempurl%
\url{https://doi.org/10.17487/RFC4632}
\showDOI{\tempurl}


\bibitem[\protect\citeauthoryear{Gilligan, Bound, Thomson, and
  Stevens}{Gilligan et~al\mbox{.}}{1999}]%
        {rfc2553}
\bibfield{author}{\bibinfo{person}{Robert~E. Gilligan}, \bibinfo{person}{Jim
  Bound}, \bibinfo{person}{Susan Thomson}, {and} \bibinfo{person}{W.~Richard
  Stevens}.} \bibinfo{year}{1999}\natexlab{}.
\newblock \bibinfo{title}{{Basic Socket Interface Extensions for IPv6}}.
\newblock \bibinfo{howpublished}{RFC 2553}.
\newblock
\urldef\tempurl%
\url{https://doi.org/10.17487/RFC2553}
\showDOI{\tempurl}


\bibitem[\protect\citeauthoryear{Grasa, Bergesio, Tarzan, Lopez, van~der Meer,
  Day, and Chitkushev}{Grasa et~al\mbox{.}}{2018}]%
        {grasa2018mobility}
\bibfield{author}{\bibinfo{person}{Edu Grasa}, \bibinfo{person}{Leo Bergesio},
  \bibinfo{person}{Miquel Tarzan}, \bibinfo{person}{Diego Lopez},
  \bibinfo{person}{Sven van~der Meer}, \bibinfo{person}{John Day}, {and}
  \bibinfo{person}{Lou Chitkushev}.} \bibinfo{year}{2018}\natexlab{}.
\newblock \showarticletitle{Mobility management in RINA networks: Experimental
  validation of architectural properties}. In
  \bibinfo{booktitle}{\emph{Wireless Communications and Networking Conference
  (WCNC), 2018 IEEE}}. IEEE, \bibinfo{pages}{1--6}.
\newblock


\bibitem[\protect\citeauthoryear{Grasa, Salvestrini, Bergesio, Tarzan,
  Vrijders, and Staessens}{Grasa et~al\mbox{.}}{2013}]%
        {grasa2013irati}
\bibfield{author}{\bibinfo{person}{Eduard Grasa}, \bibinfo{person}{Francesco
  Salvestrini}, \bibinfo{person}{Leonardo Bergesio}, \bibinfo{person}{Miquel
  Tarzan}, \bibinfo{person}{Sander Vrijders}, {and} \bibinfo{person}{Dimitri
  Staessens}.} \bibinfo{year}{2013}\natexlab{}.
\newblock \bibinfo{title}{IRATI, a RINA implementation for OS/Linux}.
\newblock
\newblock
\urldef\tempurl%
\url{https://github.com/irati/stack}
\showURL{%
\tempurl}


\bibitem[\protect\citeauthoryear{G\"ursun, Matta, and Mattar}{G\"ursun
  et~al\mbox{.}}{2010}]%
        {gurcun2010softstate}
\bibfield{author}{\bibinfo{person}{Gonca G\"ursun}, \bibinfo{person}{Ibrahim
  Matta}, {and} \bibinfo{person}{Karim Mattar}.}
  \bibinfo{year}{2010}\natexlab{}.
\newblock \showarticletitle{Revisiting A Soft-State Approach to Managing
  Reliable Transport Connections}. In \bibinfo{booktitle}{\emph{Proceedings of
  the Eighth International Workshop on Protocols for Future, Large-Scale and
  Diverse Network Transports (PFLDNeT 2010)}}.
\newblock


\bibitem[\protect\citeauthoryear{Hansen}{Hansen}{1970}]%
        {hansen1970nucleus}
\bibfield{author}{\bibinfo{person}{Per~Brinch Hansen}.}
  \bibinfo{year}{1970}\natexlab{}.
\newblock \showarticletitle{The Nucleus of a Multiprogramming System}.
\newblock \bibinfo{journal}{\emph{Commun. ACM}} \bibinfo{volume}{13},
  \bibinfo{number}{4} (\bibinfo{date}{April} \bibinfo{year}{1970}),
  \bibinfo{pages}{238--241}.
\newblock
\showISSN{0001-0782}
\urldef\tempurl%
\url{https://doi.org/10.1145/362258.362278}
\showDOI{\tempurl}


\bibitem[\protect\citeauthoryear{Hauzeur}{Hauzeur}{1986}]%
        {hauzeur1986modelnma}
\bibfield{author}{\bibinfo{person}{Bernard~M. Hauzeur}.}
  \bibinfo{year}{1986}\natexlab{}.
\newblock \showarticletitle{A Model for Naming, Addressing and Routing}.
\newblock \bibinfo{journal}{\emph{ACM Transactions on Information Systems}}
  \bibinfo{volume}{4}, \bibinfo{number}{4} (\bibinfo{date}{Dec.}
  \bibinfo{year}{1986}), \bibinfo{pages}{293--311}.
\newblock
\showISSN{1046-8188}
\urldef\tempurl%
\url{https://doi.org/10.1145/9760.9761}
\showDOI{\tempurl}


\bibitem[\protect\citeauthoryear{Hedrick}{Hedrick}{1988}]%
        {rfc1058}
\bibfield{author}{\bibinfo{person}{Chuck Hedrick}.}
  \bibinfo{year}{1988}\natexlab{}.
\newblock \bibinfo{title}{{Routing Information Protocol}}.
\newblock \bibinfo{howpublished}{RFC 1058}.
\newblock
\urldef\tempurl%
\url{https://doi.org/10.17487/RFC1058}
\showDOI{\tempurl}


\bibitem[\protect\citeauthoryear{Hintjens}{Hintjens}{2013}]%
        {hintjens2013zeromq}
\bibfield{author}{\bibinfo{person}{Pieter Hintjens}.}
  \bibinfo{year}{2013}\natexlab{}.
\newblock \bibinfo{booktitle}{\emph{ZeroMQ, Messaging for Many Applications}
  (\bibinfo{edition}{1st} ed.)}.
\newblock \bibinfo{publisher}{O'Reilly Media, Inc}.
\newblock
\showISBNx{9781449334062}


\bibitem[\protect\citeauthoryear{Holbrook, Cain, and Haberman}{Holbrook
  et~al\mbox{.}}{2006}]%
        {rfc4604}
\bibfield{author}{\bibinfo{person}{Hugh Holbrook}, \bibinfo{person}{Brad Cain},
  {and} \bibinfo{person}{Brian Haberman}.} \bibinfo{year}{2006}\natexlab{}.
\newblock \bibinfo{title}{{Using Internet Group Management Protocol Version 3
  (IGMPv3) and Multicast Listener Discovery Protocol Version 2 (MLDv2) for
  Source-Specific Multicast}}.
\newblock \bibinfo{howpublished}{RFC 4604}.
\newblock
\urldef\tempurl%
\url{https://doi.org/10.17487/RFC4604}
\showDOI{\tempurl}


\bibitem[\protect\citeauthoryear{id~Software}{id~Software}{2012}]%
        {id2012q3a}
\bibfield{author}{\bibinfo{person}{id Software}.}
  \bibinfo{year}{2012}\natexlab{}.
\newblock \bibinfo{title}{Quake III Arena GPL Source Release}.
\newblock
\newblock
\urldef\tempurl%
\url{https://github.com/id-Software/Quake-III-Arena}
\showURL{%
\tempurl}


\bibitem[\protect\citeauthoryear{IEEE}{IEEE}{1997}]%
        {ieee80211}
\bibfield{author}{\bibinfo{person}{IEEE}.} \bibinfo{year}{1997}\natexlab{}.
\newblock \bibinfo{title}{{Wireless LAN Medium Access Control (MAC) and
  Physical Layer (PHY) Specification}}.
\newblock
\newblock


\bibitem[\protect\citeauthoryear{ioquake3 community}{ioquake3
  community}{2012}]%
        {ioquake2012ioq3}
\bibfield{author}{\bibinfo{person}{ioquake3 community}.}
  \bibinfo{year}{2012}\natexlab{}.
\newblock \bibinfo{title}{Free Software FPS Game Engine Based on Quake 3 for
  Windows, Linux, and macOS}.
\newblock
\newblock
\urldef\tempurl%
\url{https://www.ioquake3.org/}
\showURL{%
\tempurl}


\bibitem[\protect\citeauthoryear{ISO X.224 (11/95)}{ISO X.224 (11/95)}{1995}]%
        {osi1995x224}
ISO X.224 (11/95) \bibinfo{year}{1995}\natexlab{}.
\newblock \bibinfo{booktitle}{\emph{{Information technology - Open Systems
  Interconnection - Protocol for providing the connection-mode transport
  service }}}.
\newblock \bibinfo{type}{Standard}. \bibinfo{institution}{International
  Organization for Standardization}, \bibinfo{address}{Geneva, CH}.
\newblock


\bibitem[\protect\citeauthoryear{ISO/IEC 15954:1999}{ISO/IEC
  15954:1999}{1999}]%
        {iso8650}
ISO/IEC 15954:1999 \bibinfo{year}{1999}\natexlab{}.
\newblock \bibinfo{booktitle}{\emph{{ Connection-mode protocol for the
  Application Service Object Association Control Service Element}}}.
\newblock \bibinfo{type}{Standard}. \bibinfo{institution}{International
  Organization for Standardization}, \bibinfo{address}{Geneva, CH}.
\newblock


\bibitem[\protect\citeauthoryear{Iyengar and Swett}{Iyengar and Swett}{2018}]%
        {ietf2018quicloss}
\bibfield{author}{\bibinfo{person}{Janardhan Iyengar} {and}
  \bibinfo{person}{Ian Swett}.} \bibinfo{year}{2018}\natexlab{}.
\newblock \bibinfo{booktitle}{\emph{{QUIC Loss Detection and Congestion
  Control}}}.
\newblock \bibinfo{type}{Internet-Draft} draft-ietf-quic-recovery-14.
  \bibinfo{institution}{Internet Engineering Task Force}.
\newblock
\urldef\tempurl%
\url{https://datatracker.ietf.org/doc/html/draft-ietf-quic-recovery-14}
\showURL{%
\tempurl}
\newblock
\shownote{Work in Progress.}


\bibitem[\protect\citeauthoryear{Jacobson}{Jacobson}{1988}]%
        {jacobson1988cac}
\bibfield{author}{\bibinfo{person}{Van Jacobson}.}
  \bibinfo{year}{1988}\natexlab{}.
\newblock \showarticletitle{Congestion Avoidance and Control}. In
  \bibinfo{booktitle}{\emph{Symposium Proceedings on Communications
  Architectures and Protocols}} \emph{(\bibinfo{series}{SIGCOMM '88})}.
  \bibinfo{publisher}{ACM}, \bibinfo{address}{New York, NY, USA},
  \bibinfo{pages}{314--329}.
\newblock
\showISBNx{0-89791-279-9}
\urldef\tempurl%
\url{https://doi.org/10.1145/52324.52356}
\showDOI{\tempurl}


\bibitem[\protect\citeauthoryear{Jain and Ramakrishnan}{Jain and
  Ramakrishnan}{1988}]%
        {jain1988congavoid}
\bibfield{author}{\bibinfo{person}{Raj Jain} {and} \bibinfo{person}{K.~K.
  Ramakrishnan}.} \bibinfo{year}{1988}\natexlab{}.
\newblock \showarticletitle{Congestion avoidance in computer networks with a
  connectionless network layer: concepts, goals and methodology}. In
  \bibinfo{booktitle}{\emph{[1988] Proceedings. Computer Networking
  Symposium}}. \bibinfo{pages}{134--143}.
\newblock
\urldef\tempurl%
\url{https://doi.org/10.1109/CNS.1988.4990}
\showDOI{\tempurl}


\bibitem[\protect\citeauthoryear{Jannotti, Gifford, Johnson, Kaashoek, and
  O'Toole}{Jannotti et~al\mbox{.}}{2000}]%
        {jannotti2000overcast}
\bibfield{author}{\bibinfo{person}{John Jannotti}, \bibinfo{person}{David~K.
  Gifford}, \bibinfo{person}{Kirk~L. Johnson}, \bibinfo{person}{M.~Frans
  Kaashoek}, {and} \bibinfo{person}{James~W. O'Toole, Jr.}}
  \bibinfo{year}{2000}\natexlab{}.
\newblock \showarticletitle{Overcast: Reliable Multicasting with on Overlay
  Network}. In \bibinfo{booktitle}{\emph{Proceedings of the 4th Conference on
  Symposium on Operating System Design \& Implementation - Volume 4}}
  \emph{(\bibinfo{series}{OSDI'00})}. \bibinfo{publisher}{USENIX Association},
  \bibinfo{address}{Berkeley, CA, USA}, Article \bibinfo{articleno}{14}.
\newblock
\urldef\tempurl%
\url{http://dl.acm.org/citation.cfm?id=1251229.1251243}
\showURL{%
\tempurl}


\bibitem[\protect\citeauthoryear{Joe, Baldine, Dutta, Finn, Ford, Jordan,
  Massey, Matta, Papadopoulos, Reiher, and Rouskas}{Joe et~al\mbox{.}}{2011}]%
        {touch2011druid}
\bibfield{author}{\bibinfo{person}{Touch. Joe}, \bibinfo{person}{Ilia Baldine},
  \bibinfo{person}{Rudra Dutta}, \bibinfo{person}{Gregory~G. Finn},
  \bibinfo{person}{Bryan Ford}, \bibinfo{person}{Scott Jordan},
  \bibinfo{person}{Dan Massey}, \bibinfo{person}{Abraham Matta},
  \bibinfo{person}{Christos Papadopoulos}, \bibinfo{person}{Peter Reiher},
  {and} \bibinfo{person}{George Rouskas}.} \bibinfo{year}{2011}\natexlab{}.
\newblock \showarticletitle{A Dynamic Recursive Unified Internet Design
  (DRUID)}.
\newblock \bibinfo{journal}{\emph{Computer Networks}} \bibinfo{volume}{55},
  \bibinfo{number}{4} (\bibinfo{year}{2011}), \bibinfo{pages}{919--935}.
\newblock
\showISSN{1389-1286}
\urldef\tempurl%
\url{https://doi.org/10.1016/j.comnet.2010.12.016}
\showDOI{\tempurl}
\newblock
\shownote{Special Issue on Architectures and Protocols for the Future
  Internet.}


\bibitem[\protect\citeauthoryear{Jungnickel}{Jungnickel}{2007}]%
        {jungnickel2007gna}
\bibfield{author}{\bibinfo{person}{Dieter Jungnickel}.}
  \bibinfo{year}{2007}\natexlab{}.
\newblock \bibinfo{booktitle}{\emph{Graphs, Networks and Algorithms}
  (\bibinfo{edition}{3rd} ed.)}.
\newblock \bibinfo{publisher}{Springer Publishing Company, Incorporated}.
\newblock
\showISBNx{3540727795, 9783540727798}


\bibitem[\protect\citeauthoryear{Kaashoek, van Renesse, van Staveren, and
  Tanenbaum}{Kaashoek et~al\mbox{.}}{1993}]%
        {kaashoek1993flip}
\bibfield{author}{\bibinfo{person}{M.~Frans Kaashoek}, \bibinfo{person}{Robbert
  van Renesse}, \bibinfo{person}{Hans van Staveren}, {and}
  \bibinfo{person}{Andrew~S. Tanenbaum}.} \bibinfo{year}{1993}\natexlab{}.
\newblock \showarticletitle{FLIP: An Internetwork Protocol for Supporting
  Distributed Systems}.
\newblock \bibinfo{journal}{\emph{ACM Trans. Comput. Syst.}}
  \bibinfo{volume}{11}, \bibinfo{number}{1} (\bibinfo{date}{Feb.}
  \bibinfo{year}{1993}), \bibinfo{pages}{73--106}.
\newblock
\showISSN{0734-2071}
\urldef\tempurl%
\url{https://doi.org/10.1145/151250.151253}
\showDOI{\tempurl}


\bibitem[\protect\citeauthoryear{Kent and Mogul}{Kent and Mogul}{1995}]%
        {kent1995harmful}
\bibfield{author}{\bibinfo{person}{Christopher~A. Kent} {and}
  \bibinfo{person}{Jeffrey~C. Mogul}.} \bibinfo{year}{1995}\natexlab{}.
\newblock \showarticletitle{Fragmentation Considered Harmful}.
\newblock \bibinfo{journal}{\emph{SIGCOMM Comput. Commun. Rev.}}
  \bibinfo{volume}{25}, \bibinfo{number}{1} (\bibinfo{date}{Jan.}
  \bibinfo{year}{1995}), \bibinfo{pages}{75--87}.
\newblock
\showISSN{0146-4833}
\urldef\tempurl%
\url{https://doi.org/10.1145/205447.205456}
\showDOI{\tempurl}


\bibitem[\protect\citeauthoryear{Kernighan and Ritchie}{Kernighan and
  Ritchie}{1988}]%
        {kernighan1988cpl}
\bibfield{author}{\bibinfo{person}{Brian~W. Kernighan} {and}
  \bibinfo{person}{Dennis~M. Ritchie}.} \bibinfo{year}{1988}\natexlab{}.
\newblock \bibinfo{booktitle}{\emph{The C Programming Language}
  (\bibinfo{edition}{2nd} ed.)}.
\newblock \bibinfo{publisher}{Prentice Hall Professional Technical Reference}.
\newblock
\showISBNx{0131103709}


\bibitem[\protect\citeauthoryear{Khademi, Ros, Welzl, Bozakov, Brunstrom,
  Fairhurst, Grinnemo, Hayes, Hurtig, Jones, Mangiante, Tuxen, and
  Weinrank}{Khademi et~al\mbox{.}}{2017}]%
        {khademi2017neat}
\bibfield{author}{\bibinfo{person}{Naeem Khademi}, \bibinfo{person}{David Ros},
  \bibinfo{person}{Michael Welzl}, \bibinfo{person}{Zdravo Bozakov},
  \bibinfo{person}{Anna Brunstrom}, \bibinfo{person}{Gorry Fairhurst},
  \bibinfo{person}{Karl-Johan Grinnemo}, \bibinfo{person}{David Hayes},
  \bibinfo{person}{Per Hurtig}, \bibinfo{person}{Tom Jones},
  \bibinfo{person}{Simone Mangiante}, \bibinfo{person}{Michael Tuxen}, {and}
  \bibinfo{person}{Felix Weinrank}.} \bibinfo{year}{2017}\natexlab{}.
\newblock \showarticletitle{NEAT: A Platform- and Protocol-Independent Internet
  Transport API}.
\newblock \bibinfo{journal}{\emph{IEEE Communications Magazine}}
  \bibinfo{volume}{55}, \bibinfo{number}{6} (\bibinfo{year}{2017}),
  \bibinfo{pages}{46--54}.
\newblock
\showISSN{0163-6804}
\urldef\tempurl%
\url{https://doi.org/10.1109/MCOM.2017.1601052}
\showDOI{\tempurl}


\bibitem[\protect\citeauthoryear{Kleinberg}{Kleinberg}{2007}]%
        {kleinberg2007geometric}
\bibfield{author}{\bibinfo{person}{Robert Kleinberg}.}
  \bibinfo{year}{2007}\natexlab{}.
\newblock \showarticletitle{Geographic Routing Using Hyperbolic Space}. In
  \bibinfo{booktitle}{\emph{IEEE INFOCOM 2007 - 26th IEEE International
  Conference on Computer Communications}}. \bibinfo{pages}{1902--1909}.
\newblock
\showISSN{0743-166X}
\urldef\tempurl%
\url{https://doi.org/10.1109/INFCOM.2007.221}
\showDOI{\tempurl}


\bibitem[\protect\citeauthoryear{Kuhn, Wattenhofer, Zhang, and Zollinger}{Kuhn
  et~al\mbox{.}}{2003}]%
        {kuhn2003geometric}
\bibfield{author}{\bibinfo{person}{Fabian Kuhn}, \bibinfo{person}{Rogert
  Wattenhofer}, \bibinfo{person}{Yan Zhang}, {and} \bibinfo{person}{Aaron
  Zollinger}.} \bibinfo{year}{2003}\natexlab{}.
\newblock \showarticletitle{Geometric Ad-hoc Routing: Of Theory and Practice}.
  In \bibinfo{booktitle}{\emph{Proceedings of the Twenty-second Annual
  Symposium on Principles of Distributed Computing}}
  \emph{(\bibinfo{series}{PODC '03})}. \bibinfo{publisher}{ACM},
  \bibinfo{address}{New York, NY, USA}, \bibinfo{pages}{63--72}.
\newblock
\showISBNx{1-58113-708-7}
\urldef\tempurl%
\url{https://doi.org/10.1145/872035.872044}
\showDOI{\tempurl}


\bibitem[\protect\citeauthoryear{Langley, Riddoch, Wilk, Vicente, Krasic,
  Zhang, Yang, Kouranov, Swett, Iyengar, Bailey, Dorfman, Roskind, Kulik,
  Westin, Tenneti, Shade, Hamilton, Vasiliev, Chang, and Shi}{Langley
  et~al\mbox{.}}{2017}]%
        {langley2017quic}
\bibfield{author}{\bibinfo{person}{Adam Langley}, \bibinfo{person}{Alistair
  Riddoch}, \bibinfo{person}{Alyssa Wilk}, \bibinfo{person}{Antonio Vicente},
  \bibinfo{person}{Charles Krasic}, \bibinfo{person}{Dan Zhang},
  \bibinfo{person}{Fan Yang}, \bibinfo{person}{Fedor Kouranov},
  \bibinfo{person}{Ian Swett}, \bibinfo{person}{Janardhan Iyengar},
  \bibinfo{person}{Jeff Bailey}, \bibinfo{person}{Jeremy Dorfman},
  \bibinfo{person}{Jim Roskind}, \bibinfo{person}{Joanna Kulik},
  \bibinfo{person}{Patrik Westin}, \bibinfo{person}{Raman Tenneti},
  \bibinfo{person}{Robbie Shade}, \bibinfo{person}{Ryan Hamilton},
  \bibinfo{person}{Victor Vasiliev}, \bibinfo{person}{Wan-Teh Chang}, {and}
  \bibinfo{person}{Zhongyi Shi}.} \bibinfo{year}{2017}\natexlab{}.
\newblock \showarticletitle{The QUIC Transport Protocol: Design and
  Internet-Scale Deployment}. In \bibinfo{booktitle}{\emph{Proceedings of the
  Conference of the ACM Special Interest Group on Data Communication}}
  \emph{(\bibinfo{series}{SIGCOMM '17})}. \bibinfo{publisher}{ACM},
  \bibinfo{address}{New York, NY, USA}, \bibinfo{pages}{183--196}.
\newblock
\showISBNx{978-1-4503-4653-5}
\urldef\tempurl%
\url{https://doi.org/10.1145/3098822.3098842}
\showDOI{\tempurl}


\bibitem[\protect\citeauthoryear{Lemon}{Lemon}{2001}]%
        {lemon2001kqueue}
\bibfield{author}{\bibinfo{person}{Jonathan Lemon}.}
  \bibinfo{year}{2001}\natexlab{}.
\newblock \showarticletitle{Kqueue - A Generic and Scalable Event Notification
  Facility}. In \bibinfo{booktitle}{\emph{Proceedings of the FREENIX Track:
  2001 USENIX Annual Technical Conference}}. \bibinfo{publisher}{USENIX
  Association}, \bibinfo{address}{Berkeley, CA, USA},
  \bibinfo{pages}{141--153}.
\newblock
\showISBNx{1-880446-10-3}
\urldef\tempurl%
\url{http://dl.acm.org/citation.cfm?id=647054.715764}
\showURL{%
\tempurl}


\bibitem[\protect\citeauthoryear{Lockwood, McKeown, Watson, Gibb, Hartke,
  Naous, Raghuraman, and Luo}{Lockwood et~al\mbox{.}}{2007}]%
        {lockwood2007netfpga}
\bibfield{author}{\bibinfo{person}{John~W. Lockwood}, \bibinfo{person}{Nick
  McKeown}, \bibinfo{person}{Greg Watson}, \bibinfo{person}{Glen Gibb},
  \bibinfo{person}{Paul Hartke}, \bibinfo{person}{Jad Naous},
  \bibinfo{person}{Ramanan Raghuraman}, {and} \bibinfo{person}{Jianying Luo}.}
  \bibinfo{year}{2007}\natexlab{}.
\newblock \showarticletitle{NetFPGA--An Open Platform for Gigabit-Rate Network
  Switching and Routing}. In \bibinfo{booktitle}{\emph{2007 IEEE International
  Conference on Microelectronic Systems Education (MSE'07)}}.
  \bibinfo{pages}{160--161}.
\newblock
\urldef\tempurl%
\url{https://doi.org/10.1109/MSE.2007.69}
\showDOI{\tempurl}


\bibitem[\protect\citeauthoryear{Loewenstern and Norberg}{Loewenstern and
  Norberg}{2008}]%
        {loewenstern2008bep5}
\bibfield{author}{\bibinfo{person}{Andrew Loewenstern} {and}
  \bibinfo{person}{Arvind Norberg}.} \bibinfo{year}{2008}\natexlab{}.
\newblock \bibinfo{title}{Bittorrent Enhancement Proposal (BEP) 5, Bittorrent
  DHT Protocol}.
\newblock
\newblock
\urldef\tempurl%
\url{http://www.bittorrent.org/beps/bep_0005.html}
\showURL{%
\tempurl}


\bibitem[\protect\citeauthoryear{Maffione}{Maffione}{2015}]%
        {maffione2018rlite}
\bibfield{author}{\bibinfo{person}{Vincenzo Maffione}.}
  \bibinfo{year}{2015}\natexlab{}.
\newblock \bibinfo{title}{rlite: A light RINA implementation}.
\newblock
\newblock
\urldef\tempurl%
\url{https://github.com/rlite/rlite}
\showURL{%
\tempurl}


\bibitem[\protect\citeauthoryear{Maymounkov and Mazieres}{Maymounkov and
  Mazieres}{2002}]%
        {maymounkov2002kademlia}
\bibfield{author}{\bibinfo{person}{Petar Maymounkov} {and}
  \bibinfo{person}{David Mazieres}.} \bibinfo{year}{2002}\natexlab{}.
\newblock \showarticletitle{Kademlia: A peer-to-peer information system based
  on the xor metric}. In \bibinfo{booktitle}{\emph{International Workshop on
  Peer-to-Peer Systems}}. Springer, \bibinfo{pages}{53--65}.
\newblock


\bibitem[\protect\citeauthoryear{McCanne and Jacobson}{McCanne and
  Jacobson}{1993}]%
        {mccanne1993bsd}
\bibfield{author}{\bibinfo{person}{Steven McCanne} {and} \bibinfo{person}{Van
  Jacobson}.} \bibinfo{year}{1993}\natexlab{}.
\newblock \showarticletitle{The BSD Packet Filter: A New Architecture for
  User-level Packet Capture.}. In \bibinfo{booktitle}{\emph{USENIX winter}},
  Vol.~\bibinfo{volume}{46}.
\newblock


\bibitem[\protect\citeauthoryear{McIlroy, Pinson, and Tague}{McIlroy
  et~al\mbox{.}}{1978}]%
        {mcilroy1978unix}
\bibfield{author}{\bibinfo{person}{Malcolm~Douglas McIlroy},
  \bibinfo{person}{Elliot~N. Pinson}, {and} \bibinfo{person}{Berk~A. Tague}.}
  \bibinfo{year}{1978}\natexlab{}.
\newblock \showarticletitle{UNIX Time-Sharing System: Foreword}.
\newblock \bibinfo{journal}{\emph{Bell Sys. Tech. J.}} \bibinfo{volume}{57},
  \bibinfo{number}{6} (\bibinfo{year}{1978}), \bibinfo{pages}{1899--1904}.
\newblock


\bibitem[\protect\citeauthoryear{McKenzie}{McKenzie}{2011}]%
        {mckenzie2011inwg}
\bibfield{author}{\bibinfo{person}{Alexander McKenzie}.}
  \bibinfo{year}{2011}\natexlab{}.
\newblock \showarticletitle{INWG and the Conception of the Internet: An
  Eyewitness Account}.
\newblock \bibinfo{journal}{\emph{IEEE Annals of the History of Computing}}
  \bibinfo{volume}{33}, \bibinfo{number}{1} (\bibinfo{date}{Jan.}
  \bibinfo{year}{2011}), \bibinfo{pages}{66--71}.
\newblock
\showISSN{1058-6180}
\urldef\tempurl%
\url{https://doi.org/10.1109/MAHC.2011.9}
\showDOI{\tempurl}


\bibitem[\protect\citeauthoryear{Meyer and Lewis}{Meyer and Lewis}{2013}]%
        {meyer2013locator}
\bibfield{author}{\bibinfo{person}{David Meyer} {and} \bibinfo{person}{Darrel
  Lewis}.} \bibinfo{year}{2013}\natexlab{}.
\newblock \showarticletitle{The locator/Id separation protocol (LISP)}.
\newblock \bibinfo{journal}{\emph{IETF RFC 6830}} (\bibinfo{year}{2013}).
\newblock


\bibitem[\protect\citeauthoryear{Mockapetris}{Mockapetris}{1987a}]%
        {rfc1034}
\bibfield{author}{\bibinfo{person}{Paul~V. Mockapetris}.}
  \bibinfo{year}{1987}\natexlab{a}.
\newblock \bibinfo{title}{{Domain names - concepts and facilities}}.
\newblock \bibinfo{howpublished}{RFC 1034}.
\newblock
\urldef\tempurl%
\url{https://doi.org/10.17487/RFC1034}
\showDOI{\tempurl}


\bibitem[\protect\citeauthoryear{Mockapetris}{Mockapetris}{1987b}]%
        {rfc1035}
\bibfield{author}{\bibinfo{person}{Paul~V. Mockapetris}.}
  \bibinfo{year}{1987}\natexlab{b}.
\newblock \bibinfo{title}{{Domain names - implementation and specification}}.
\newblock \bibinfo{howpublished}{RFC 1035}.
\newblock
\urldef\tempurl%
\url{https://doi.org/10.17487/RFC1035}
\showDOI{\tempurl}


\bibitem[\protect\citeauthoryear{Moy}{Moy}{1998}]%
        {rfc2328}
\bibfield{author}{\bibinfo{person}{John Moy}.} \bibinfo{year}{1998}\natexlab{}.
\newblock \bibinfo{title}{{OSPF Version 2}}.
\newblock \bibinfo{howpublished}{RFC 2328}.
\newblock
\urldef\tempurl%
\url{https://doi.org/10.17487/RFC2328}
\showDOI{\tempurl}


\bibitem[\protect\citeauthoryear{Nagle}{Nagle}{1984}]%
        {nagle1984cc}
\bibfield{author}{\bibinfo{person}{John Nagle}.}
  \bibinfo{year}{1984}\natexlab{}.
\newblock \showarticletitle{Congestion Control in IP/TCP Internetworks}.
\newblock \bibinfo{journal}{\emph{SIGCOMM Comput. Commun. Rev.}}
  \bibinfo{volume}{14}, \bibinfo{number}{4} (\bibinfo{date}{Oct.}
  \bibinfo{year}{1984}), \bibinfo{pages}{11--17}.
\newblock
\showISSN{0146-4833}
\urldef\tempurl%
\url{https://doi.org/10.1145/1024908.1024910}
\showDOI{\tempurl}


\bibitem[\protect\citeauthoryear{Needham}{Needham}{1993}]%
        {needham1993names}
\bibfield{author}{\bibinfo{person}{Roger~M. Needham}.}
  \bibinfo{year}{1993}\natexlab{}.
\newblock \showarticletitle{Distributed Systems (2nd Ed.)}.
\newblock \bibinfo{publisher}{ACM Press/Addison-Wesley Publishing Co.},
  \bibinfo{address}{New York, NY, USA}, Chapter Names, \bibinfo{pages}{315 --
  327}.
\newblock
\showISBNx{0-201-62427-3}


\bibitem[\protect\citeauthoryear{Open~Group}{Open~Group}{2018}]%
        {opengroup2017posix}
\bibfield{author}{\bibinfo{person}{The Open~Group}.}
  \bibinfo{year}{2018}\natexlab{}.
\newblock \bibinfo{booktitle}{\emph{IEEE Standard for Information
  Technology--Portable Operating System Interface (POSIX\textregistered) Base
  Specifications, Issue 7}}.
\newblock \bibinfo{type}{{T}echnical {R}eport}. \bibinfo{institution}{IEEE}.
  \bibinfo{pages}{1--3951} pages.
\newblock
\urldef\tempurl%
\url{https://doi.org/10.1109/IEEESTD.2018.8277153}
\showDOI{\tempurl}


\bibitem[\protect\citeauthoryear{Ousterhout, Cherenson, Douglis, Nelson, and
  Welch}{Ousterhout et~al\mbox{.}}{1988}]%
        {ousterhout1988sprite}
\bibfield{author}{\bibinfo{person}{John~K. Ousterhout},
  \bibinfo{person}{Andrew~R. Cherenson}, \bibinfo{person}{Frederick Douglis},
  \bibinfo{person}{Michael~N. Nelson}, {and} \bibinfo{person}{Brent~B. Welch}.}
  \bibinfo{year}{1988}\natexlab{}.
\newblock \showarticletitle{The Sprite Network Operating System}.
\newblock \bibinfo{journal}{\emph{Computer}} \bibinfo{volume}{21},
  \bibinfo{number}{2} (\bibinfo{date}{Feb.} \bibinfo{year}{1988}),
  \bibinfo{pages}{23--36}.
\newblock
\showISSN{0018-9162}
\urldef\tempurl%
\url{https://doi.org/10.1109/2.16}
\showDOI{\tempurl}


\bibitem[\protect\citeauthoryear{Papadimitriou and Ratajczak}{Papadimitriou and
  Ratajczak}{2005}]%
        {papadimitriou2005geometric}
\bibfield{author}{\bibinfo{person}{Christos~H. Papadimitriou} {and}
  \bibinfo{person}{David Ratajczak}.} \bibinfo{year}{2005}\natexlab{}.
\newblock \showarticletitle{On a conjecture related to geometric routing}.
\newblock \bibinfo{journal}{\emph{Theoretical Computer Science}}
  \bibinfo{volume}{344}, \bibinfo{number}{1} (\bibinfo{year}{2005}),
  \bibinfo{pages}{3--14}.
\newblock
\showISSN{0304-3975}
\urldef\tempurl%
\url{https://doi.org/10.1016/j.tcs.2005.06.022}
\showDOI{\tempurl}
\newblock
\shownote{Algorithmic Aspects of Wireless Sensor Networks.}


\bibitem[\protect\citeauthoryear{Pauly, Trammell, Brunstrom, Fairhurst,
  Perkins, Tiesel, and Wood}{Pauly et~al\mbox{.}}{2018}]%
        {ietf2018tapsarch}
\bibfield{author}{\bibinfo{person}{Tommy Pauly}, \bibinfo{person}{Brian
  Trammell}, \bibinfo{person}{Anna Brunstrom}, \bibinfo{person}{Gorry
  Fairhurst}, \bibinfo{person}{Colin Perkins}, \bibinfo{person}{Phillip~S.
  Tiesel}, {and} \bibinfo{person}{Christopher~A. Wood}.}
  \bibinfo{year}{2018}\natexlab{}.
\newblock \bibinfo{booktitle}{\emph{{An Architecture for Transport Services}}}.
\newblock \bibinfo{type}{Internet-Draft} draft-ietf-taps-arch-01.
  \bibinfo{institution}{Internet Engineering Task Force}.
\newblock
\urldef\tempurl%
\url{https://datatracker.ietf.org/doc/html/draft-ietf-taps-arch-01}
\showURL{%
\tempurl}
\newblock
\shownote{Work in Progress.}


\bibitem[\protect\citeauthoryear{Perkins, Alpert, and Woolf}{Perkins
  et~al\mbox{.}}{1997}]%
        {perkins1997mobile}
\bibfield{author}{\bibinfo{person}{Charles~E Perkins},
  \bibinfo{person}{Sherman~R Alpert}, {and} \bibinfo{person}{Bobby Woolf}.}
  \bibinfo{year}{1997}\natexlab{}.
\newblock \bibinfo{booktitle}{\emph{Mobile IP; Design Principles and
  Practices}}.
\newblock \bibinfo{publisher}{Addison-Wesley Longman Publishing Co., Inc.}
\newblock


\bibitem[\protect\citeauthoryear{Perlman}{Perlman}{1985}]%
        {perlman1985stp}
\bibfield{author}{\bibinfo{person}{Radia Perlman}.}
  \bibinfo{year}{1985}\natexlab{}.
\newblock \showarticletitle{An Algorithm for Distributed Computation of a
  Spanningtree in an Extended LAN}.
\newblock \bibinfo{journal}{\emph{SIGCOMM Comput. Commun. Rev.}}
  \bibinfo{volume}{15}, \bibinfo{number}{4} (\bibinfo{date}{Sept.}
  \bibinfo{year}{1985}), \bibinfo{pages}{44--53}.
\newblock
\showISSN{0146-4833}
\urldef\tempurl%
\url{https://doi.org/10.1145/318951.319004}
\showDOI{\tempurl}


\bibitem[\protect\citeauthoryear{Pike}{Pike}{2000}]%
        {pike2000sri}
\bibfield{author}{\bibinfo{person}{Rob Pike}.} \bibinfo{year}{2000}\natexlab{}.
\newblock \bibinfo{title}{Systems Software Research is Irrelevant. Invited talk
  at University of Utah}.
\newblock
\newblock


\bibitem[\protect\citeauthoryear{Plummer}{Plummer}{1982}]%
        {rfc826}
\bibfield{author}{\bibinfo{person}{David Plummer}.}
  \bibinfo{year}{1982}\natexlab{}.
\newblock \bibinfo{title}{{An Ethernet Address Resolution Protocol}}.
\newblock \bibinfo{howpublished}{RFC 826}.
\newblock
\urldef\tempurl%
\url{https://doi.org/10.17487/RFC0826}
\showDOI{\tempurl}


\bibitem[\protect\citeauthoryear{Postel}{Postel}{1980}]%
        {rfc768}
\bibfield{author}{\bibinfo{person}{Jonathan~B. Postel}.}
  \bibinfo{year}{1980}\natexlab{}.
\newblock \bibinfo{title}{{User Datagram Protocol}}.
\newblock \bibinfo{howpublished}{RFC 768}.
\newblock
\urldef\tempurl%
\url{https://doi.org/10.17487/RFC0768}
\showDOI{\tempurl}


\bibitem[\protect\citeauthoryear{Postel}{Postel}{1981a}]%
        {rfc791}
\bibfield{author}{\bibinfo{person}{Jonathan~B. Postel}.}
  \bibinfo{year}{1981}\natexlab{a}.
\newblock \bibinfo{title}{{Internet Protocol}}.
\newblock \bibinfo{howpublished}{RFC 791}.
\newblock
\urldef\tempurl%
\url{https://doi.org/10.17487/RFC0791}
\showDOI{\tempurl}


\bibitem[\protect\citeauthoryear{Postel}{Postel}{1981b}]%
        {rfc793}
\bibfield{author}{\bibinfo{person}{Jonathan~B. Postel}.}
  \bibinfo{year}{1981}\natexlab{b}.
\newblock \bibinfo{title}{{Transmission Control Protocol}}.
\newblock \bibinfo{howpublished}{RFC 793}.
\newblock
\urldef\tempurl%
\url{https://doi.org/10.17487/RFC0793}
\showDOI{\tempurl}


\bibitem[\protect\citeauthoryear{Postel and Reynolds}{Postel and
  Reynolds}{1994}]%
        {rfc1700}
\bibfield{author}{\bibinfo{person}{Jonathan~B. Postel} {and}
  \bibinfo{person}{Joyce~K. Reynolds}.} \bibinfo{year}{1994}\natexlab{}.
\newblock \bibinfo{title}{{Assigned Numbers}}.
\newblock \bibinfo{howpublished}{RFC 1700}.
\newblock
\urldef\tempurl%
\url{https://doi.org/10.17487/RFC1700}
\showDOI{\tempurl}


\bibitem[\protect\citeauthoryear{Postel, Sunshine, and Cohen}{Postel
  et~al\mbox{.}}{1981}]%
        {postel1981arpa}
\bibfield{author}{\bibinfo{person}{Jonathan~B. Postel},
  \bibinfo{person}{Carl~A. Sunshine}, {and} \bibinfo{person}{Danny Cohen}.}
  \bibinfo{year}{1981}\natexlab{}.
\newblock \showarticletitle{The ARPA internet protocol}.
\newblock \bibinfo{journal}{\emph{Computer Networks (1976)}}
  \bibinfo{volume}{5}, \bibinfo{number}{4} (\bibinfo{year}{1981}),
  \bibinfo{pages}{261--271}.
\newblock
\showISSN{0376-5075}
\urldef\tempurl%
\url{https://doi.org/10.1016/0376-5075(81)90003-9}
\showDOI{\tempurl}


\bibitem[\protect\citeauthoryear{Pouzin}{Pouzin}{1974}]%
        {pouzin1974catenet}
\bibfield{author}{\bibinfo{person}{Louis Pouzin}.}
  \bibinfo{year}{1974}\natexlab{}.
\newblock \showarticletitle{{A proposal for interconnecting packet switching
  networks}}. In \bibinfo{booktitle}{\emph{Proceedings of Eurocomp}}.
  \bibinfo{publisher}{IEEE}, \bibinfo{pages}{1023--1036}.
\newblock
\showISSN{0090-6778}
\urldef\tempurl%
\url{https://doi.org/10.1109/TCOM.1980.1094702}
\showDOI{\tempurl}


\bibitem[\protect\citeauthoryear{Presotto and Winterbottom}{Presotto and
  Winterbottom}{1995}]%
        {presotto1995il}
\bibfield{author}{\bibinfo{person}{Dave Presotto} {and} \bibinfo{person}{Phil
  Winterbottom}.} \bibinfo{year}{1995}\natexlab{}.
\newblock \bibinfo{title}{The IL protocol}.
\newblock
\newblock
\urldef\tempurl%
\url{http://doc.cat-v.org/plan_9/4th_edition/papers/il/}
\showURL{%
\tempurl}


\bibitem[\protect\citeauthoryear{Quinn and Almeroth}{Quinn and
  Almeroth}{2001}]%
        {rfc3170}
\bibfield{author}{\bibinfo{person}{Bob Quinn} {and} \bibinfo{person}{Kevin~C.
  Almeroth}.} \bibinfo{year}{2001}\natexlab{}.
\newblock \bibinfo{title}{{IP Multicast Applications: Challenges and
  Solutions}}.
\newblock \bibinfo{howpublished}{RFC 3170}.
\newblock
\urldef\tempurl%
\url{https://doi.org/10.17487/RFC3170}
\showDOI{\tempurl}


\bibitem[\protect\citeauthoryear{Rekhter, Hares, and Li}{Rekhter
  et~al\mbox{.}}{2006}]%
        {rfc4271}
\bibfield{author}{\bibinfo{person}{Yakov Rekhter}, \bibinfo{person}{Susan
  Hares}, {and} \bibinfo{person}{Tony Li}.} \bibinfo{year}{2006}\natexlab{}.
\newblock \bibinfo{title}{{A Border Gateway Protocol 4 (BGP-4)}}.
\newblock \bibinfo{howpublished}{RFC 4271}.
\newblock
\urldef\tempurl%
\url{https://doi.org/10.17487/RFC4271}
\showDOI{\tempurl}


\bibitem[\protect\citeauthoryear{Rizzo}{Rizzo}{2012}]%
        {rizzo2012netmap}
\bibfield{author}{\bibinfo{person}{Luigi Rizzo}.}
  \bibinfo{year}{2012}\natexlab{}.
\newblock \showarticletitle{Netmap: a novel framework for fast packet I/O}. In
  \bibinfo{booktitle}{\emph{21st USENIX Security Symposium (USENIX Security
  12)}}. \bibinfo{pages}{101--112}.
\newblock


\bibitem[\protect\citeauthoryear{R{\"u}ngeler, T{\"u}xen, and
  Rathgeb}{R{\"u}ngeler et~al\mbox{.}}{2009}]%
        {rungeler2009ccsctp}
\bibfield{author}{\bibinfo{person}{Irene R{\"u}ngeler},
  \bibinfo{person}{Michael T{\"u}xen}, {and} \bibinfo{person}{Erwin~P.
  Rathgeb}.} \bibinfo{year}{2009}\natexlab{}.
\newblock \showarticletitle{Congestion and Flow Control in the Context of the
  Message-Oriented Protocol SCTP}. In \bibinfo{booktitle}{\emph{NETWORKING
  2009}}, \bibfield{editor}{\bibinfo{person}{Luigi Fratta},
  \bibinfo{person}{Henning Schulzrinne}, \bibinfo{person}{Yutaka Takahashi},
  {and} \bibinfo{person}{Otto Spaniol}} (Eds.). \bibinfo{publisher}{Springer},
  \bibinfo{address}{Berlin, Heidelberg}, \bibinfo{pages}{468--481}.
\newblock
\showISBNx{978-3-642-01399-7}


\bibitem[\protect\citeauthoryear{Russell}{Russell}{2013}]%
        {russell2013internetwasnt}
\bibfield{author}{\bibinfo{person}{Andrew~L. Russell}.}
  \bibinfo{year}{2013}\natexlab{}.
\newblock \showarticletitle{The internet that wasn't}.
\newblock \bibinfo{journal}{\emph{IEEE Spectrum}} \bibinfo{volume}{50},
  \bibinfo{number}{8} (\bibinfo{date}{Aug.} \bibinfo{year}{2013}),
  \bibinfo{pages}{39--43}.
\newblock
\showISSN{0018-9235}
\urldef\tempurl%
\url{https://doi.org/10.1109/MSPEC.2013.6565559}
\showDOI{\tempurl}


\bibitem[\protect\citeauthoryear{Saltzer}{Saltzer}{1978}]%
        {saltzer1978naming}
\bibfield{author}{\bibinfo{person}{Jerome~H. Saltzer}.}
  \bibinfo{year}{1978}\natexlab{}.
\newblock \showarticletitle{Naming and binding of objects}.
\newblock In \bibinfo{booktitle}{\emph{Bayer R., Graham R.M., Seegmüller G.
  (eds), Operating Systems, Lecture Notes in Computer Science (LNCS), vol.
  60}}. \bibinfo{publisher}{Springer}, \bibinfo{address}{Berlin, Heidelberg},
  \bibinfo{pages}{99--208}.
\newblock
\showISBNx{978-3-540-08755-7}
\urldef\tempurl%
\url{https://doi.org/10.1007/3-540-08755-9_4}
\showDOI{\tempurl}


\bibitem[\protect\citeauthoryear{Shoch}{Shoch}{1978}]%
        {shoch1978naming}
\bibfield{author}{\bibinfo{person}{John~F. Shoch}.}
  \bibinfo{year}{1978}\natexlab{}.
\newblock \showarticletitle{Inter-network naming, addressing, and routing}. In
  \bibinfo{booktitle}{\emph{Proceedings of IEEE Computer Conference, COMPCON}}.
  \bibinfo{publisher}{IEEE}, \bibinfo{pages}{72--79}.
\newblock


\bibitem[\protect\citeauthoryear{Sriraman, Butler, McDaniel, and
  Raghavan}{Sriraman et~al\mbox{.}}{2007}]%
        {sriraman2007ipaddr}
\bibfield{author}{\bibinfo{person}{Anousha Sriraman}, \bibinfo{person}{Kevin
  R.~B. Butler}, \bibinfo{person}{Patrick~D. McDaniel}, {and}
  \bibinfo{person}{Padma Raghavan}.} \bibinfo{year}{2007}\natexlab{}.
\newblock \showarticletitle{Analysis of the IPv4 Address Space Delegation
  Structure}. In \bibinfo{booktitle}{\emph{2007 12th IEEE Symposium on
  Computers and Communications}}. \bibinfo{pages}{501--508}.
\newblock
\showISSN{1530-1346}
\urldef\tempurl%
\url{https://doi.org/10.1109/ISCC.2007.4381538}
\showDOI{\tempurl}


\bibitem[\protect\citeauthoryear{Staessens and Vrijders}{Staessens and
  Vrijders}{2017}]%
        {staessens2017ouroboros}
\bibfield{author}{\bibinfo{person}{Dimitri Staessens} {and}
  \bibinfo{person}{Sander Vrijders}.} \bibinfo{year}{2017}\natexlab{}.
\newblock \bibinfo{title}{Ouroboros: a simple decentralized IPC subsystem}.
\newblock
\newblock
\urldef\tempurl%
\url{https://ouroboros.ilabt.imec.be}
\showURL{%
\tempurl}


\bibitem[\protect\citeauthoryear{Tanenbaum and Bos}{Tanenbaum and Bos}{2014}]%
        {tanenbaum2014mos}
\bibfield{author}{\bibinfo{person}{Andrew~S. Tanenbaum} {and}
  \bibinfo{person}{Herbert Bos}.} \bibinfo{year}{2014}\natexlab{}.
\newblock \bibinfo{booktitle}{\emph{Modern Operating Systems}
  (\bibinfo{edition}{4th} ed.)}.
\newblock \bibinfo{publisher}{Prentice Hall Press}, \bibinfo{address}{Upper
  Saddle River, NJ, USA}.
\newblock
\showISBNx{013359162X, 9780133591620}


\bibitem[\protect\citeauthoryear{Tanenbaum and Wetherall}{Tanenbaum and
  Wetherall}{2010}]%
        {tanenbaum2010cn}
\bibfield{author}{\bibinfo{person}{Andrew~S. Tanenbaum} {and}
  \bibinfo{person}{David~J. Wetherall}.} \bibinfo{year}{2010}\natexlab{}.
\newblock \bibinfo{booktitle}{\emph{Computer Networks} (\bibinfo{edition}{5th}
  ed.)}.
\newblock \bibinfo{publisher}{Prentice Hall Press}, \bibinfo{address}{Upper
  Saddle River, NJ, USA}.
\newblock
\showISBNx{0132126958, 9780132126953}


\bibitem[\protect\citeauthoryear{Touch and Pingali}{Touch and Pingali}{2008}]%
        {touch2008rnametaprotocol}
\bibfield{author}{\bibinfo{person}{Joseph~D. Touch} {and}
  \bibinfo{person}{Venkata~K. Pingali}.} \bibinfo{year}{2008}\natexlab{}.
\newblock \showarticletitle{The RNA Metaprotocol}. In
  \bibinfo{booktitle}{\emph{2008 Proceedings of 17th International Conference
  on Computer Communications and Networks}}. \bibinfo{pages}{1--6}.
\newblock
\showISSN{1095-2055}
\urldef\tempurl%
\url{https://doi.org/10.1109/ICCCN.2008.ECP.46}
\showDOI{\tempurl}


\bibitem[\protect\citeauthoryear{Touch, Wang, and Pingali}{Touch
  et~al\mbox{.}}{2006}]%
        {touch2006rna}
\bibfield{author}{\bibinfo{person}{Joseph~D. Touch}, \bibinfo{person}{Yu-Shun
  Wang}, {and} \bibinfo{person}{Venkata Pingali}.}
  \bibinfo{year}{2006}\natexlab{}.
\newblock \bibinfo{booktitle}{\emph{A Recursive Network Architecture}}.
\newblock \bibinfo{type}{{T}echnical {R}eport}. \bibinfo{institution}{USC/ISI}.
\newblock


\bibitem[\protect\citeauthoryear{Trouva, Grasa, Day, Matta, Chitkushev, Phelan,
  Ponce De~Leon, and Bunch}{Trouva et~al\mbox{.}}{2011}]%
        {trouva2011rina}
\bibfield{author}{\bibinfo{person}{Eleni Trouva}, \bibinfo{person}{Eduard
  Grasa}, \bibinfo{person}{John Day}, \bibinfo{person}{Ibrahim Matta},
  \bibinfo{person}{Lubomir~T. Chitkushev}, \bibinfo{person}{Patrick Phelan},
  \bibinfo{person}{Miguel Ponce De~Leon}, {and} \bibinfo{person}{Steve Bunch}.}
  \bibinfo{year}{2011}\natexlab{}.
\newblock \showarticletitle{Is the Internet an unfinished demo? Meet RINA!}. In
  \bibinfo{booktitle}{\emph{TERENA Networking Conference (TNC) 2011}}.
  \bibinfo{address}{Prague}, \bibinfo{pages}{12}.
\newblock


\bibitem[\protect\citeauthoryear{Turing}{Turing}{1936}]%
        {turing1936a}
\bibfield{author}{\bibinfo{person}{Alan~M. Turing}.}
  \bibinfo{year}{1936}\natexlab{}.
\newblock \showarticletitle{On Computable Numbers, with an Application to the
  {E}ntscheidungsproblem}.
\newblock \bibinfo{journal}{\emph{Proceedings of the London Mathematical
  Society}} \bibinfo{volume}{2}, \bibinfo{number}{42} (\bibinfo{year}{1936}),
  \bibinfo{pages}{230--265}.
\newblock
\urldef\tempurl%
\url{http://www.cs.helsinki.fi/u/gionis/cc05/OnComputableNumbers.pdf}
\showURL{%
\tempurl}


\bibitem[\protect\citeauthoryear{Vrijders, Staessens, Capitani, and
  Maffione}{Vrijders et~al\mbox{.}}{2018}]%
        {vrijders2018rumba}
\bibfield{author}{\bibinfo{person}{Sander Vrijders}, \bibinfo{person}{Dimitri
  Staessens}, \bibinfo{person}{Marco Capitani}, {and} \bibinfo{person}{Vincenzo
  Maffione}.} \bibinfo{year}{2018}\natexlab{}.
\newblock \showarticletitle{Rumba: A python framework for automating
  large-scale Recursive Internet Experiments on GENI and FIRE+}. In
  \bibinfo{booktitle}{\emph{IEEE INFOCOM 2018 - IEEE Conference on Computer
  Communications Workshops (INFOCOM WKSHPS)}}. \bibinfo{publisher}{IEEE},
  \bibinfo{address}{Honolulu, HI}, \bibinfo{pages}{324--329}.
\newblock
\urldef\tempurl%
\url{https://doi.org/10.1109/INFCOMW.2018.8406981}
\showDOI{\tempurl}


\bibitem[\protect\citeauthoryear{Vrijders, Staessens, Colle, Salvestrini,
  Grasa, Tarzan, and Bergesio}{Vrijders et~al\mbox{.}}{2014}]%
        {vrijders2014irati}
\bibfield{author}{\bibinfo{person}{Sander Vrijders}, \bibinfo{person}{Dimitri
  Staessens}, \bibinfo{person}{Didier Colle}, \bibinfo{person}{Francesco
  Salvestrini}, \bibinfo{person}{Eduard Grasa}, \bibinfo{person}{Miquel
  Tarzan}, {and} \bibinfo{person}{Leonardo Bergesio}.}
  \bibinfo{year}{2014}\natexlab{}.
\newblock \showarticletitle{Prototyping the recursive internet architecture:
  the IRATI project approach}.
\newblock \bibinfo{journal}{\emph{IEEE Network}} \bibinfo{volume}{28},
  \bibinfo{number}{2} (\bibinfo{date}{March} \bibinfo{year}{2014}),
  \bibinfo{pages}{20--25}.
\newblock
\showISSN{0890-8044}
\urldef\tempurl%
\url{https://doi.org/10.1109/MNET.2014.6786609}
\showDOI{\tempurl}


\bibitem[\protect\citeauthoryear{Vrijders, Trouva, Day, Grasa, Staessens,
  Colle, Pickavet, and Chitkushev}{Vrijders et~al\mbox{.}}{2013}]%
        {vrijders2013migr}
\bibfield{author}{\bibinfo{person}{Sander Vrijders}, \bibinfo{person}{Eleni
  Trouva}, \bibinfo{person}{John Day}, \bibinfo{person}{Eduard Grasa},
  \bibinfo{person}{Dimitri Staessens}, \bibinfo{person}{Didier Colle},
  \bibinfo{person}{Mario Pickavet}, {and} \bibinfo{person}{Lou Chitkushev}.}
  \bibinfo{year}{2013}\natexlab{}.
\newblock \showarticletitle{Unreliable inter process communication in Ethernet:
  migrating to RINA with the shim DIF}. In \bibinfo{booktitle}{\emph{2013 5th
  International Congress on Ultra Modern Telecommunications and Control Systems
  and Workshops (ICUMT)}}. \bibinfo{pages}{215--221}.
\newblock
\showISSN{2157-023X}
\urldef\tempurl%
\url{https://doi.org/10.1109/ICUMT.2013.6798429}
\showDOI{\tempurl}


\bibitem[\protect\citeauthoryear{Watson}{Watson}{1981a}]%
        {watson1981deltat}
\bibfield{author}{\bibinfo{person}{Richard~W. Watson}.}
  \bibinfo{year}{1981}\natexlab{a}.
\newblock \bibinfo{booktitle}{\emph{{Delta-t protocol specification}}}.
\newblock \bibinfo{type}{{T}echnical {R}eport}. \bibinfo{institution}{Lawrence
  Livermore Laboratories}.
\newblock


\bibitem[\protect\citeauthoryear{Watson}{Watson}{1981b}]%
        {watson1981timerbased}
\bibfield{author}{\bibinfo{person}{Richard~W. Watson}.}
  \bibinfo{year}{1981}\natexlab{b}.
\newblock \showarticletitle{Timer-based mechanisms in reliable transport
  protocol connection management}.
\newblock \bibinfo{journal}{\emph{Computer Networks (1976)}}
  \bibinfo{volume}{5}, \bibinfo{number}{1} (\bibinfo{year}{1981}),
  \bibinfo{pages}{47--56}.
\newblock
\showISSN{0376-5075}
\urldef\tempurl%
\url{https://doi.org/10.1016/0376-5075(81)90031-3}
\showDOI{\tempurl}


\bibitem[\protect\citeauthoryear{Welzl}{Welzl}{2005}]%
        {welzl2005cc}
\bibfield{author}{\bibinfo{person}{Michael Welzl}.}
  \bibinfo{year}{2005}\natexlab{}.
\newblock \bibinfo{booktitle}{\emph{Network congestion control : managing
  internet traffic}}.
\newblock \bibinfo{publisher}{John Wiley}, \bibinfo{address}{Chichester}.
\newblock


\bibitem[\protect\citeauthoryear{Williams}{Williams}{2017}]%
        {williams2017interposer}
\bibfield{author}{\bibinfo{person}{Matt Williams}.}
  \bibinfo{year}{2017}\natexlab{}.
\newblock \bibinfo{title}{Interposer between BSD Sockets/POSIX and RINA}.
\newblock
\newblock
\urldef\tempurl%
\url{https://github.com/matt-williams/rina-interposer}
\showURL{%
\tempurl}


\bibitem[\protect\citeauthoryear{Zhang, Deering, Estrin, Shenker, and
  Zappala}{Zhang et~al\mbox{.}}{1993}]%
        {zhang1993rsvp}
\bibfield{author}{\bibinfo{person}{Lixia Zhang}, \bibinfo{person}{Stephen
  Deering}, \bibinfo{person}{Deborah Estrin}, \bibinfo{person}{Scott Shenker},
  {and} \bibinfo{person}{Daniel Zappala}.} \bibinfo{year}{1993}\natexlab{}.
\newblock \showarticletitle{RSVP: a new resource ReSerVation Protocol}.
\newblock \bibinfo{journal}{\emph{IEEE Network}} \bibinfo{volume}{7},
  \bibinfo{number}{5} (\bibinfo{date}{Sept} \bibinfo{year}{1993}),
  \bibinfo{pages}{8--18}.
\newblock
\showISSN{0890-8044}
\urldef\tempurl%
\url{https://doi.org/10.1109/65.238150}
\showDOI{\tempurl}


\bibitem[\protect\citeauthoryear{Zhang, Fall, and Meyer}{Zhang
  et~al\mbox{.}}{2007}]%
        {rfc4984}
\bibfield{author}{\bibinfo{person}{Lixia Zhang}, \bibinfo{person}{Kevin Fall},
  {and} \bibinfo{person}{David Meyer}.} \bibinfo{year}{2007}\natexlab{}.
\newblock \bibinfo{title}{{Report from the IAB Workshop on Routing and
  Addressing}}.
\newblock \bibinfo{howpublished}{RFC 4984}.
\newblock
\urldef\tempurl%
\url{https://doi.org/10.17487/RFC4984}
\showDOI{\tempurl}


\bibitem[\protect\citeauthoryear{Zimmermann}{Zimmermann}{1980}]%
        {zimmermann1980osi}
\bibfield{author}{\bibinfo{person}{Hubert Zimmermann}.}
  \bibinfo{year}{1980}\natexlab{}.
\newblock \showarticletitle{{OSI reference model: The ISO model of architecture
  for open systems interconnection}}.
\newblock \bibinfo{journal}{\emph{IEEE Transactions on communications}}
  \bibinfo{volume}{28}, \bibinfo{number}{4} (\bibinfo{date}{April}
  \bibinfo{year}{1980}), \bibinfo{pages}{425--432}.
\newblock


\end{thebibliography}

\end{document}